\documentclass[aps,amsfonts,pra,twocolumn,showpacs]{revtex4-1}
\usepackage{epsfig,amsmath,amssymb,bm,epsf,graphics,psfrag,verbatim,subfigure}

\def\beq{\begin{equation}}
\def\eeq{\end{equation}}
\def\bea{\begin{eqnarray}}
\def\eea{\end{eqnarray}}

\def\mass{M}

\def\sbK{K_{{\mathrm{0}}}}

\def\bc{2\pi(m + n) = \sbK R}
\def\ncomponent{{p}}
\def\mydim{d}
\def\contact{\overline{U}}
\def\fullHam{{\cal H}}
\def\atomgap{{\omega_{A}}}
\def\atomHam{H_{\mathrm{at}}}
\def\rgscale{B}
\def\newrgscale{\mathcal{B}}
\def\rgdenom{\overline{r} + \newrgscale^2}

\begin{document}
%-------------

%-------------
\title{% 
Atom-light crystallization of BECs in multimode cavities: 
Nonequilibrium classical and quantum phase transitions, 
emergent lattices, supersolidity, and frustration 
%%%
}

%-------------

\author{Sarang Gopalakrishnan$^{1,2}$}

\author{Benjamin L.~Lev$^{1}$}

\author{Paul M.~Goldbart$^{1,2}$}

\affiliation{% 
$^1$Department of Physics and 
$^2$Institute for Condensed Matter Theory,
University of Illinois at Urbana-Champaign,
1110 West Green Street, Urbana, Illinois 61801, U.S.A.}

%-------------
 \date{July 5, 2010} 

\begin{abstract}
The self-organization of a Bose-Einstein condensate in a transversely pumped optical cavity is a process akin to crystallization: when pumped by a laser of sufficient intensity, the coupled matter and light fields evolve, spontaneously, into a spatially modulated pattern, or crystal, whose lattice structure is dictated by the geometry of the cavity.
In cavities having multiple degenerate modes, the quasi-continuum of possible lattice arrangements, and the continuous symmetry breaking associated with the adoption of a particular lattice arrangement, give rise to phenomena such as phonons, defects, and frustration, which have hitherto been unexplored in ultracold atomic settings involving neutral atoms. 
The present work develops a nonequilibrium field-theoretic approach to explore the self-organization of a BEC in a pumped, lossy optical cavity.  
We find that the transition is well described, in the regime of primary interest, by an effective equilibrium theory. 
At nonzero temperatures, the self-organization occurs via a fluctuation-driven first-order phase transition of the Brazovskii class; this transition persists to zero temperature, and crosses over into a quantum phase transition of a new universality class.
We make further use of our field-theoretic description to investigate the role of nonequilibrium fluctuations on the self-organization transition, as well as to explore the nucleation of ordered-phase droplets, the nature and energetics of topological defects, supersolidity in the ordered phase, and the possibility of frustration controlled by the cavity geometry.
In addition, we discuss the range of experimental parameters for which we expect the phenomena described here to be observable, along with possible schemes for detecting ordering and fluctuations via either atomic correlations or the correlations of the light emitted from the cavity.

\end{abstract}

%-------------
\maketitle
%-------------

\section{Introduction}\label{sec:intro}

Over the course of the past fifteen years, many phenomena of long-standing interest in condensed-matter physics have been realized in ultracold atomic settings~\cite{bloch:review}. Such realizations are considerably different from condensed-matter systems: in particular, ultracold atomic systems are highly controllable---i.e., they are isolated from the environment and are governed by thoroughly understood microscopic Hamiltonians---and tunable---i.e., the interaction strength, lattice depth, etc.\ are governed by quantities such as laser intensities, which are easy to alter. The high degree of control and tunability has made it possible both to explore emergent phenomena in a simpler setting than is typical in condensed matter and to address hitherto experimentally inaccessible questions, such as the dynamics of ordering in systems that are quenched past a quantum critical point~\cite{greiner:quench}.  To date, most ultracold atomic realizations have focused on simulating the physics of electrons propagating through \textit{static }lattices (via, e.g., realizations of the Hubbard model~\cite{moritz}) or on constructing novel quantum fluids (e.g., Tonks-Girardeau gases~\cite{bloch:tonks} or unitary Fermi gases~\cite{ufg}). Areas of condensed matter such as soft matter, supersolidity~\cite{supersolids}, and glassiness---which involve  \textit{emergent, compliant} lattices capable of exhibiting dynamics, defects, melting etc.---have proven inaccessible to ultracold atomic physics because the lasers that create the lattice potentials in typical experiments are essentially insensitive to the atomic motion in those potentials.  Aspects of such condensed matter phenomena remain unsettled in their traditional settings (e.g., the dynamics of glassy media and of supersolids), and therefore ultracold atomic realizations of them are especially desirable.

A possible approach to realizing phenomena dependent on the emergent, compliant character of the lattice is to have the atoms interact with a potential created by dynamical, \textit{responsive} quantum light, instead of static lasers. Exploring precisely such interactions has been the central theme of cavity QED~\cite{walls}. Traditionally, cavity QED has aimed to realize systems involving a single atom coupled to a single mode of the electromagnetic field; however, the physics of \textit{many} atoms coupled to one (or more) electromagnetic modes, i.e., \textit{many-body} cavity QED, has also been studied extensively in recent work~\cite{ritschprl, ritschpra, vuletic:prl, ringcav, domokos08, morigi08, morigi10, esslinger06, baumann}. In particular, it was predicted in in Refs.~\cite{ritschprl, ritschpra} that a cloud of atoms confined in an optical cavity would exhibit collective effects such as self-organization; these effects were subsequently observed~\cite{vuletic:prl}.  The atoms considered in these works were essentially classical, thermal particles: however, the dynamics of many \textit{quantum} atoms (e.g., a Bose-Einstein condensate or a Mott insulator) confined in a cavity has since been explored both theoretically~\cite{domokos08, morigi08, morigi10} and experimentally~\cite{esslinger06, baumann}. Much of the work in this area, to date, has explored the novel implications of the atom-cavity coupling for standard ultracold-atomic phenomena such as the superfluid-Mott insulator transition~\cite{morigi08, morigi10} or the collective excitations of a Bose-Einstein condensate~\cite{domokos08, esslinger06}. The objective of the present work is to suggest that a quite different class of condensed-matter problems, involving the emergence and dynamics of spatially ordered states, can be realized and explored using ultracold atoms confined in optical cavities. Elements of this work were reported in Ref.~\cite{us}; related issues, involving the simulation of phonons in optical lattices, were previously raised in Ref.~\cite{lewenstein:multimode}.

A central idea in the extant literature as well as the present work is the idea of cavity-induced self-organization~\cite{ritschprl, ritschpra}, which may be explained as follows: Consider $N$ two-level atoms in a \textit{single-mode} optical cavity, interacting with the cavity mode and a pump laser oriented transverse to the cavity axis (see Fig.~\ref{fig:modestruc}). The atoms coherently scatter light between the pump and cavity modes.
Atoms arranged at every other antinode of the cavity field (i.e., one cavity-mode wavelength $\lambda$ apart) emit in phase; therefore, $\lambda$-period fluctuations of the atomic density increase the number of photons in the cavity, thus drawing the atoms into $\lambda$-spaced wells at either the even or odd antinodes. This leads to greater constructive interference in the emitted light, stronger atomic trapping, and so on.  The system reaches a spatially modulated steady state when the energetic gain from the atom-light interaction is balanced by the cost, in kinetic energy or repulsive interactions, of confining the atoms to either the even or the odd sites of the emergent lattice.

Although self-organization in a single-mode cavity results in the spontaneous breaking of a \textit{discrete} symmetry, the locations of the antinodes are not themselves emergent, but are fixed by the cavity geometry. In other words, the non-crystalline state does not possess continuous translational invariance; thus self-organization resembles, e.g., a phase transition between crystal structures, rather than true crystallization. With multimode cavities, by contrast, self-organization results in the breaking of \textit{continuous symmetries}, both in the collective choice of which mode(s) to populate with photons and in the choice of relative phases between the modes. For example, in the case of the ring cavity (which consists of two counter-propagating traveling-wave modes~\cite{ringcav}), the atoms must collectively choose the (continuous) relative phase between the two counter-propagating modes, thus setting the location of the antinodes of the cavity field.  As with real solid-state crystallization, the breaking of this continuous symmetry induces rigidity with respect to lattice deformations. With larger families of modes, one can envision realizing such characteristically crystalline notions as dislocations and geometrical frustration.

In a previous article~\cite{us}, we developed a field-theoretic description of the interacting many-atom, many-mode system, which we applied to the case of a transversely pumped concentric cavity.  We observed: 
(i)~that a quasi-two-dimensional atomic cloud undergoes a weakly first-order transition into a spatially ordered state, and that this transition becomes a quantum phase transition at $T = 0$; and 
(ii)~that for the case of a strongly layered three-dimensional cloud, inter-layer frustration precludes global ordering, and the system instead breaks up into inhomogeneous, static domains.
In addition, we suggested that both fluctuation phenomena and signatures of supersolidity may be observed via the spatial and temporal correlations of the light emitted from the cavity.  The formalism developed in Ref.~\cite{us} assumed, however, that because the flux of energy through the atom-cavity system is negligible in the regime of interest, one could adopt a quasi-equilibrium description of this phase transition. One of our objectives in the present work is to justify this assumption within a general nonequilibrium formalism; furthermore, we compute the leading corrections to the effective theory of Ref.~\cite{us} that arise because of nonequilibrium effects.

In the present work, our approach towards meeting these objectives is as follows: 
(i)~we develop a fully nonequilibrium, field-theoretic description of the atom-cavity system, using the Schwinger-Keldysh functional-integral formalism~\cite{schwinger, keldysh}; 
(ii)~we show that the nonequilibrium description can be reduced, in the regime in which the cavity's photon decay time is longer than the timescales for atomic motion, to an effective equilibrium description; 
(iii)~we analyze this effective equilibrium theory using diagrammatic and renormalization-group techniques to establish the nature of the self-organization transition, in the specific case of a concentric cavity; and 
(iv)~we reintroduce the nonequilibrium effects, due to the leakage of photons out of the cavity, using perturbation theory, and account for their effects on critical behavior near the self-organization transition. 
Our analysis of the equilibrium theory extends that used in our previous work~\cite{us} via an adaptation of Shankar's renormalization-group treatment of the Fermi liquid~\cite{shankar}, as well as an analogy with the $O(\ncomponent)$-invariant vector model of magnetism, to establish that for an interacting Bose-Einstein condensate the self-organization transition is always discontinuous in the concentric cavity. Furthermore, we show that the chief consequence of nonequilibrium effects is to decohere quantum correlations on a timescale related to the linewidth of the cavity; thus, the self-organization transition is always \textit{classical} on the longest timescales. (For sufficiently high-finesse cavities, there should, however, be an extended crossover regime of timescales for which the phase transition appears quantum.)

After presenting our analysis of the phase transition itself, we turn to the properties of the ordered (i.e., self-organized) state.  We show that the ordered state has low-lying excitations, associated with the continuous symmetry-breaking, that resemble the excitations of smectic liquid crystals.  We also expand on our previous observation~\cite{us} that, for a Bose-condensed atomic cloud, the ordered state would be a ``supersolid''\ (or, more accurately, a ``super-smectic'') in that it simultaneously possesses emergent (liquid-)crystalline and superfluid order.  The properties and even the existence of supersolidity in $^4$He are much-discussed topics in the condensed-matter literature~\cite{supersolids, toner08, shevchenko, west09, davis:superglass}; the appeal of ultracold-atomic realizations is that one can explore the characteristic \textit{phenomenology} of supersolids in contexts where it is less challenging to establish that they do in fact possess supersolidity.  In order to explore the relevant phenomenology, it is necessary that the solidity be associated with a broken \textit{continuous} spatial symmetry; for this purpose a continuum supersolid such as the one explored in the present work should serve as a more suitable setting than would, e.g., the lattice supersolids proposed in Refs.~\cite{lew02, sun07}.  The present scheme has the added advantage of being more readily realizable.  The self-organization of a BEC in a cavity was, in fact, recently demonstrated experimentally by Baumann et al.~\cite{baumann} for the case of a \textit{single-mode} cavity; generalizing this experiment to the multimode case, in which one has continuous symmetry-breaking, should be technically straightforward. Continuing with the theme of supersolidity, we propose a scheme for detecting such as state, and develop a schematic phase diagram for the system (Fig.~\ref{fig:phasediag}), which exhibits three phases: the supersolid, the normal solid, and the uniform superfluid.  The phase diagram for a multimode cavity differs from that for a single-mode cavity in that the \textit{multimode} case features a direct transition from the uniform superfluid to the normal solid, whereas in the single-mode case there is always a supersolid regime separating the uniform superfluid and the normal solid.

This paper is organized as follows.  
In Sec.~\ref{sec:model} we describe the microscopic model of the atom-cavity system that is used in the rest of the paper, and in Sec.~\ref{sec:expectations} we discuss the qualitative behavior one might expect from this model. 
In the next three sections we construct and analyze a nonequilibrium field-theoretic formulation of this model:  Sec.~\ref{sec:keldysh} introduces the relevant field-theoretic formalism, 
Sec.~\ref{sec:atomsonly} applies this formalism to derive an atoms-only action, and 
Sec.~\ref{sec:eqm} describes the quasi-equilibrium limit of the atoms-only action. 
In Sec.~\ref{sec:landautheory} we derive an effective Ginzburg-Landau free energy, valid near the phase transition, which realizes a version of Brazovskii's model~\cite{brazovskii} of ordering at a finite wavelength, 
and analyze the effects of fluctuations on the self-organization transition in both the classical ($ T > 0$) and quantum ($T = 0$) cases. 
In Sec.~\ref{sec:neqc} we turn to the effects of departures from equilibrium, both on the fluctuations near the transition and (following the work of Hohenberg and Swift~\cite{swift:bubbles}) on the nucleation of ordered states.
In the next two sections we focus the properties of the ordered state: 
Sec.~\ref{sec:elasticity} reviews the properties and elementary excitations of the ordered state, and Sec.~\ref{sec:supersolidity} discusses the supersolid aspects of the ordered state.
In Sec.~\ref{sec:expt} we discuss the experimental feasibility of the phenomena that we are investigating, showing that most of them should be readily detectable in the laboratory. 
In Sec.~\ref{sec:glassiness} we briefly consider the case of strongly layered three-dimensional systems, which have the feature that geometrical factors tend to frustrate the development of globally coherent long-range order.
Finally, in Sec.~\ref{sec:conclusions} we summarize the results of this work, and discuss its relationship with other problems involving phase transitions and related collective effects in atom-light systems.
Various supplementary issues are addressed in four appendices.

\section{Model}\label{sec:model}

The system analyzed in this paper consists of
$N$ two-level atoms confined in a multimode optical cavity,
together with the electromagnetic modes of the cavity.
The system is pumped by an external laser, which is oriented transverse to the cavity axis, as shown in Fig.~\ref{fig:modestruc}.
In addition, the system is coupled to a set of extracavity electromagnetic modes, which constitute a bath for the system.  The complete Hamiltonian $\fullHam$ governing the system and bath comprises three elements---viz., $\atomHam$, the atoms-only Hamiltonian; $H_{\mathrm{em}}$, the light-only Hamiltonian; and $H_{\mathrm{int}}$, the atom-light interaction Hamiltonian---which we discuss in detail in the rest of this section.

\subsubsection{Two-level atoms}\label{sec:twolevelatoms}
The atoms are described by the Hamiltonian
\beq
\atomHam=
\sum_{n=1}^N
\left(
\frac{\mathbf{p}_n^2}{2\mass} +
\frac{\hbar\omega_A}{2} \big(1+\sigma^z_n\big)
\right)
\,+\,  \contact \!\!\!\! \sum_{1\le n<n'\le N}\!\!\!\!\!
\delta(\mathbf{x}_n - \mathbf{x}_{n'}),
\eeq
where $N$ is the number of atoms (indexed by $n=1,\ldots,N$),
each of which has mass $\mass$;
the position and momentum of atom $n$ are, respectively, $\mathbf{x}_n$ and $\mathbf{p}_n$.
The operator $\sigma_z$ denotes the Pauli operator, which acts on the internal state of the atoms,
which, for the sake of simplicity, we take to be two-state atoms, with $\hbar \omega_A$ being the
energy splitting between the ground (g) and excited (e) states.
(Our conclusions do not hinge in any essential way on this restriction to two states.)\thinspace\
We model the interaction between atoms via a repulsive contact potential,
which is parametrized in terms of the (positive) parameter $\contact$.
To ensure the correct handling of the Bose-Einstein quantum statistics of the atoms,
we employ the framework of second quantization.  Thus, $\atomHam$ becomes
\begin{eqnarray}\label{eq:atomH}
\atomHam \! \! & = & \! \! -
\frac{\hbar^2}{2\mass} \!
\int \! d^{\mydim}x \!
\left\{
\Psi_g^\dagger(\mathbf{x}) \nabla^2 \Psi^{\phantom{\dagger}}_g(\mathbf{x}) +
\Psi_e^\dagger(\mathbf{x}) \nabla^2 \Psi^{\phantom{\dagger}}_e(\mathbf{x})
\right\} \\
&&
+\hbar\atomgap\int d^{\mydim}x \,\Psi_e^\dagger(\mathbf{x})\Psi^{\phantom{\dagger}}_e(\mathbf{x})
\nonumber \\
&&
+\frac{\contact}{2}\int\! d^{\mydim}x \!
\left(
 \Psi_{g}^{\dagger}\Psi_{g}^{\dagger}\Psi_{g}^{\phantom{\dagger}}\Psi_{g}^{\phantom{\dagger}}+
 \Psi_{e}^{\dagger}\Psi_{e}^{\dagger}\Psi_{e}^{\phantom{\dagger}}\Psi_{e}^{\phantom{\dagger}}+
2\Psi_{g}^{\dagger}\Psi_{e}^{\dagger}\Psi_{e}^{\phantom{\dagger}}\Psi_{g}^{\phantom{\dagger}}\right),\nonumber
\end{eqnarray}
in which the pair of bosonic field operators, $\Psi_g(\mathbf{x})$ and $\Psi_e(\mathbf{x})$, respectively represent the ground- and excited-state atoms. In the last line of Eq.~(\ref{eq:atomH}), all the coordinate indices are $\mathbf{x}$, as the interaction is taken to be local in space.

Next, we assume that the density of excited-state atoms is sufficiently low that collisions between such atoms can be neglected.  This assumption necessarily holds in the low-temperature regime, which is the regime of primary interest to us, because the rate of spontaneous decay (which is proportional to  the number of excited atoms) must be kept low in order to avoid the heating associated with it.  With this regime in mind, we approximate the interaction term as
\begin{equation}\label{eq:intenenergy}
\frac{\contact}{2}\int d^{\mydim}x \,
\left(
 \Psi_{g}^{\dagger}\Psi_{g}^{\dagger}\Psi_{g}^{\phantom{\dagger}}\Psi_{g}^{\phantom{\dagger}}+
2\Psi_{g}^{\dagger}\Psi_{e}^{\dagger}\Psi_{e}^{\phantom{\dagger}}\Psi_{g}^{\phantom{\dagger}}
\right).
\end{equation}
In what follows we shall replace $\contact$ by the \textit{frequency} $U \equiv \contact / \hbar$, in order that its magnitude be expressed in the same units as those conventionally used to describe the atom-cavity coupling.

\subsubsection{Electromagnetic modes}\label{sec:emmodes}

Optical cavities typically consist of two or more mirrors that support localized modes of the electromagnetic field between them~\cite{siegman}. The modes might be standing waves, as in the single-mode cavity, or the traveling waves that one can have in the ring cavity. In general, we shall be concerned with transverse electromagnetic (TEM) modes. The vector character of the electromagnetic field can be absorbed into the effective atom-mode coupling~\cite{walls}; hence, such modes can effectively be described in terms of harmonic solutions to the scalar wave equation~\cite{siegman}.   Furthermore, as the mirrors are not perfectly reflective, these modes are in fact weakly coupled to the extracavity modes, and thus cavity-mode photons tend to \lq\lq leak\rq\rq\ out of the cavity at some nonzero rate $\kappa$ (which sets the intrinsic linewidth of the cavity).  These considerations lead us to model the electromagnetic field---both intracavity and extracavity---in terms of the Hamiltonian
\bea
H_{\mathrm{em}} & = & 
\sum_{\mathrm{\alpha\, in\, cav.}} \!\!\!\!
\hbar\omega_\alpha a^\dagger_\alpha a_\alpha +
\!\!\! \sum_{\mathrm{\varepsilon \, in \, env.}} \!\!\!\!
\hbar\omega_\varepsilon A^\dagger_\varepsilon A_\varepsilon \nonumber \\
&& \quad +
\left(
\sum_{\alpha, \varepsilon} \!
\hbar\kappa_{\alpha \varepsilon} A^\dagger_\varepsilon a_\alpha + \mathrm{h.c.}
 \right),
\eea
where $a_{\alpha}$ and $A_{\varepsilon}$ respectively represent the intracavity and extracavity photons,
$\omega_{\alpha}$ and $\omega_{\varepsilon}$ are the intracavity and extracavity mode frequencies, and
$\kappa_{\alpha, \varepsilon}$ describes the coupling between the intracavity mode $\alpha$ and the extracavity mode $\varepsilon$.  We assume that the intracavity-extracavity coupling is weak enough (i.e., the cavity is of sufficiently high finesse) that it is meaningful to separate the modes into intracavity and extracavity ones.

\begin{figure*}
	\centering
		\includegraphics{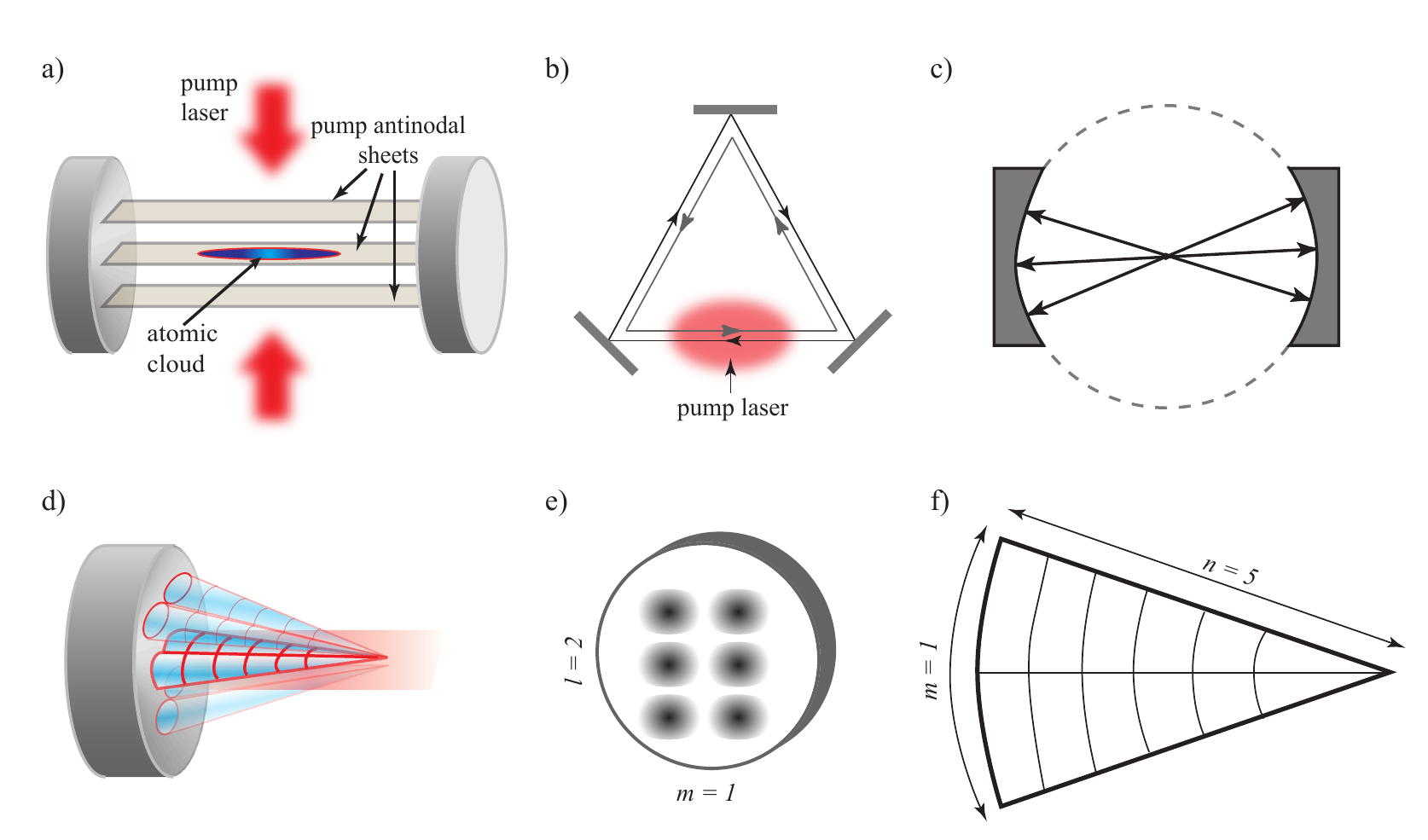}
	\caption{(a) The transversely pumped, quasi-two-dimensional geometry primarily discussed in this paper. (b) Ring cavity geometry. The pump laser beam is perpendicular to the plane defined by the three mirrors, as indicated in the figure. (c) Schematic representation of a concentric cavity, showing the partial rotational symmetry that such a cavity inherits from the sphere of which both cavity mirrors are arcs. (d) Three-dimensional view of a representative mode function for the concentric cavity. This mode function is labeled by $(l,m,n) = (2,1,5)$, or alternatively by $\mathrm{TEM}_{21}$. The three numbers enumerate the nodes (one fewer than the number of lobes) in the pump ($z$), angular, and radial directions, respectively. The axial mode index $n$ is fixed by the requirement that $l + m + n$ be constant for a family of degenerate modes, and can therefore be suppressed. (e) The intensity profile of the representative mode TEM$_{21}$ at one of the cavity's end mirrors. (f) The intensity profile of the mode TEM$_{21}$ in the equatorial (i.e., $z = 0$) plane of the cavity.}
	\label{fig:modestruc}
\end{figure*}

The modes of a generic standing-wave cavity are not frequency-degenerate: the typical frequency spacing, or \textit{longitudinal} ``free spectral range,''\ is of order $c/L$, where $L$ is the linear dimension of the cavity: e.g., for a $1$~cm cavity, the free spectral range is about $15$~GHz. For such a cavity, the frequency spacing between higher-order \textit{transverse} modes is an appreciable fraction of this number; thus there are no degenerate modes.
However, certain specific cavity geometries do support degenerate modes. The simplest of these is the ring cavity, which is a three-mirror arrangement that supports two counterpropagating traveling-wave modes (see Fig.~\ref{fig:modestruc}a). Even larger degeneracies are possible, e.g., in confocal or concentric cavities.  Let us label the cavity-mode structure by the integers $(l,m,n)$, where $n$ is the number of nodes along the cavity's axial direction, and $(l,m)$ are the numbers of nodes along the transverse directions; see Fig.~\ref{fig:modestruc}. (The corresponding mode functions are approximately sinusoidal in the axial direction, and Hermite-Gaussian (or Laguerre-Gaussian) in the transverse directions, although this approximation breaks down in the limit of a concentric cavity.)\thinspace\ In the confocal cavity, the condition for frequency degeneracy is that $n+((l + m)/2)=M$ (see, e.g., Ref.~\cite{siegman}) for some fixed integer $M$; in the concentric cavity, the condition becomes $l+m+n=M$.  In principle, these conditions imply that there are of order $M^2$ degenerate modes, where $M$ is roughly the number of optical wavelengths across the cavity, commonly~$10^4$ or more. In practice, higher-order modes are increasingly lossy, because their profiles (i.e., spot sizes) are larger, and therefore more of their light leaks out of the sides of the cavity mirrors, which typically occupy only a modest amount of solid angle. We approximately account for this effect by assuming that $M_0$ of the modes have a common loss rate $\kappa$, and that the other modes are perfectly lossy (i.e., for them $\kappa = \infty$).

\subsubsection{Atom-light interactions}\label{sec:interactions}

In the dipole and rotating-wave approximations~\cite{walls}, the atom-light coupling has the generic form
$i \sigma_- \, a^\dagger_n \, g_n(\mathbf{x}) \, -$ h.c.,
where the $\sigma$ operators raise or lower the atomic internal state;
$a_n$ is either an intracavity or extracavity mode;
and $g_n(\mathbf{x}) \equiv g \,\Xi_n(\mathbf{x})$,
in which $g$ is the coupling between the atom and the mode,
and $\Xi_n$ is an appropriately normalized mode function. Note that $g$ is proportional to $1/\sqrt{V}$, where $V$ is the system volume~\cite{walls}; hence $g^2 N$ stays finite in the thermodynamic limit. 
In the special case of the pump laser mode, we shall treat the corresponding $a_n$ as a classical variable, so that the atom-laser coupling takes the form
$\Omega(\mathbf{x},t)\sigma_- +{\rm h.c.}$,
where $\Omega(\mathbf{x}) \equiv \Omega(\mathbf{x}) e^{-i\omega_L t}$, with $\Omega(\mathbf{x})$ being the local Rabi frequency (which is proportional to the local amplitude) of the laser.  In our analysis, we shall largely neglect the term that governs spontaneous atomic emission into extracavity modes (i.e., spontaneous decay); this term is proportional to the linewidth of the transition, which is denoted as $\gamma$. The reason we can neglect spontaneous decay processes is not that such processes are always weak.  Rather, it is that their effects amount to the heating up of the sample via the random impulses they give to the atoms, and therefore the timescale for spontaneous decay [which is given by $1/R_\gamma = \Delta_A^2 / (\gamma \Omega^2)$] acts as an upper limit on the duration of an experiment:
quantum dynamics on timescales slower than $R_\gamma$ is likely to be washed out as a result of spontaneous decay.
Neglecting, then, spontaneous decay processes, the  Hamiltonian governing the atom-light interactions is given by
\beq\label{eq:intH}
H_{\mathrm{int}}\!\! = \!\!
i\hbar\!\int\! d^\mydim x\,
\Psi^\dagger_e(\mathbf{x})
\Psi_g(\mathbf{x})\!
\left[
\sum_{\alpha \in \mathrm{cav.}}\!\!\! g^{\phantom{\dagger}}_\alpha(\mathbf{x})\,a^\dagger_\alpha + \Omega(\mathbf{x},t)
\right] -
\mathrm{h.c.}
\eeq

\section{Physical expectations}\label{sec:expectations}

\subsection{Single-mode case: semiclassical picture}\label{sec:semiclassical}

As discussed in the Introduction, the atom-cavity system is unstable towards crystallization when transversely pumped at sufficient intensity.  A more quantitative, albeit semiclassical, picture of the crystallization is as follows. Assume that the atoms are pumped by a pair of counterpropagating lasers perpendicular to the equatorial plane of the cavity (see Fig.~\ref{fig:modestruc}a), so that the electric field due to the lasers is given by $E_L \cos k z \mathbf{\hat{y}}$, and that the field in the cavity mode, which is a standing wave, is given by $E_C \cos kx \mathbf{\hat{y}}$. The polarizations of the two modes, being parallel, can be neglected for the present purposes. Atoms in the plane $z = 0$ are subject to an effective electric field of intensity
$E_L^2 + 2 E_L E_C \cos k x + E_C^2 \cos^2 k x$.
The first term is spatially constant; of the spatially varying terms, for small $E_C / E_L$ the second is the dominant one. The potential energy of the (high-field-seeking) atoms in this field can therefore be taken to be
\beq\label{eq:physex1}
\mathcal{E} \sim -E_C E_L \int dx\,n(x) \cos kx.
\eeq
Furthermore, the magnitude of $E_C$ is set by the atomic distribution, i.e.,
\beq\label{eq:physex2}
E_C \sim E_L \int dx\,n(x) \cos kx \exp(i\omega_L t + \phi),
\eeq
where $\phi$ is the pump laser-cavity phase difference.  From Eqs.~(\ref{eq:physex1}) and (\ref{eq:physex2}) it follows that
$\mathcal{E} \sim - \left( \int dx\, n(x) \cos kx  \right)^2 \equiv - n_k^2$.
When $\mathcal{E}$ exceeds the kinetic-energy cost of self-organizing, the atoms undergo self-organization.
As $\mathcal{E}$ depends only on $n_k^2$, it is invariant under $n_k \rightarrow - n_k$, i.e., under translating the distribution of atoms through half a mode wavelength (e.g., from even to odd antinodes) along with changing the sign of the cavity field.  This is the even-odd symmetry (which is an inversion symmetry, provided the cavity has an even number of total wavelengths) that is broken when the atom-light system crystallizes. This crystallization transition can be thought of in one of two equivalent ways: either as a crystallization of the atoms or as a locking of the phases of the laser and the cavity modes. This phase-locking is associated with the condensation of photons into a cavity mode; analogously, conventional crystallization can be thought of as a condensation of phonons. 

Note that, for a single-mode cavity, the presence of the term in $\mathcal{E}$ that goes as $E_C^2 \cos^2 kx $ does not qualitatively affect the picture just outlined, unless $E_C$ is of the same order of magnitude as $E_L$.  If $E_C > E_L$, the \textit{energetics} would still favor symmetry breaking; however, the dynamics of ordering would then be complicated by the fact that there are local electric-field minima at the ``minority''\ antinodes---i.e., the ones at which the atomic density is supposed to be \textit{low}---and these local minima can trap atoms, as discussed in Ref.~\cite{ritschpra}. We shall not discuss this regime further in the present work.

\subsection{Multimode case}\label{sec:multimodeselforg}

The question that arises when one attempts to extend the idea of atom-light self-organization to the setting of multimode cavities is this: Which mode(s) do the atoms crystallize into?  In this section, we offer some heuristic general considerations aimed at addressing this issue.  We shall return to this question in the specific context of the concentric cavity, once we have developed the relevant field-theoretic techniques.

\subsubsection{Traveling waves}
The argument given in Sec.~\ref{sec:semiclassical} on self-organization for a single-mode cavity proceeds similarly for the case of multimode cavities, except that the individual mode-function $\cos kx$ must be replaced by the (as yet unspecified) set of cavity mode functions $\{ g_\alpha(\mathbf{x}) \}$.  There are, however, two crucial differences.  The first is pertinent whenever the cavity supports traveling-wave modes (e.g., the ring cavity).  In this case, the potential energy is given by
\beq
\mathcal{E} \sim -
\int d^\mydim x\,
g_\alpha(\mathbf{x})\,
n(\mathbf{x})
\int d^\mydim x'\,
g^{\ast}_\alpha(\mathbf{x}')\,
n(\mathbf{x}'),
\eeq
and the dynamics of each mode is coupled to that of its partner under time-reversal.  This has an important consequence, which is easiest to illustrate in the case of the ring cavity. Here, $g_\alpha = \exp ikx$, and $\mathcal{E}$ is invariant under the transformation $n(x) \rightarrow n(x + \epsilon)$, which involves shifting the atomic distribution along the cavity axis (and adjusting the antinodes of the cavity mode accordingly). Therefore, crystallization in multimode cavities having traveling-wave modes necessarily involves the spontaneous breaking of a continuous translational symmetry.

\subsubsection{Mode selection}
If two cavity mode functions $g_\alpha(\mathbf{x})$ and $g_{\alpha'}(\mathbf{x})$ are associated with a pair of frequency-degenerate harmonic solutions of the wave equation for the same (homogeneous) boundary conditions, any linear combination
$C_{\alpha}\,g_{\alpha}(\mathbf{x}) + C_{\alpha'} \,g_{\alpha'}(\mathbf{x})$
is also a legitimate cavity mode.  Therefore, it might seem that, in an $N$-fold degenerate cavity, any normalized mode of the form $\sum_\alpha C_\alpha \, g_{\alpha}(\mathbf{x})$ would be an ``equally good''\ arrangement for crystallization, i.e., there is an $N$-dimensional degenerate subspace. This is \textit{not} generally true, as a result of terms in the energy, omitted so far in the present section, that lift this degeneracy, such as the interatomic contact repulsion. Consider the extreme simplification involving two modes having the respective mode functions $\cos kx$ and $\cos ky$ (as would arise, e.g., from two cavities, perpendicular to one another and to the laser):
any function of the form
$(C_\alpha,C_{\alpha'}) \equiv C_{\alpha}\cos(kx) + C_{\alpha'}\cos(ky)$,
with $C_{\alpha}^2 + C_{\alpha'}^2 = 1$,
is a legitimate mode function.  If the atoms are self-organized in the state $(C_\alpha,C_{\alpha'})$,
the expectation value of the atomic field is given by
$\langle\psi(\mathbf{x})\rangle \sim
A + B \left[C_{\alpha}\cos(k x) + C_{\alpha'}\cos(ky)\right]$,
and the interaction energy, Eq.~(\ref{eq:intH}), goes as
$\int d^\mydim x\,\vert\Psi(\mathbf{x})\vert^4$. This can readily be checked to be smallest when either $C_{\alpha}= 0$ or $C_{\alpha'} = 0$,
i.e., for a stripe-like arrangement along either the $x$ axis or the $y$ axis.

A similar effect arises from the mode-mode scattering term (i.e., the $E_C^2$ term),
\beq
\int d^\mydim x\,
n(\mathbf{x})\,g_{\alpha}(\mathbf{x})\,g_{\alpha'}(\mathbf{x}).
\eeq
In the two-mode example discussed in the previous paragraph, in which the modes are at right angles to one another, this term is essentially diagonal in the mode indices for either of the stripe-like states. Suppose, however, that the two cavities lie at a small angle $\theta$ rather than at right angles to one another, so that the modes are $\cos(\mathbf{k \cdot x})$ and $\cos(\mathbf{k' \cdot x})$ with $\mathbf{k} \approx \mathbf{k}'$.  In this case, atomic density fluctuations of wave-vector $|\mathbf{k} - \mathbf{k'}| \approx |\mathbf{k}| \theta$---which, for small $\theta$, could be excited either thermally or quantum-mechanically---would suffice to mix the cavity modes. The effect of such mixing would be to lock the relative phases of the two modes.

\section{Field-theoretic formulation}\label{sec:keldysh}

Our objective in this and the subsequent two sections, Sec.~\ref{sec:atomsonly} and Sec.~\ref{sec:eqm}, is to construct a useful field-theoretic formulation that will enable us to explore the quantum statistical mechanics of correlated many-atom, many-photon systems in multimode cavities.  Having done that, in Sec.~\ref{sec:landautheory} we employ this formulation to address issues such as the emergence and nature of the spatial structure and spatio-temporal atomic and photonic correlation properties of such systems, focusing on the vicinity of the transition to the self-organized state.  The atom-cavity systems of interest here are neither closed nor in thermal equilibrium, because they are driven by an external pump laser (which adds energy) and leak photons into the continuum of modes that lie outside the cavity (hence losing energy); thus, even in its steady states there is a flux of energy through the system.  For these reasons, the structure and correlation properties must be computed within a nonequilibrium formalism.  The one we employ is the closed-time-path formalism, due to Schwinger~\cite{schwinger} and Keldysh~\cite{keldysh}, which enables the use of diagrammatic methods as well as renormalization-group techniques to analyze fluctuations. (For a discussion of the differences between the equilibrium and nonequilibrium formalisms, see Sec.~\ref{sec:eqm} and Ref.~\cite{kamenev}.)

Although, as we have just discussed, a full analysis of the problem demands a nonequilibrium approach, we find that, for systems that are near the threshold for self-organization and in the dispersive regime (i.e., the pump laser is far-detuned from the atomic resonance), an effective equilibrium description is valid, to a reasonable approximation.  As we shall see, e.g.~in Sec.~\ref{sec:neqc}, the description of this regime can be improved, and other regimes (such as the strongly organized regime) can be analyzed, using the full machinery of the Schwinger-Keldysh nonequilibrium approach.

\subsection{Schwinger-Keldysh functional integral}
\label{sec:keldysh1}

The quantities of interest in quantum many-body dynamics are the expectation values of observables and their response and correlation functions. Formally, the task of computing these may be stated as follows: suppose that we know the state of the system in the infinite past, as described by its density matrix $\rho(t = -\infty)$, when it is taken to be isolated and noninteracting.  The system is then coupled to an environment (or environments), which generically force the composite of system and environment(s) to be out of equilibrium, and the intra-system interactions are adiabatically switched on.  The question then becomes: What are the expectation values and response and correlation functions of the various \textit{system} observables, once the system and environment have relaxed to a steady state?

Let us first consider the case of a single harmonic oscillator with Hamiltonian
$H = \hbar \omega_0\, b^\dagger \,b$, i.e., a free bosonic degree of freedom having the characteristic frequency $\omega_0$. Suppose, as a simple example, that we are interested in the expectation value of some observable $A(t)$ at time $t$, given that the system was at some time $t_i < t$ in a thermal state at temperature $T$, i.e., governed by the density matrix $\rho(t_i) \propto \exp(- \hbar \omega_0 b^\dagger b / k_B T)$.  Thus, we wish to compute the quantity

\beq\label{eq:expval}
\langle A(t) \rangle \equiv \mathrm{Tr} \left[ \rho(t_i) A(t) \right].
\eeq
One can expand the trace in terms of bosonic coherent states~\cite{altland} at the times $t_i$ and $t_f > t$, thus arriving at the expression
\def\otherw{\tilde{w}}
\begin{widetext}
\beq\label{eq:pi0}
\frac{1}{(2\pi)^4}
\int d\otherw^{\phantom{*}}_f d\otherw^*_f d\otherw^{\phantom{*}}_i d\otherw^*_i dw^{\phantom{*}}_i dw^*_i dw^{\phantom{*}}_f dw^*_f e^{-|\otherw_f|^2 - |\otherw_i|^2 - |w_i|^2 - |w_f|^2} \langle w_i | \rho(t_i) | \otherw_i \rangle \, \times \, \langle \otherw_i | \otherw_f \rangle \, \times \, \langle \otherw_f | 1 | w_f \rangle \times \langle w_f | A(t) | w_i \rangle.
\eeq
Note that the primary motivation for inserting the complete set of states $| w_f \rangle$ is to make the above expression more symmetric between initial and final times.
 The relevant matrix element of the initial density matrix is given by $\exp(-\beta \hbar \omega_0 w_i^* \otherw_i)$, the overlap $\langle \otherw_f | 1 | w_f \rangle$ is given by $\exp(-\otherw_f^* w_f^{\phantom{*}})$, and the two other expressions, which are transition amplitudes, can be rewritten as coherent-state path integrals:

\beq
\langle \otherw_i | \otherw_f \rangle \langle w_f | A(t) | w_i \rangle =
\int_{z_+(t_i) = \otherw_i}^{z_+(t_f) = \otherw_f} D(z_+,z_+^*) e^{iS[z_+,z^*_+]}
\int_{z_-(t_i) = w_i}^{z_-(t_f) = w_f} D(z_-,z_-^*) A(z_-,z^*_-) e^{-iS[z_-,z_-^*]},
\eeq
in which the $\pm$ signs indicate whether the final state is the ket ($+$) or the bra ($-$), and the action $S$ is given by
\beq\label{eq:action}
S[z_\pm,z^*_\pm] = \int_{t_i}^{t_f} dt \, z_\pm^* (i \partial_t - \omega_0) z_\pm.
\eeq
One can thus rewrite Eq.~(\ref{eq:expval}) as follows:
\begin{eqnarray}\label{eq:pi}
\langle A(t) \rangle & = &
\int d(\otherw^{\phantom{*}}_f \otherw^*_f \otherw^{\phantom{*}}_i \otherw^*_i w^{\phantom{*}}_i w^*_iw^{\phantom{*}}_f w^*_f)\,
e^{-|\otherw_f|^2 - |\otherw_i|^2 - |w_f|^2 - |w_i|^2} e^{-\beta \hbar \omega_0 w_i^* \otherw^{\phantom{*}}_i} e^{-w_f^{\phantom{*}} \otherw^*_f} \nonumber \\ & & \quad
\int_{z_+(t_i) = \otherw_i}^{z_+(t_f) = \otherw_f} D(z_+,z_+^*)
\int_{z_-(t_i) = w_i}^{z_-(t_f) = w_f} D(z_-,z_-^*)
\frac{\delta}{\delta \xi_-(t)}\Bigg\vert_{\xi_\pm= 0}
\!\!\!\!
e^{iS[z_+,z_+^*] - i S[z_-,z_-^*] + \int dt \xi_-(t) A_-(t) + \xi_+(t) A_+(t)}.
\end{eqnarray}
\end{widetext}
If we omit the functional derivative ${\delta}/{\delta \xi_-(t)}$ from the right hand side of Eq.~(\ref{eq:pi}), the remaining formula is the quantity commonly denoted as $Z$ (by analogy with the partition function).  $Z$ is a generating functional for the correlation functions of the oscillator: it can be differentiated repeatedly with respect to either of the two source functions $\xi_\pm(t)$ in order to generate all requisite correlation functions of $A$.  By differentiating with respect to the $+$ and $-$ sources appropriately, one can compute expectation values that involve various orderings of the operators, e.g., the time-ordered, retarded, and advanced Green functions.  By contrast, in the zero-temperature and Matsubara nonzero temperature equilibrium formalisms, the only correlation functions that can be computed directly are the time-ordered ones; the (physically relevant) retarded Green functions are then to be inferred using identities that hold in equilibrium or at zero temperature.

Because, in Eq.~(\ref{eq:pi}), the initial and final values of the paths are integrated over (with an exponential measure), the path integral in Eq.~(\ref{eq:pi}) is in effect an \textit{unconstrained} path integral over the two sets of paths $z_\pm$.
Moreover, these paths are uncoupled from one another \textit{except} at the two endpoints of the path integrals. For the oscillator in question one can write the action in the form
\beq\label{eq:spi}
S = \left( \begin{array}{cc} z^*_+ & z^*_- \end{array} \right)
\left( \begin{array}{cc} S_{++} & S_{+-} \\ S_{-+} & S_{--} \end{array} \right)
\left(\begin{array}{c} z_+ \\ z_- \end{array} \right),
\eeq
in which integration is implied over time.
We shall sometimes denote as $\mathbf{S}$ the matrix comprising the four block matrices $S_{\pm\pm}$. The diagonal blocks $S_{++}$ and $S_{--}$ are given by Eq.~(\ref{eq:action}); the off-diagonal blocks $S_{+-}$ and $S_{-+}$ are zero, except at the time-endpoints; they cannot, however, be neglected, because the presence of such off-diagonal terms changes the \textit{inverse} of $\mathbf{S}$, denoted $\mathbf{G}$, which contains the two-point correlation functions and has the following structure:
\begin{subequations}
\label{eq:ManyGs}
\begin{eqnarray}
G_{+-}(t,t') & = & n_B(\omega_0) e^{-i\omega_0(t - t')} \nonumber\\
&&\qquad\qquad\qquad
= \langle z(t) z^\ast(t') \rangle, \\
G_{-+}(t,t') & = & (n_B(\omega_0) + 1)\, e^{-i\omega_0(t - t')} \nonumber\\
&&\qquad\qquad\qquad
= \langle z^\ast(t) z(t') \rangle, \\
G_{++}(t,t') & = & \Theta(t - t') G_{+-} + \Theta(t' - t) G_{-+},\\
G_{--}(t,t') & = & \Theta(t - t') G_{-+} + \Theta(t' - t) G_{+-},
\end{eqnarray}
\end{subequations}
where $n_B(\omega_0)$ is the Bose-Einstein distribution function.  Evidently, $G_{++}$ and $G_{--}$ are the time-ordered and anti-time-ordered Green functions.  It is convenient for our purposes to rotate $z_\pm$ into its ``classical'' and ``quantum'' components, defined as follows:
$z_{c} \equiv (z_+ + z_-) / 2$, $z_q \equiv (z_+ - z_-)/2$, in Eq.~(\ref{eq:spi}).
This has the advantage of reducing the number of independent Green functions by one:

\beq
\mathbf{G} = \left( \begin{array}{cc} G_K & G_R \\ G_A & 0 \end{array} \right),
\eeq
where
\begin{subequations}
\begin{eqnarray}
\!\!\!\!\!\!\! G_K \equiv \langle z^*_c(t) z^{\phantom{*}}_c(t') \rangle & \!\!=\!\! & -i
\big(2 n_B(\omega_0) + 1\big)
e^{-i\omega_0(t - t')}\!, \\
\!\!\!\!\!\!\! G_R \equiv \langle z^*_c(t) z^{\phantom{*}}_q(t') \rangle & \!\!=\!\! & -i \Theta(t - t') e^{-i\omega_0 (t - t')}, \\
\!\!\!\!\!\!\! G_A \equiv \langle z^*_q(t) z^{\phantom{*}}_c(t') \rangle & \!\!=\!\! & \phantom{-}i \Theta(t' - t) e^{-i\omega_0 (t - t')}.
\end{eqnarray}
\end{subequations}
$G_R$ and $G_A$ are the retarded and advanced Green functions---which describe the \textit{response} of the system to an external perturbation---whereas $G_K$, the ``Keldysh Green function\rlap,'' depends on the system's initial density matrix and its \textit{correlations}.  For a system in equilibrium with a bath, the correlations and response are related by the fluctuation-dissipation theorem; by contrast, for an isolated, noninteracting (and hence nonequilibrating) system such as the harmonic oscillator, the two properties---response and correlation---are  independent. In \textit{interacting} systems that are away from equilibrium, there is in general a complicated interplay between the correlations and the response; therefore, all the Green functions contain information about both correlations and response.  The nature of this relation, however, varies from system to system, and hence the information contained in the three Green functions is not redundant.

\subsection{Application to atom-photon system}
\label{sec:keldysh2}

The prescription for proceeding from the second-quantized Hamiltonian for a generic set of bosonic degrees of freedom $\{ \phi_i(t) \}$ to the appropriate coherent-state path integral (in real time) is as follows:
\bea\label{eq:hamtoaction}
S  & = &  \int dt\, \sum\limits_r \phi^*_{r,+}(t) \, i\partial_t \phi_{r,+}(t) -
H\big(\{ \phi^*_{r,+}(t), \phi^{\phantom{*}}_{r,+}(t)\}\big)
\nonumber \\
& & \qquad\qquad + \sum\limits_r\mu_r |\phi_{r,+}(t)|^2 - (+ \leftrightarrow - ).
\eea
In this expression, $\mu_r$ is the chemical potential for field $r$---in the present case, the chemical potential for the photons is zero, whereas that for the atoms determines the atomic density---and the symbol $(+ \leftrightarrow - )$ indicates a formally identical set of terms in which $+$ fields are replaced by the corresponding $-$ fields.
This prescription can be carried out for each of the terms in the Hamiltonian $\mathcal{H}$ introduced in Sec.~\ref{sec:model}, and generates all the $++$ and $--$ components of the action. The off-diagonal ($+-$ and $-+$) blocks in the bare (i.e., microscopic) theory depend on the appropriate Bose-Einstein distributions of the free photonic modes and the free atomic modes.  Because these blocks are infinitesimal (as they arise from the end-point couplings discussed in the previous section), it is more useful to write down the relevant blocks in the inverse action, viz., the bare Green functions $G_{+-}$ and $G_{-+}$.
For the cavity photon modes these have precisely the forms in Eqs.~(\ref{eq:ManyGs}), i.e.,

\bea
G_{+-}(t,t') & = & n_B(\omega_C) e^{-i\omega_C(t - t')},\\
G_{-+}(t,t') & = & \big(n_B(\omega_C) + 1\big) e^{-i\omega_C(t - t')},
\eea
whereas for bosonic atoms (in their internal ground state) and in a single-particle eigenstate of the kinetic energy having eigenvalue $E$, these have the form

\bea
G_{+-}(t,t') & = & n_B(E - \mu) e^{-i (E - \mu)(t - t') / \hbar} \\
G_{-+}(t,t') & = & \big(n_B(E - \mu) + 1\big) e^{-i(E - \mu)(t - t') / \hbar}.
\eea
Other degrees of freedom can be treated similarly.

\section{Constructing the atoms-only action}\label{sec:atomsonly}

In this section we derive an effective action, involving the ground-state atoms, which we shall use to determine expectation values and correlators involving atoms and/or intracavity photons.  This is accomplished by integrating out all other degrees of freedom---a task that is straightforward, owing to the fact that they appear quadratically in the complete action.

\subsection{Eliminating the atomic excited state}
\label{sec:excitedstate}
\begin{widetext}

As we see from Eqs.~(\ref{eq:hamtoaction}), (\ref{eq:atomH}), and (\ref{eq:intH}), the complete action involves the atomic excited state in the following terms:
\begin{eqnarray}\label{eq:sec5eq1}
% S_\pm^{at} & = &
&&
\int dt \, d^d x \Psi_{e,\pm}^\dagger(\mathbf{x},t) \left( i \partial_t + \frac{\hbar \nabla^2}{2\mass} - \omega_A \right) \Psi_{e,\pm}(\mathbf{x},t)
%&& \quad + \Psi_{g,\pm}^\dagger(\mathbf{x},t) \left(i \partial_t + \frac{\hbar \nabla^2}{2\mass}\right) \Psi_{g,\pm}(\mathbf{x},t) \nonumber \\
+ i \Big[ \sum_\alpha g_\alpha(\mathbf{x}) \Psi_{g,\pm}(\mathbf{x},t) \Psi^\dagger_{e,\pm}(\mathbf{x},t) a_{\alpha,\pm}(t) - \mathrm{h.c.} \Big]
\nonumber\\
&& \qquad - \Big. U |\Psi_{g,\pm}(\mathbf{x},t)|^2 |\Psi_{e,\pm}(\mathbf{x},t)|^2+ \cdots.
\end{eqnarray}
The functional integral over the quadratically occurring $\Psi_{e,\pm}$ in Eq.~(\ref{eq:sec5eq1}) can be performed exactly.  In the regime of interest, the atom-laser detuning $\hbar\Delta_A \equiv\hbar\omega_A - \hbar\omega_L$ is much greater than the energy scale associated with atomic motion; therefore, one can simplify matters by dropping the gradient term for $\Psi_e$. (Put heuristically, excited-state atoms, being short-lived and massive, ``decay''\ before they have time to move, so that the interactions they mediate are local in space and time.)\thinspace\ Thus, one can integrate out the excited state, determining the necessary kernel via the standard technique of solving the classical equations of motion for $\Psi_{e,\pm}$ and $\Psi_{e,\pm}^{\dagger}$ to obtain
\beq\label{eq:sec5eq2}
\Psi_{e,\pm}(\mathbf{x},t) = i \frac{\sum\nolimits_\alpha g_\alpha(\mathbf{x}) a_\alpha(t) \Psi_{g,\pm}(\mathbf{x},t)}{\Delta_A + U |\Psi_{g,\pm}(\mathbf{x},t)|^2},
\eeq
where, for convenience, we have made the change of photon variables $a_\alpha \rightarrow a_\alpha\, e^{-i\omega_L t}$ to enable us later to exploit the approximate degeneracy of the laser and cavity modes.  By inserting the classical solutions, Eq.~(\ref{eq:sec5eq2}) into the action, Eq.~(\ref{eq:sec5eq1}), we arrive at the following contributions to the action:

%\begin{widetext}

\begin{eqnarray}
&&
\int dt\, d^\mydim x \, \Psi_{g,\pm}(\mathbf{x},t) \left( i\partial_t + \frac{\hbar \nabla^2}{2\mass} - U |\Psi_{g,\pm}(\mathbf{x},t)|^2 + \mu \right) \Psi_{g,\pm}(\mathbf{x},t) + \sum_{\alpha} \frac{g_\alpha(\mathbf{x}) \Omega}{\Delta_A + U |\Psi_{g,\pm}(\mathbf{x},t)|^2} |\Psi(\mathbf{x},t)_\pm|^2 a^\dagger_{\alpha,\pm}(t) + \mathrm{h.c.} \nonumber \\
&& \qquad + \sum_{\alpha\beta} \frac{g^{\phantom{*}}_\alpha(\mathbf{x}) g^*_\beta(\mathbf{x})}{\Delta_A + U |\Psi_{g,\pm}(\mathbf{x},t)|^2} |\Psi_{g,\pm}(\mathbf{x},t)|^2 a^\dagger_{\alpha,\pm}(t) a_{\beta,\pm}(t).
\end{eqnarray}
In what follows we shall approximate $\Delta_A + U |\Psi_{g,\pm}(\mathbf{x},t)|^2$ by $\Delta_A$, using the fact that the interatomic interaction is typically many orders of magnitude weaker than $\Delta_A$.

\subsection{Eliminating the photon states}\label{sec:photonstates}
	
Next, we integrate out the photon states, doing so in two steps:
(i)~by integrating out the environment modes to arrive at an effective action for the cavity photons, and
(ii)~by integrating out the cavity photon modes to arrive at an effective action for ground-state atoms alone.
To achieve step~(i) we use the result of Caldeira and Castro-Neto
[i.e., Eqs.~(36) of Ref.~\cite{castroneto} in the limit $\hbar \omega_C \gg k_B T$]
for the path integral over extracavity modes, and thus we identify the following contributions
to the action that involve the cavity modes:

%\begin{widetext}

\begin{eqnarray}
&&\sum_\alpha \int dt
\left( \begin{array}{cc} a^*_{\alpha, +}(t) & a^*_{\alpha, -}(t) \end{array} \right)
\left( \begin{array}{cc} i \partial_t - \omega_C + i\kappa & 0 \\ 2i \kappa & - i\partial_t + \omega_C + i \kappa  \end{array} \right)
\left( \begin{array}{c} a_{\alpha, +}(t) \\ a_{\alpha,-}(t) \end{array} \right) \\
 &&\quad
 +\frac{1}{\Delta_A} \left[ \sum_\alpha \int dt \, d^dx ( \Omega e^{i\omega_L t} g_\alpha(\mathbf{x})n_+(\mathbf{x},t) a^*_{+,\alpha}(t) + \mathrm{h.c.})
 + \sum_{\alpha,\beta} \int dt \, d^dx \, a^*_{\alpha,+}(t)a_{\beta,+}(t) g_\alpha(\mathbf{x}) g^*_\beta (\mathbf{x}) n_+(\mathbf{x},t) - (+ \leftrightarrow -)  \right], \nonumber
\end{eqnarray}
where $n_\pm(\mathbf{x},t)\equiv \Psi^*_\pm(\mathbf{x},t) \Psi_\pm(\mathbf{x},t)$
is the local atomic density.  Note that we have dropped the atomic internal-state index $g$
(i.e., we have made the relabeling $\Psi_{g,\pm}\to\Psi_{\pm}$).

To achieve step~(ii), we observe that the action is quadratic in the cavity photon modes, so they too can be integrated out, to produce the desired atom-only action which, for convenience, we express in terms of the classical-quantum (i.e., $c-q$) basis for the fields rather than the $\pm$ basis (see Sec.~\ref{sec:keldysh}):

%\begin{subequations}
\begin{eqnarray}
S^{\mathrm{eff}} & = & \int dt\, d^dx \Psi^*_c(\mathbf{x},t)  \left( i \partial_t + \frac{\hbar \nabla^2}{2M} + \mu \right) \Psi_q(\mathbf{x,t}) - U \Psi^*_c (\mathbf{x},t) \Psi^*_q (\mathbf{x},t) [\Psi_c(\mathbf{x},t)^2 + \Psi_q(\mathbf{x},t)^2 ] + \mbox{h.c.} \\
& + & \frac{1}{2} \mathrm{Tr} \ln \mathbf{M} + \int d^dx\, d^dx'\, d\omega\, d\omega' \sum_\alpha \frac{\Omega^2 g_\alpha^2(\mathbf{x})}{\Delta_A^2}
\left( \begin{array}{cc} n_1(\mathbf{x}, \omega) & n_2(\mathbf{x}, \omega) \end{array} \right)
\big[ \mathbf{M}(\omega, \omega'; \alpha, \beta) \big]^{-1}
\left( \begin{array}{c} n_1(\mathbf{x'},\omega') \\ n_2(\mathbf{x'},\omega') \end{array} \right),\nonumber
\end{eqnarray}
in which $\mathbf{M}$ is the matrix

\beq
\mathbf{M}(\omega, \omega'; \alpha, \beta) \equiv \left( \begin{array}{cc} 0 & \omega - \omega_C + i\kappa \\
\omega - \omega_C - i\kappa & 2 i\kappa
\end{array} \right) \delta(\omega - \omega') \, \delta_{\alpha\beta}
+ \left( \begin{array}{cc} 0 & D_1(\omega' - \omega; \alpha,\beta) \\ D_1^*(\omega' - \omega;\alpha,\beta) & D_2(\omega' - \omega;\alpha,\beta) \end{array} \right),
\eeq
\end{widetext}
and the quantities $D_i$ are defined as follows:
\beq
D_i \equiv \frac{1}{\Delta_A}\int d^d x g_\alpha(\mathbf{x}) \, g_\beta(\mathbf{x}) \, n_i(\mathbf{x}, \omega' - \omega),
\eeq
where the Keldysh components of the atomic density are given by

\begin{eqnarray}
n_1(\mathbf{x},\omega) & \equiv & \int d\omega'\, \Psi^*_c(\mathbf{x},\omega') \Psi_q(\mathbf{x},\omega - \omega') + \mathrm{h.c.}, \\
n_2(\mathbf{x},\omega) & \equiv & \int d\omega'\,
\Big\{
\Psi^*_c(\mathbf{x},\omega') \Psi_c(\mathbf{x},\omega - \omega')
\nonumber\\
& & \quad \qquad +
\Psi^*_q(\mathbf{x},\omega') \Psi_q(\mathbf{x},\omega - \omega')
\Big\}.
\end{eqnarray}
%\end{subequations}
%
How one should proceed from here depends on the relative magnitudes of $\kappa$, $\Delta_C$, and $g^2 N / \Delta_A$.
Our objective in this paper is to analyze the self-organization transition in a multimode cavity. Physically, this transition is associated with the laser-cavity interference term, as discussed in Sec.~\ref{sec:expectations}, and is most straightforward to analyze when $E_L \gg E_C$ (i.e., when $\Omega \gg g_\alpha$); moreover, as discussed in Ref.~\cite{ritschprl} (and as we shall show), the steady-state temperature of the system is proportional to $\hbar \kappa$. Therefore, in order to explore self-organization at low-temperatures, it is natural to take $\Delta_C \gg \kappa, g^2 N / \Delta_A$
\cite{Note1}.
We may therefore expand $\ln\mathbf{M}$ and  $\mathbf{M}^{-1}$ in powers of
$\kappa/\Delta_C$ and $g^2 N / (\Delta_A \Delta_C)$,
thus arriving at a simplified atom-only effective action:
\beq\label{eq:Seffective}
S_{\mathrm{eff}} = S_0 + S_\zeta + S_\xi + S_U + S_\kappa,
\eeq
where the five terms---which are labeled by their corresponding coupling constants---respectively account for:
$S_0$, the kinetic energy of the atoms;
$S_\zeta$, the $\lambda$-periodic interaction caused by the scattering of photons between the laser and cavity modes;
$S_\xi$, the interaction due to the scattering of photons between cavity modes;
$S_U$ the contact repulsion between the atoms; and
$S_\kappa$, dissipative processes due to the leakage of photons through the cavity mirrors.
The terms have the following explicit forms:

\begin{eqnarray}
S_0 & = & \int d\omega \, d^dx \, \Psi_c^* \left( \omega + \frac{\hbar \nabla^2}{2\mass} + \mu \right) \Psi_q + \mathrm{h.c.},
\\
S_\zeta & = & \zeta \Delta_C \sum_\alpha \int d\omega \, d^dx \, d^dx'\, \Xi_\alpha(\mathbf{x}) \Xi_\alpha(\mathbf{x}')
\nonumber \\
& & \qquad \times \left\{ \frac{n_1(\mathbf{x},\omega) \, n_2(\mathbf{x}',\omega)}{\omega - \Delta_C + i \kappa} + \frac{n_1(\mathbf{x},\omega) \, n_2(\mathbf{x}',\omega)}{\omega - \Delta_C - i \kappa} \right\},
\nonumber \\
S_\xi & = & \xi\Delta_C \sum_{\alpha\beta} \int d^dx \, d^dx'\,  \Xi_\alpha(\mathbf{x})\, \Xi_\beta(\mathbf{x})\,\Xi_\alpha(\mathbf{x}') \,\Xi_\beta(\mathbf{x}') \nonumber \\
&& \qquad \times \int d\omega\, d\omega' \frac{n_1(\mathbf{x},\omega) \, n_2(\mathbf{x}',\omega)}{(\omega + \omega' - \Delta_C + i \kappa)(\omega - \Delta_C - i \kappa)} , \nonumber \\
S_U & = & U \int d^dx \, dt \, \Psi^*_c (\mathbf{x}t) \Psi^*_q (\mathbf{x}t) [\Psi_c(\mathbf{x}t)^2 + \Psi_q(\mathbf{x}t)^2 ] + \mathrm{h.c.},
\nonumber \\
S_\kappa & = & \int d\omega \frac{1}{(\omega - \omega_C)^2 + \kappa^2} \int d\omega \,d^dx \,d^dx' \,g_\alpha(\mathbf{x})\, g_\alpha(\mathbf{x}')
\nonumber \\
& & \qquad \times i \,\zeta\, \kappa \,\Delta_C \,\coth(\hbar\kappa/k_B T)\, n_1(\mathbf{x}\omega)\, n_1(\mathbf{x}\omega).
\nonumber
\end{eqnarray}
To streamline the notation we have introduced the coupling constants
$\zeta \equiv g^2 \Omega^2 / (\Delta_A^2 \Delta_C)$ and
$\xi \equiv g^4 / (\Delta_A^2 \Delta_C)$.
For the regime in which $\Delta_C$ is large compared with the other frequencies in the effective action (e.g., the typical atomic kinetic energy), a further simplification is possible: the integrals over $\omega$ and $\omega'$ can be expanded in a gradient expansion in terms of $(1/\Delta_C) \partial_t$. In what follows, we shall keep only the zeroth order term in this expansion, thus arriving at the following ``instantaneous'' forms of $S_\zeta$ and $S_\xi$, in which we have expanded $n_1$ and $n_2$ in terms of the atomic field operators:

\bea
S_\zeta \! &\! =\! & \! \zeta \sum_\alpha \int dt \,d^dx\, d^dx'\, \Xi_\alpha(\mathbf{x}) \,\Xi_\alpha(\mathbf{x}') \\
&& \times \Psi^*_c (\mathbf{x}t) \Psi^*_q (\mathbf{x'}t) [\Psi_c(\mathbf{x}t) \Psi_c (\mathbf{x}'t)\! +\! \Psi_q(\mathbf{x}t) \Psi_q(\mathbf{x'}t)]\! + \! \mathrm{h.c.} \nonumber \\
S_\xi \! & \! = \! & \! \xi \sum_{\alpha\beta}\int dt \,d^dx \,d^dx'\, \Xi_\alpha(\mathbf{x})\, \Xi_\beta(\mathbf{x}) \,\Xi_\alpha(\mathbf{x}')\, \Xi_\beta(\mathbf{x}') \nonumber \\
 && \times \Psi^*_c (\mathbf{x}t) \Psi^*_q (\mathbf{x'}t) [\Psi_c(\mathbf{x}t) \Psi_c (\mathbf{x}'t)\! +\! \Psi_q(\mathbf{x}t) \Psi_q(\mathbf{x'}t)] \! + \!  \mathrm{h.c.} \nonumber
\eea
Note that if we neglect the effects due to $S_\kappa$, the nonequilibrium character of the theory would apparently disappear: the nonequilibrium laser- and cavity-mediated interactions $S_\zeta$ and $S_\xi$ have precisely the same form in terms of Keldysh and time indices as the contact repulsion $S_U$; thus, an effective equilibrium description of these terms should be possible. In the rotating-wave approximation (see Sec.~\ref{sec:eqm} below and Ref.~\cite{walls}), the laser, although a nonequilibrium element, only influences the density matrix by raising the system's ``apparent''\ energy eigenvalues by $\hbar\omega_L$. As explained in the following section, we can use this fact to develop an effective equilibrium description of the self-organization transition, to which the effects of a nonzero $S_\kappa$ can later be added as a perturbative correction.

\section{Effective equilibrium theory}\label{sec:eqm}

In this section we offer some considerations on self-organization in pumped cavities in general and, in particular, on the quasi-equilibrium character of this transition in the high-finesse limit.  The fundamental difference between the zero-temperature and nonequilibrium formalisms is the assumption---valid in the former case---that the initial state (in this case, the ground state) of the system without interactions evolves adiabatically into a pure energy eigenstate (in this case, the ground state) of the interacting system, provided the interactions are switched on sufficiently slowly. This assumption does not, in general, hold away from equilibrium, but there are certain nonequilibrium systems---e.g., an atom pumped by a far-off-resonant laser---for which it does hold: in this example, Fermi's golden rule implies that the laser does not stimulate \textit{real} transitions between the atomic levels unless the laser is resonant with the atomic transition. (For a broadened atomic level, Fermi's golden rule implies that real transitions are negligible as long as the laser-atom detuning, $\Delta_A$, exceeds the linewidth, $\gamma$, of the atomic transition.)

In the present case, the excitation gap between the trivial ground state (i.e., the photon-free cavity and a uniform distribution of atoms) and the lowest excited state, which has photons and a $\lambda$-periodic atomic density modulation (a ``phonon''), consists of two parts: (i) the energy cost $\hbar \Delta_C$ of adding a cavity photon and (ii) the energy gain due to the photon-phonon coupling, which in second-order perturbation theory would have the form $\{\hbar^2 (\Omega g / \Delta_A)^2 \} / \{ \hbar^2 \sbK^2 / 2 \mass \}$, where $\sbK \equiv 2\pi/\lambda$. Thus the total gap must have the form:

\beq
\delta = \Delta_C \left\{1-
\mathrm{const.} \times \frac{\hbar\zeta }{\hbar^2 K^2 / 2 M}
\right\}.
\eeq
For the system at hand, the assumption of adiabaticity holds as long as the cavity is of sufficiently high finesse, i.e., $\delta \geq \kappa$.  As we shall see in Sec.~\ref{sec:expt}, the self-organization transition is weakly discontinuous, and the range of values of the control parameter $\Omega$ over which the ordered and disordered phases coexist can be made larger than $\kappa$.  Under these conditions, the self-organization transition should be well-described by the effective equilibrium theory sketched in Ref.~\cite{us}.  Only for large values of $\kappa$ (i.e., bad cavities) would nonequilibrium effects have a chance of playing a leading role. For an alternative construction of such an equilibrium theory, which illustrates the relation between its quantum phase transitions and conventional equilibrium quantum phase transitions, see App.~\ref{app:toymodel}.

The assumption of adiabatic switching implies that if interactions are turned on adiabatically in the distant past and turned off adiabatically in the distant future, an initial energy eigenstate would evolve into itself, up to a (physically unimportant) phase factor~\cite{agd}.  This consequence in turn implies that a \textit{single} path integral can capture all the dynamical information; the second path integral in Eq.~(\ref{eq:action}), which was demanded by the necessity of summing over all final states, would be superfluous because the final state would be known.  In terms of Green functions, this implies that the components $G_{+-}$ and $G_{-+}$ become redundant; indeed, it can be proved directly, in terms of Keldysh diagrammatics~\cite{biroli}, that the expansion for $G_{++}$ (and $G_{--}$) is closed, containing all information about correlations and response.

Formally, the case of nonzero temperatures is less straightforward because the adiabatic switching assumption is not available for mixed states, and therefore the derivation of the imaginary-time Matsubara formalism from the real-time Keldysh formalism is, in general, nontrivial.   In this case, we are guided by the following consideration: if the fluctuation-dissipation theorem holds for the \textit{exact} Green functions, the Matsubara formalism is valid; in the present case, we can use the Keldysh diagrammatic technique (see Sec.~\ref{sec:neqc}) to show that violations of the fluctuation-dissipation theorem are \textit{small}, and hence that the Matsubara technique is approximately valid.  This is what one expects on physical grounds: the chief effect of the laser photons is to mediate an effective atom-atom interaction, rather than to cause an energy flux.

Our strategy in the next section, Sec.~\ref{sec:landautheory}, will be to analyze the equilibrium critical behavior of the system at both $T = 0$ and $T \neq 0$, using established results from statistical mechanics and quantum field theory, and neglecting dissipative processes.  After that, we shall return to the Keldysh formalism, in Sec.~\ref{sec:neqc}, to reinstate the dissipative processes and explore their consequences.

\section{Quasi-equilibrium Landau-Wilson description}\label{sec:landautheory}

In this section we derive coarse-grained, Landau-Wilson forms of the atom-only action, Eq.~(\ref{eq:Seffective}), both for zero and nonzero temperatures, valid in the quasi-equilibrium regime.  These Landau-Wilson actions are closely related to the one first introduced by Brazovskii~\cite{brazovskii}; we exploit this relationship in order to describe the impact of collective fluctuations on self-organization in multimode cavities.  Finally, we discuss how the correlations of these fluctuations can be detected via the light emitted from the cavity.

In an effective equilibrium theory, the prescription for going from the Keldysh to the Matsubara or zero-temperature formalisms is to trace, in reverse, the steps one would have taken to go from the equilbrium to the Keldysh formalism: i.e., keep the $(++)$ component of the action and drop the Keldysh indices~\cite{kamenev}.  It is, furthermore, convenient to reformulate the action in terms of an order parameter, i.e., a quantity that is zero in the uniform phase and nonzero in the self-organized phase.
The considerations of Sec.~\ref{sec:expectations} suggest that the appropriate order parameter for detecting crystallization into cavity mode $\alpha$ should be given by

\beq\label{eq:OPdef}
\rho_\alpha(t) \equiv \int d^d x\,\rho(\mathbf{x})\,g_\alpha(\mathbf{x}).
\eeq
The details of this procedure are dependent on the cavity geometry; we shall focus in this work on the case of the concentric cavity, as it is the most straightforward case.

\subsection{Mode structure of the concentric cavity}\label{sec:modestructure}

A concentric cavity can be thought of as consisting of two mirrors that cover antipodal regions of the same sphere.  Its mode structure is derived from the solutions of the Helmholtz equation inside the sphere.  The effects of the edges of the cavity mirrors can be hard to compute accurately; the standard technique is to solve the Helmholtz problem approximately, by requiring the electric field to vanish at the the mirrors edges. (As with the subsequent approximations that we shall make, this one works best for large cavities.)\thinspace\ As the atomic distribution is quasi-two-dimensional, being confined near the equatorial plane, it breaks the spherical symmetry of the cavity, and it is therefore convenient to employ cylindrical coordinates $(r,\theta,z)$. Thus, the planar dependence of the cavity modes takes the form $J_m (k_n r) \cos(m \theta)$, where $J_m$ is a Bessel function, $m$ is quantized by the requirement that $\cos(m \theta_0) = 0$ (see Fig.~\ref{fig:modestruc}), and $n$ by the requirement that the field should vanish at the mirrors.  The quantization of solutions along the $z$ direction (i.e., the pump laser axis; see Fig.~\ref{fig:modestruc}) yields a third mode index $l$.  For large cavities, modes having a fixed value of $l + m + n$ are frequency-degenerate.  Because, as discussed in Sec.~\ref{sec:model}, the free spectral range of the cavity is larger than the energy scales relevant to self-organization, we can restrict ourselves to cavity modes having a certain fixed value of $\Lambda_0\equiv l + m + n$.

The atomic density, which is quasi-two-dimensional, can be expanded over a similar set of mode functions, indexed by $(m,n)$, provided one retains all such modes and not just those satisfying $m + n = \Lambda_0$.  In addition, the boundary conditions on these mode functions are not in general the same as those on the cavity modes, as the atoms are confined by an external, confining laser field rather than by the cavity mirrors. For sufficiently large traps, however, this distinction is not expected to have important effects, and we shall neglect it.

The cavity modes for which $l = 0$ are expected to be favored for crystallization, as they have the highest amplitude in the equatorial plane of the cavity, to which the atoms are confined. However, as we shall see, modes having $l > 0$ are also of importance in setting the \textit{range} of the effective atom-atom interaction, and thus in determining, e.g., the extent of the fluctuation-dominated regime, as well as the size and stability properties of droplets of the ordered phase~\cite{Note2}.

In the $(m,n)$ basis for the atomic density, the order parameter, Eq.~(\ref{eq:OPdef}), is given by $\rho_{mn}$.  Note, also, that provided that the atomic density is spread out over a large number of optical wavelengths, the following asymptotic result holds:
\beq\label{eq:momentumcons}
\int d\mathbf{x}\,\prod\nolimits_i \Xi_{m_i n_i}(\mathbf{x}) \approx
\hat{\delta}_{\sum\nolimits_i m_i}\,\hat{\delta}_{\sum\nolimits_i n_i}\,.
\eeq
In this expression, we have introduced the sign-insensitive Kronecker delta
$\hat{\delta}_\Sigma m_i\equiv \frac{1}{4} \sum\nolimits_\pm \delta_{\sum (-1)^\pm m_i}$.
Eq.~(\ref{eq:momentumcons}) is exact for the angular ($m$) component, and holds approximately for the radial ($n$) component, if $n > m$; this is the regime of interest because modes for which $m \geq n$ have large diffractive losses and do not couple to the atoms. Eq.~(\ref{eq:momentumcons}) is closely analogous to momentum conservation, and simplifies the structure of quartic and higher-order terms in the action.

It is useful, at this point, to specialize to two cases.
The first addresses an ultracold gas of bosonic atoms, \textit{without contact interactions}, which may or may not be Bose-Einstein condensed.
We have dealt with this case, which involves the introduction of an auxiliary field, in some detail in the Supplementary Information of Ref.~\cite{us}; we revisit this case below in Sec.~\ref{sec:idealgas}.
This approach can, in principle, be generalized to the case of an interacting gas (whether Bose-Einstein condensed or not), but is unwieldy for such systems, in which there are non-cavity-mediated interactions, as it involves the introduction of multiple auxiliary fields. Thus, we treat the \textit{second} case of interest, which is that of an interacting Bose-Einstein condensate (BEC) at temperatures well below the condensation temperature, using an alternative approach that does not involve introducing auxiliary fields.  Instead, we exploit the off-diagonal long-range order of the BEC and use a correspondingly modified form of the order parameter.
In a BEC that is well below its condensation temperature, the atomic density factorizes to leading order:
$\langle \psi^\dagger (\mathbf{r})\,\psi(\mathbf{r}')\rangle \approx
\langle \psi^\dagger (\mathbf{r})\rangle\,\langle \psi(\mathbf{r}')\rangle$.
It is clear that, at temperatures much lower than the self-organization energy scale $\hbar^2 \sbK^2 / 2 M$, the low-energy modes---near the self-organization transition---are those corresponding to two widely separated regimes of \lq\lq momentum\rq\rq\ $m + n$, viz.~$m + n \approx 0$ and $m + n \approx \sbK R/2\pi$.  In the case where the system is Bose-condensed, one can use the presence of off-diagonal long-range order to exchange the order parameter, Eq.~(\ref{eq:OPdef}), for the condensate amplitude $\langle \psi_{mn} \rangle$---because self-organization then involves the macroscopic occupation of a mode with $m + n = \sbK R/2\pi$~\cite{Note3}.  This exchange considerably simplifies the structure of the theory, and makes it possible to treat the effects of both the contact interaction between the atoms and the cavity-mediated interactions between them in a relatively transparent way.
The final structure of the theory is, as we shall see, the same in both cases (i.e., the ultracold noninteracting system and the Bose-Einstein condensed interacting system).

\begin{widetext}

\subsection{Ideal Bose gas}\label{sec:idealgas}

In the ideal-gas case, we proceed as follows. We note that the cavity-mediated interaction term can be written as
$\int d\tau \sum_{mn} \zeta_{mn} \rho_{mn}(\tau)\, \rho_{mn}(\tau)$,
where $\zeta_{mn}$ is the cavity-mediated interaction favoring atomic modulation at wavenumber $(m + n)$.
As mentioned in Sec.~\ref{sec:modestructure}, the coupling $\zeta_{mn}$ is to be considered as being peaked about modes obeying $m + n = K_0 R/2\pi$.
Next, we perform a Hubbard-Stratonovich transformation, which consists of introducing an additional, Gaussian functional integral into the partition function via the identity~\cite{altland}

\beq
\exp\left( \int d\tau \, \zeta_{mn} \,\rho_{mn}(\tau)\, \rho_{-mn}(\tau) \right) = \int D\hat{\rho}_{mn}\, \exp\left( - \int d\tau \frac{(k_B T)^2}{\hbar^2 \zeta_{mn}} \hat{\rho}_{mn}(\tau)\, \hat{\rho}_{mn}(\tau) + 2 \frac{k_B T}{\hbar} \rho_{mn}(\tau)\, \hat{\rho}_{mn}(\tau) \right),
\eeq
in order to render the action quadratic in the $\Psi$ variables. The partition function can then be rewritten

\beq
Z = \int D(\Psi^*,\Psi) D\hat{\rho} \, e^{-S'},
\eeq
where $S'$ is given by %
\beq
S' = \int d\tau \int d^d x\, \Psi^*(\mathbf{x}, \tau) \left( \partial_\tau -\frac{\hbar \nabla^2}{2M} - \frac{\mu}{\hbar} + 2\sum_{mn} \frac{k_B T}{\hbar} \hat{\rho}_{mn}(\tau)\,\Xi_{mn}(\mathbf{x}) \right) \Psi(\mathbf{x},\tau) \nonumber + \int d\tau \sum_{mn} \frac{(k_B T)^2}{\hbar^2 \zeta_{mn}} \, \hat{\rho}_{mn}(\tau) \,\hat{\rho}_{mn}(\tau).
\eeq
Provided the laser strength is below the self-organization threshold and the gas is Bose-condensed, the field operators $\Psi(\mathbf{x},\tau)$ can be expressed in terms of condensate and non-condensate parts as $\Psi(\mathbf{x},\tau) = \sqrt{N_0/\mathcal{A}} + \Phi(\mathbf{x},\tau)$, where $N_0(T)$ is the equilibrium condensate fraction at temperature $T$ and $\mathcal{A}$ is the area occupied by the atoms.  Transforming the Bose fields to the basis of $(m,n)$ mode functions and Matsubara frequencies, in which the kinetic energy is diagonal, one has
$\Psi_{mn}(\omega_\nu) =
\sqrt{N_0} \, \delta_{m,0}\, \delta_{n,0} \, \delta_{\nu, 0} +
\Phi_{mn}(\omega_\nu)$.
Integrating out $(\Phi^\ast, \Phi)$, one arrives at the action
\beq
S'' = \frac{1}{2} \mathrm{Tr} \ln(\mathbf{M}) + \frac{k_B T}{\hbar} \sum_{mn\nu} \left[ \frac{1}{\zeta_{mn}} \hat{\rho}_{mn\nu} \, \hat{\rho}_{mn-\nu} - N_0 \, \hat{\rho}_{mn\nu} \, (\mathbf{M}^{-1})_{mn\nu, m'n'\nu'} \, \hat{\rho}_{m'n'\nu'} \right],
\eeq
where the (infinite-dimensional) matrix $\mathbf{M}$ is defined by

\beq
\mathbf{M}_{mn\nu, m'n'\nu'} \equiv \left[-i \omega_\nu + \frac{\hbar (m + n)^2}{2 M R^2} \delta_{mn, m'n'} \right] \delta_{\nu\nu'} + \frac{2 k_B T}{\hbar} \sum_{p q\nu''} \hat{\rho}_{p q \nu''} \, \delta_{m m' + p} \, \delta_{n n' + q}\, \delta_{\nu + \nu'', \nu}.
\eeq
Below threshold, so that self-organization is not present and $\langle\hat{\rho}_{mn\nu}\rangle=0$,
it is useful to  expand $\mathbf{M}$ in powers of $\hat{\rho}$;
the quadratic term in the action is then given by
\beq
\sum_{mn\lambda} \hat{\rho}_{mn\lambda}\, \hat{\rho}_{-mn-\lambda} \left[ \frac{k_B T}{\hbar \zeta_{mn}} -\frac{N_0 k_B T}{-i \omega_\lambda + \frac{\hbar^2(m + n)^2}{2 M R^2}} - \frac{(k_B T)^2}{2 \hbar^2} \sum_{p q \nu} \frac{1}{\left(i\omega_\nu - \frac{\hbar(p + q)^2}{2 M R^2} - \mu\right)\left(-i\omega_\nu - \frac{\hbar(m + n - (p + q))^2}{2 M R^2} - \mu\right)} \right].
\eeq
The last term can be usefully rearranged if one recalls that $(k_B T/\hbar) \sum_{mn\nu} (i \omega_\nu - \frac{\hbar (m + n)^2}{2 M R^2} - \mu)^{-1} = N - N_0$ for a Bose-Einstein condensate~\cite{altland}.
(This statement also holds for a non-Bose-condensed gas, if one sets $N_0 = 0$.)\thinspace\ We now use the fact that the atoms in a Bose-Einstein condensate typically have energies that are small compared with the recoil energy to approximate the last term, and find that at low temperatures the quadratic part of action is then given by

\beq
\frac{k_B T}{\hbar} \sum_{mn} \hat{\rho}_{mn\nu} \, \hat{\rho}_{-mn-\nu} \left[ \frac{1}{\zeta_{mn}} - \frac{N}{- i\omega_\nu + \hbar(m + n)^2/2 M R^2}\right],
\eeq
in which the vanishing of the coefficient for any $(m,n,\nu)$ signals, at the mean-field level, the instability of the gas towards self-organization in the corresponding mode. (Note that self-organization is therefore possible only in the $\nu = 0$ sector.)
To enable an analytic treatment of the transition, we replace the (analytically inaccessible) exact form of $\zeta_{mn}$ by the convenient approximate form
$\zeta_{mn}=\zeta [1 - \chi (m + n - \Lambda_0)^2]$,
where the parameter $1/\chi$ represents the extent to which the coupling to $l \neq 0$ modes is suppressed (see App.~\ref{app:chi} for further discussion of this point).
This approximation captures the fact that the coupling of the atoms to the cavity modes is strongest for the modes that obey $m + n =\Lambda_0$ and otherwise simplifies the structure of the theory without making any drastic modifications to it.
(For $\chi = \infty$ the $l \neq 0$ modes are entirely suppressed;
for $\chi = 0$ the atoms couple equally strongly to all cavity modes, in which case there is no preferred lengthscale for self-organization.)\thinspace\

Continuing with the expansion of $\ln (\mathbf{M})$ and $\mathbf{M}^{-1}$ in powers of $\hat{\rho}$, assembling the two contributions, and retaining the zeroth-order term in a gradient expansion, we arrive at the following form for the quartic-order term:

\beq
\frac{(k_B T)^2 N}{\hbar^4 K_0^4 / 4 M^2} \sum_{m_i n_i} \hat{\rho}_{m_1 n_1} \,\hat{\rho}_{m_2 n_2} \,\hat{\rho}_{m_3 n_3} \,\hat{\rho}_{m_4 n_4} \,\hat{\delta}_{\sum m_i} \,\hat{\delta}_{\sum n_i} .
\eeq
Finally, we make the rescaling
$\hat{\rho} \rightarrow \hat{\rho} \sqrt{\hbar\zeta / k_B T \chi}$;
in terms of these rescaled fields, the action, to quartic order, assumes the following Landau-Wilson form:

\begin{eqnarray}
S_{\mathrm{LW}} & = & \sum_{mn\nu} \left[ \frac{1}{\chi} \left( 1 - \frac{N \zeta}{\hbar K_0^2 / 2M} + \frac{i \omega_\nu \zeta N}{(\hbar \sbK^2 / 2 M)^2} \right) + \left(m + n - (K_0 R/2\pi)\right)^2 \right] \hat{\rho}_{mn\nu}\, \hat{\rho}_{-mn-\nu} \\
&+ & \frac{\zeta^2 N}{\chi^2 \hbar^2 K_0^4 / 4 M^2} \sum_{m_i n_i \nu_i} \hat{\rho}_{m_1 n_1 \nu_1} \,\hat{\rho}_{m_2 n_2 \nu_2} \,\hat{\rho}_{m_3 n_3 \nu_3} \,\hat{\rho}_{m_4 n_4 \nu_4} \,\delta_{\sum m_i, 0} \, \delta_{\sum \nu_i,0} \, \delta_{n_1 + n_2, n_3 + n_4}.\nonumber
\end{eqnarray}
If $T > 0$, we can restrict ourselves to the $\nu = 0$
sector of the order-parameter theory, as this is the only sector of the theory that plays
an important role for thermal phase transitions. In this case, $S_{\mathrm{LW}}$ is an instance of Brazovskii's free energy~\cite{brazovskii}.
We shall return to this ``action'' (which is, in effect, a free energy rescaled by $k_B T$) and discuss its implications for the character of the self-organization transition in Sec.~\ref{sec:cbraz}, after first deriving a closely analogous effective action in the case of the BEC with contact interactions. The case of $T = 0$ requires extending Brazovskii's analysis to \textit{quantum} phase transitions; we discuss this case in Sec.~\ref{sec:qbraz}.

\end{widetext}

\subsection{Interacting BEC}\label{sec:interactingbec}

For the interacting BEC, one begins with $S_{\mathrm{eff}}$ [Eq.~(\ref{eq:Seffective})], as in the previous subsection (Sec.~\ref{sec:idealgas}), but proceeds differently. It would be inconvenient to apply the auxiliary-field technique to the present case because the action contains three quartic terms, each having a different ``momentum''-space structure; decoupling the action would therefore require the introduction of three auxiliary fields.
Instead, we exploit the fact that the ``momentum''-space structure of the action simplifies considerably for the low-energy modes, which are the modes of interest because they are the ones  that provide the critical fluctuations. This simplification is analogous to that which arises in Fermi-liquid theory owing to the constraint that all low-energy excitations must have momenta that are approximately equal to the Fermi momentum~\cite{shankar}.
In the present case, the operative constraint is that all values of $m + n$ must be either approximately zero or approximately $\Lambda_0 \equiv \sbK R/2\pi$.  We denote the $m + n \approx 0$ components of the atomic field by $\phi$, and the $m + n \approx \Lambda_0$ components by $\Phi$. ``Momentum conservation\rlap,\rq\rq\ i.e., Eq.~(\ref{eq:momentumcons}), then implies that the following kinds of quartic terms are admissible:
(i)~four $\phi$ fields;
(ii)~two $\phi$ and two $\Phi$ fields; and
(iii)~four $\Phi$ fields.
(Processes involving three $\Phi$ and one $\phi$ fields are suppressed because at least one of the $\Phi$ fields would have to have $m \geq n$, which would imply large diffractive losses~\cite{siegman}.)\thinspace\
For terms of type~(ii), it is clear that the only kinds of processes that survive to arbitrarily low energies are those in which the two $\Phi$'s and the two $\phi$'s have the same values of $(m,n)$.
[In principle, terms involving pairs $(\Phi,\phi)$ having the same value of $n/m$ should also survive to arbitrarily low energies, but they can be shown to have negligible phase space, compared with the other terms mentioned.]\thinspace\
Similarly, for terms of type~(iii), the only sets of $(m,n)$ that survive to arbitrarily low energies are forward- and backward-scattering processes, viz., those for which $(m_1,n_1) = (m_2,n_2)$ and $(m_3,n_3) = (m_4,n_4)$.  The other processes are said to be ``irrelevant at tree level\rlap,\rq\rq~\cite{shankar} because they become progressively less important at lower energies.

Applying the arguments just given to the three quartic terms in $S_{\mathrm{eff}}$, we find:
(1)~that at low energies the mode-mode scattering term $S_\xi$ is irrelevant for $\phi$'s
(except for the term in which two of the incoming momenta are zero),
whereas it \textit{does} survive for $\Phi$'s; and
(2)~that the contact repulsion separates into three parts, and can be written as follows
(note that all interactions are local in time; for this reason we have suppressed the time arguments):

\begin{widetext}

\bea
\frac{S_U}{U} & = & \sum_{m_i n_i} \phi^*_{m_1 n_1} \phi^*_{m_3 n_3} \phi^{\phantom{*}}_{m_2 n_2} \phi^{\phantom{*}}_{m_4 n_4} \hat{\delta}_{\Sigma m_i} \hat{\delta}_{\Sigma n_i}
+  \sum_{m_i n_i} \Phi^*_{m_1 n_1} \Phi^*_{m_3 n_3} \Phi^{\phantom{*}}_{m_2 n_2} \Phi^{\phantom{*}}_{m_4 n_4}\hat{\delta}_{\Sigma m_i} \hat{\delta}_{\Sigma n_i} \\
&& \quad  + 2 \left(\sum_{mn} \phi^*_{mn} \phi^{\phantom{*}}_{mn}\right) \left(\sum_{m'n'} \Phi^*_{m'n'} \Phi^{\phantom{*}}_{m'n'}\right) + \left(\sum_{mn} \phi^*_{mn} \phi^*_{mn}\right) \left(\sum_{m'n'} \Phi_{m'n'} \Phi_{m'n'}\right) + \mathrm{h.c.}
\eea
As in Eq.~(\ref{eq:momentumcons}), we have used the sign-insensitive Kronecker delta, $\hat{\delta}_\Sigma m_i\equiv \frac{1}{4} \sum\nolimits_\pm \delta_{\sum (-1)^\pm m_i}$.
We can now integrate out the $\phi$ modes, provided we first render the action quadratic in these modes; this can be done either by making the Bogoliubov approximation (see, e.g., Ref.~\cite{agd}) or, more generally, by exploiting the fact (which follows from Goldstone's theorem~\cite{altland}) that the low-lying modes of a BEC are linearly dispersing phonons.  If this is done, the action for the $\phi$ fields assumes the form

\beq
S_{\phi} =
\int d\omega
\sum_{ mn}
\sum_{ m'n'}
 \begin{array}{cc} \big(\phi^{\phantom{*}}_{\omega mn} & \phi^*_{-\omega mn} \big)\\ & {} \end{array}
\left( \begin{array}{cc} \mathcal{S} & \mathcal{T} \\ \mathcal{T}^* & \mathcal{S}^* \end{array} \right)
\left( \begin{array}{c} \phi^*_{\omega' m'n'} \\ \phi_{-\omega' m'n'} \end{array} \right).
\eeq
This form of the action is, strictly speaking, only appropriate for zero-temperature;
for $T > 0$ the integral over $\omega$ becomes the discrete sum
$k_B T\,\sum\nolimits_{\omega_\nu}$.
For compactness, we shall present only the expressions for $T = 0$, except when the two cases differ substantively. The blocks $\mathcal{S}$ and $\mathcal{T}$ are given as follows
[note that both are diagonal in the $(m,n)$ index;
the appropriate delta functions have been omitted for compactness]:

\begin{subequations}
\bea
\mathcal{S}_{mn}(\omega,\omega') & = & \left[ i\omega + \frac{\hbar (m + n)^2}{2M R^2} + U n_0  \right] \delta_{\omega\omega'} + \sum_{m',n'} [U + \zeta(m' + m, n' + n)]\, n_{m'n'} (\omega - \omega'), \\
\mathcal{T}_{mn}(\omega,\omega') & = & U n_0 \delta_{-\omega\omega'} +
\sum_{m',n'} [U + \zeta(m' + m, n' + n)]\,\nu_{m'n'} (- \omega - \omega'),
\eea
where $n_0$ denotes the number-density of particles in the condensate, and

\beq
n_{mn}(\Omega)\equiv \int d\omega\, \Phi_{mn}^*(\Omega + \omega) \, \Phi_{mn}(\omega), \qquad
\nu_{mn}(\Omega)\equiv \int d\omega\, \Phi_{mn}(\Omega - \omega) \, \Phi_{mn}(\omega).
\eeq
\end{subequations}
We now integrate out the $\phi$ fields to arrive at the following effective action in terms of $\Phi$:

\beq
S[\Phi,\Phi^*] = \int d\omega \sum_{mn}
\begin{array}{cc} \big( \Phi^*_{mn}(\omega) & \Phi_{mn}(-\omega)\big) \\ & \end{array}
\left(\begin{array}{cc} i\omega + \frac{\hbar (m + n)^2}{2 \mass R^2} + \mathcal{W} + \mathcal{V} &
- \mathcal{W} + \mathcal{V} \\ -\mathcal{W} + \mathcal{V} & -i\omega + \frac{\hbar(m+n)^2}{2 \mass R^2} + \mathcal{W} + \mathcal{V} \end{array} \right)
\left( \begin{array}{c} \Phi_{mn}(\omega) \\ \Phi^*_{mn}(-\omega) \end{array} \right) + \cdots,
\label{eq:SPhiTotal}
\eeq
\end{widetext}
where
\begin{subequations}
\bea
\mathcal{V}_{mn}&\equiv&U n_0 - \zeta_{mn} N_0,\\
\mathcal{W}_{mn}&\equiv&U (n - n_0) +
\int d\omega \frac{-\zeta_{mn} U n_0}{\omega^2 + U \frac{\hbar (m + n - \Lambda)^2}{2 \mass R^2} },
\eea
\end{subequations}
with $N_0$ being the total number of particles in the condensate.

This action, Eq.~(\ref{eq:SPhiTotal}), can, once again, be addressed by means of a Bogoliubov transformation.
Upon performing such a transformation, we find that, for sufficiently low energies, the quadratic term $S_2$ in the action can be expressed as follows in terms of the \textit{real} field
$\widehat{\Phi}(\omega)\equiv \Phi(\omega) + \Phi^*(-\omega)$:

\beq\label{eq:qbraz}
S_2 \!=\!\! \int\! d\omega \sum_{ mn}
\widehat{\Phi}_{\omega mn}
\frac{\tau + \omega^2 + \frac{\hbar \sbK^2}{2\mass} \zeta N {\chi}
(m + n - \Lambda_0)^2}{\hbar \sbK^2 / 2 \mass R^2} \widehat{\Phi}_{-\omega mn},
\eeq
in which ${\chi}$ again represents the coupling to modes having $l > 0$,
but the control parameter $\tau$ is now given by

\beq
\tau = \frac{\hbar \sbK^2}{2\mass} \left( \frac{\hbar \sbK^2}{2\mass} + U - \zeta N \right).
\eeq
As for the term quartic in $\Phi$,
its leading-order gradient expansion is a contact interaction among the $\widehat{\Phi}$'s, proportional to

\beq
S_4 = \mathfrak{U} \prod_{i = 1}^4
\int d\omega_{i}\sum_{M_i,N_i}
\widehat{\Phi}_{M_i,N_i,\omega_I}
\delta_{\Sigma M_i}\,
\delta_{\Sigma N_i}\,
\delta({\Sigma \omega_i}).
\label{eq:Squartic}
\eeq
The coefficient $\mathfrak{U}$ accompanying this term receives contributions arising in two ways:
(1)~from terms in $S_{\mathrm{eff}}$ that are quartic in $\Phi$, and
(2)~via the integrating out of $\phi$.
In principle, these each give rise to three terms, one associated with each of the quartic terms the microscopic action, Eq.~(\ref{eq:Seffective}).
However, owing to the ``momentum''-space structure of this action,
$S_{\zeta}$ contributes no terms of type~(1), and
$S_{\xi}$ contributes no terms of type~(2).
On the other hand,
$S_{U}$ contributes both types of term.
Of these, the term of type~(2) is proportional to the above-the-condensate density, which is expected to be small at the relevant temperatures; hence, this term is subleading, compared with the term of type~(1).
%%%
%%%
Putting these facts together, we find that the quartic term, Eq.~(\ref{eq:Squartic}), has the coefficient

\beq
\mathfrak{U} = U - \xi N + \frac{\zeta^2 (N - N_0)^2}{U}.
\eeq
At nonzero temperatures and near the self-organization transition, the sum over Matsubara frequencies is dominated by the $\omega = 0$ sector; under these circumstances, the action $S_2 + S_4 + \cdots$ [Eq.~(\ref{eq:SPhiTotal})] has the same form as that for the noninteracting case, and they are both variants of the free-energy functional first analyzed by Brazovskii.  At $T = 0$,  the situation is somewhat different; we shall return to this case in Sec.~\ref{sec:qbraz}.

That the actions have the same form is due to the phenomenon of ``universality\rq\rq\ near phase transitions: the structure of any theory sufficiently close to a critical point depends only on the symmetries of the order parameter and the free energy (or ``action''). In the present case, there are two salient features:
(1)~that there is a strip of degenerate, low-lying atomic density modes around $\bc$; and
(2)~that terms cubic in the order parameter, which would be allowed by symmetry, are forbidden because they would involve at least one mode having $m > n$;
such modes have high diffractive losses and cannot, therefore, be effectively populated with photons.
These constraints are sufficient to force the action to have the above form near the phase transition.

\subsection{Classical Brazovskii transition}
\label{sec:cbraz}

In this section, we briefly outline our adaptation of Brazovskii's self-consistent analysis of the eponymous model~\cite{brazovskii} to the present setting.  Brazovskii's analysis begins with the following free-energy functional in terms of the real order-parameter field $\psi(\mathbf{x})$ and its Fourier-space counterpart
$\psi_{\mathbf{k}}\equiv\int d^{d}x\,\exp(i\mathbf{k}\cdot\mathbf{x})\,\psi(\mathbf{x})$:

\beq\label{eq:brazovskii}
F = \int d^d k\,
\psi_{\mathbf{k}} \left[ \mathcal{R} + (|\mathbf{k}| - k_c)^2 \right] \psi_{\mathbf{-k}} + \mathcal{U} \int d^d x \left( \psi(\mathbf{x}) \right)^4,
\eeq
where the bare phenomenological parameters $\mathcal{R}$ and $\mathcal{U}$ are, respectively, the control parameter for the transition and the interaction parameter.  At the mean-field level, $F$ is minimized for $\mathcal{R} > 0$ by the uniform configuration $\psi(\mathbf{x}) = 0$, and for $\mathcal{R} < 0$ by $\psi_{\mathbf{k}}$ having the nonzero
value $\sqrt{ - \mathcal{R} / 2 \mathcal{U}}$ for any one of the momenta $\mathbf{k}$ having magnitude $k_c$.

\begin{figure}
	\centering
		\includegraphics{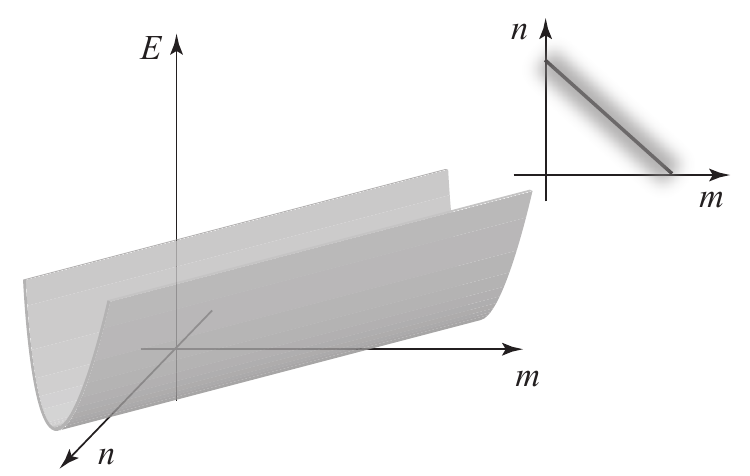}
	\caption{Dispersion relation for low-energy atomic excitations, i.e., those that approximately satisfy $\bc$; as discussed in the text, the trough-like form of this dispersion enhances fluctuation effects. The inset shows a ``top view''\ of the dispersion: the black line represents modes at the minimum of the trough, which exactly satisfy $\bc$; self-organization results in the macroscopic occupation of one of these modes.}
	\label{fig:dispers2}
\end{figure}

In order to adapt Brazovskii's analysis to the present case, we must replace all instances of the momenta $\mathbf{k}$ by the sets of positive mode numbers $(m,n)$.  In particular, we must replace the expression $(|\mathbf{k}| - k_c)^2$ by $R^{-2}(m + n - \Lambda_0)^2$, where $R$ is the radius of the concentric cavity.  Therefore, the low-lying excitations in the present geometry do not lie on a circular or spherical shell in momentum-space, as they do in the original Brazovskii problem; instead, they lie on a linear ribbon along $m + n = \Lambda_0$ in mode-space (see Fig.~\ref{fig:dispers2}).  This difference does not affect Brazovskii's argument, except for some numerical factors of order unity, as we shall see below.

\begin{figure}
	\centering
		\includegraphics{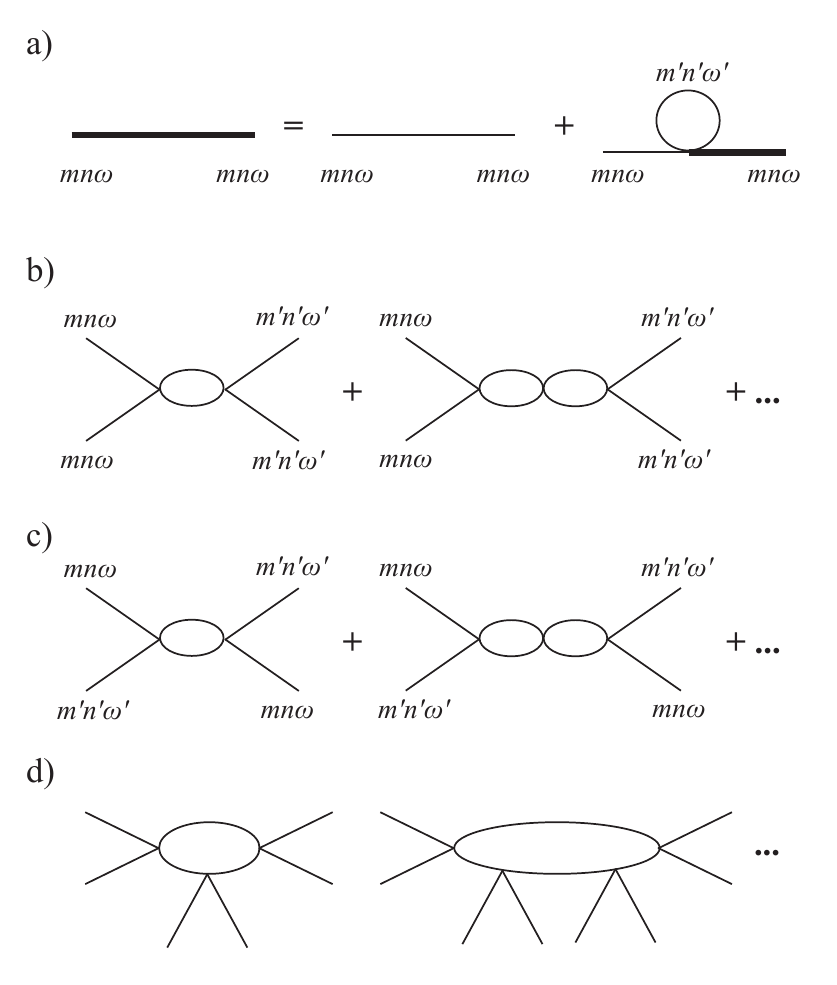}
	\caption{(a)~Dyson equation for the self-energy at one loop order
(i.e., the leading fluctuation correction to $r$).
(b)~A geometric series of corrections to the vertex (i.e., to $u$), which constitute the primary fluctuation corrections for $(m',n',\omega')\ne(m,n,\omega)$. (For  the classical case, $\omega'=\omega=0$.)\thinspace\
(c)~A geometric series of corrections to $u$ that contribute only when $(m',n',\omega')\approx(m,n,\omega)$.
It is these contributions that change the sign of $u$, thus causing a first-order transition.
(d)~Higher-order vertices that emerge under coarse-graining.}
	\label{fig:brazdiag}
\end{figure}

The primary consequence of order-parameter fluctuations, to leading (i.e., one-loop) order, is to renormalize the bare parameters $\mathcal{R}$ and $\mathcal{U}$ in the free energy Eq.~(\ref{eq:brazovskii}); thus, the fluctuation-corrected free-energy has the same form as Eq.~(\ref{eq:brazovskii}), but with corrected parameters $r$ and $u$ instead of $\mathcal{R}$ and $\mathcal{U}$ respectively. 
For the former, one must evaluate the Feynman diagram in Fig.~\ref{fig:brazdiag}a; the result is that $r$ is implicitly given by
\beq\label{eq:r0}
r = \mathcal{R} + \alpha \, \sbK \, \mathcal{U}\, r^{-1/2},
\eeq
in which the coefficient $\alpha$ in the \lq\lq self-energy\rq\rq\ term
is a geometrical factor that, in the present case, is approximately given by the expression

\beq\label{eq:alpha}
\alpha \approx 4 \sqrt{2} n_{\mathrm{modes}} \lambda / L,
\eeq
with $n_{\mathrm{modes}}$ being the number of modes having finesse high enough to be populated.
(As one might expect, fluctuation corrections are more important in cavities having a larger number of degenerate modes.)\thinspace\
Note that, regardless of the sign of $\mathcal{R}$, the corrected value of $r$ is \textit{positive};
hence the apparent second-order transition out of the disordered state is precluded by fluctuations.

\begin{figure}
	\centering
		\includegraphics{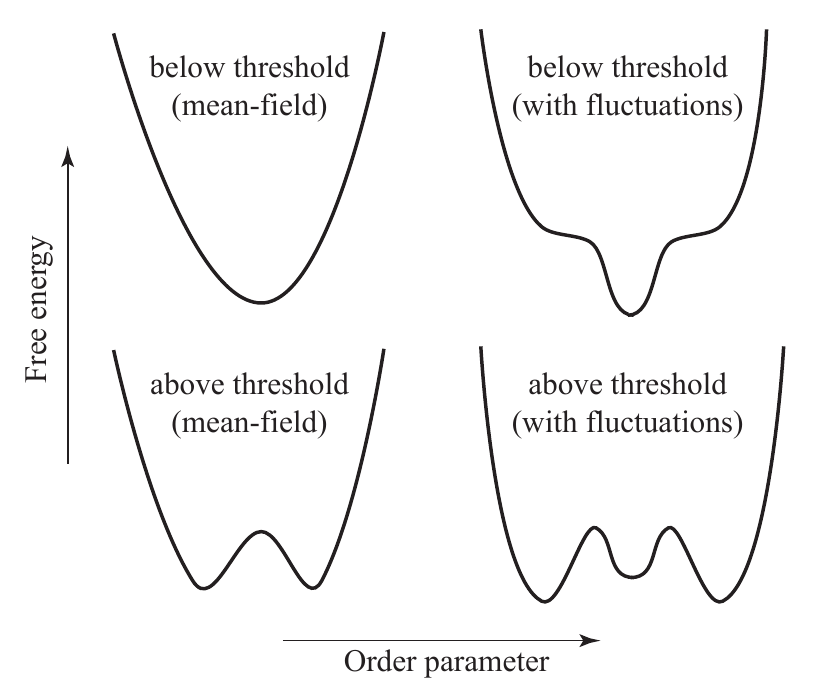}
	\caption{Schematic form of the free energy as a function of the order parameter, both above and below threshold, indicating how fluctuations change the character of the phase transition.}
	\label{fig:freeenergy}
\end{figure}

The leading corrections to $\mathcal{U}$ are given, for generic values of $m,n,m',n'$, by Fig.~\ref{fig:brazdiag}b. For $(m,n)=(m',n')$ one must also consider the diagrams in Fig.~\ref{fig:brazdiag}c.  Summing up these series of diagrams, one finds that the corrected value $u$, for $(m,n)=(m',n')$, is of the form

\beq
u = \mathcal{U}\,\frac{1 - (\mathcal{U}/r^{3/2})}{1 + (\mathcal{U}/r^{3/2})}.
\eeq
Evidently, $u$ turns negative for sufficiently small $r$. Na{\"\i}vely this would mean that the free energy becomes unbounded below; however, perturbative corrections encoded in diagrams such as Fig.~\ref{fig:brazdiag}d generate a positive six-point coupling, associated with a coupling denoted as $w$, which stabilizes the action and gives it the profile with multiple minima shown in Fig.~\ref{fig:freeenergy}. This profile for the free energy suggests that any phase transition that the system undergoes is likely to be first-order (i.e., discontinuous).

To determine when the transition becomes energetically favorable \textit{in equilibrium}, one should compare the free energy of the uniform state with that of the possible self-organized states. (We shall revisit this point for the nonequilibrium case in Sec.~\ref{sec:neqc}.)\thinspace\ Let us consider, first, the state in which all the atoms are self-organized in a single mode, so that
$\langle \psi_{mn} \rangle =
A \,
\delta_{m,\widetilde{m}}\,
\delta_{n,\widetilde{n}}$,
the coefficient $A$ being the amplitude of the order parameter.  If $\Phi_0$ is the bulk free energy of the disordered state and $F_1$ that of the ordered state, one can formally write
\beq
F_1 - F_0 = \int_0^A dA\,\frac{\partial F}{\partial A} =
\int dA \sum_{mn} \frac{\delta F}{\delta \psi_{mn}} \frac{\partial \psi_{mn}}{\partial A}.
\eeq
The motivation for this rewriting is that $h_{mn}\equiv\delta F / \delta \psi_{mn} $ is the biasing field that would render a certain order-parameter configuration stable.  Both the disordered and ordered states are locally stable at zero field; therefore, $h$ should go to zero at both ends but should be nonzero between, so as to ``drag''\ the system from one phase to the other.  The advantage of integrating $h$, as opposed to computing the free energies directly, is that one avoids having to compute terms in the free energy that are the same in both phases. The leading contributions to $h$ can be written as follows:
\beq
h_{mn} = \frac{1}{6} u A^3 -
\frac{1}{2} r A +
\frac{1}{2} u A \sum_{mn} \langle \psi_{mn} \psi_{mn} \rangle.
\eeq
The last term should, in principle, be computed to the same order as fluctuations have been computed in the disordered state, viz.~to one-loop order.  At this order there are two diagrams that need to be computed: Fig.~\ref{fig:brazdiag}a and Fig.~\ref{fig:brazdiag}b.
The corrections are substantially different for $(m,n)=(\widetilde{m},\widetilde{n})$ (i.e., the longitudinal component)
and $(m,n)\ne(\widetilde{m},\widetilde{n})$ (i.e., the transverse, or Brazovskii's anomalous, component).
The \textit{longitudinal} corrections have essentially the same form as those we computed in the disordered phase.
The \textit{transverse} corrections, however, diverge with system size, as a consequence of the Mermin-Wagner theorem (see the following section).  Swift and Hohenberg~\cite{swift:rbc} have shown that \textit{for a finite system} in two spatial dimensions, these corrections are small as long as $\sbK R \leq u^{-2/5}$, a condition that is met for sufficiently weak coupling~\cite{Note4}. Neglecting these contributions, one finds that the free-energy difference between the disordered and ordered states is given by

\beq
\Delta \Phi = \frac{\alpha}{2} (\sqrt{r_A} - \sqrt{r}) - \frac{1}{2 \mathcal{R}} (r_A^2 + r^2),
\eeq
where $r_A$ is defined implicitly via

\beq
r_A = \mathcal{R} + \frac{\alpha \mathcal{U}}{\sqrt{r_A}} + \mathcal{U} A^2
\eeq
and $r$ is given via Eq.~(\ref{eq:r0}).  Solving this pair of equations, one finds that $\Delta \Phi$ changes sign---and the equilibrium phase transition therefore occurs---when $\mathcal{R} \approx -(\alpha \mathcal{U})^{2/3}$.

    \subsection{Relevance of the Mermin-Wagner Theorem}\label{sec:MWTrelevance}

A well-known result in the theory of phase transitions, the Mermin-Wagner theorem (see, e.g., Ref.~\cite{chaikin}), states that long-range order is impossible in two dimensions for any thermodynamic system with a continuous symmetry (and short-ranged interactions). This result is a consequence of the large phase space associated with long-wavelength fluctuations of the \textit{direction} of ordering---in the present case, to fluctuations of the phase of the superfluid order parameter and/or the direction of ordering~\ref{sec:basicproperties}. Therefore, one would not expect an infinitely large sample to exhibit true long-range order at finite temperatures. This result is not, however, particularly relevant to the case at hand, for the following reason. At $T = 0$, the system is effectively three-dimensional rather than two-dimensional, owing to the additional dimension that corresponds to imaginary time; at sufficiently low temperatures, therefore, one expects the distances over which \textit{spatial} fluctuations destroy long-range order to exceed the system size for relatively small systems such as a typical BEC~\cite{Note5}. Even at higher temperatures, one can suppress long-wavelength fluctuations by using a system having multiple layers, so as to increase the effective stiffness against fluctuations of the order parameter.

    \subsection{Quantum Brazovskii transition}\label{sec:qbraz}

The {\it quantum\/} case of the Brazovskii transition, which occurs at $T = 0$, differs from the classical case in that the quadratic part $S_2$ of the action governing it has the form given in Eq.~(\ref{eq:qbraz}), viz.,
\beq
S_2\! =\! \int\! d\omega \sum_{ mn}
\widehat{\Phi}_{\omega mn}
\left[ \mathcal{R} + \omega^2 + \frac{\hbar (m + n - \Lambda_0)^2}{2 \mass R^2} \right] \widehat{\Phi}_{-\omega mn},
\eeq
in which the frequency integration variable $\omega$ has been rescaled to absorb certain dimensionful factors. The quartic term in the action, Eq.~(\ref{eq:Squartic}), also includes frequency integrals.
The presence of these frequency integrals, absent from the classical case, changes the spectrum of fluctuations. Qualitatively, this is because there are now \textit{two} dimensions transverse to the critical surface; therefore, instead of a ribbon of critical modes, one must consider a tube. If one suitably adapts the Brazovskii diagrammatic procedure, one arrives at the following implicit expressions for the fluctuation-corrected parameters:
\begin{subequations}
\begin{eqnarray}
r & = & \mathcal{R} + [\alpha \mathcal{U} \ln(\rgscale / r)], \\
u & = & \mathcal{U} \frac{1 - [\alpha \mathcal{U} / r]}{1 + [\alpha \mathcal{U} / r]}.
\end{eqnarray}
\end{subequations}
In these equations, $\rgscale$ is a high-energy cutoff (which would be of order $\Delta_A$ in the physical system).
It is tempting to interpret them as follows:
as $\mathcal{R} \rightarrow -\infty$, we have that $r \rightarrow 0$;
therefore, $r \sim \rgscale \exp(-|\mathcal{R}|/\alpha\mathcal{U})$,
and thus $r$ is always positive,
although it does become exponentially small in the $\mathcal{R} \rightarrow -\infty$ limit.
This would seem to suggest that criticality is not restored at zero temperature.
Furthermore, the fluctuation-corrected vertex has the approximate form
\beq
u \approx \mathcal{U}
\frac
{\rgscale - \alpha \mathcal{U}^2 \exp(|\mathcal{R}|/\alpha\mathcal{U})}
{\rgscale + \alpha \mathcal{U}^2 \exp(|\mathcal{R}|/\alpha\mathcal{U})},
\eeq
which suggests (cf. Sec.~\ref{sec:cbraz}) that metastability should set in when
$|\mathcal{R}| = \alpha\mathcal{U} \log(\rgscale/\alpha \mathcal{U}^2)$.

\begin{figure*}
	\centering
		\includegraphics{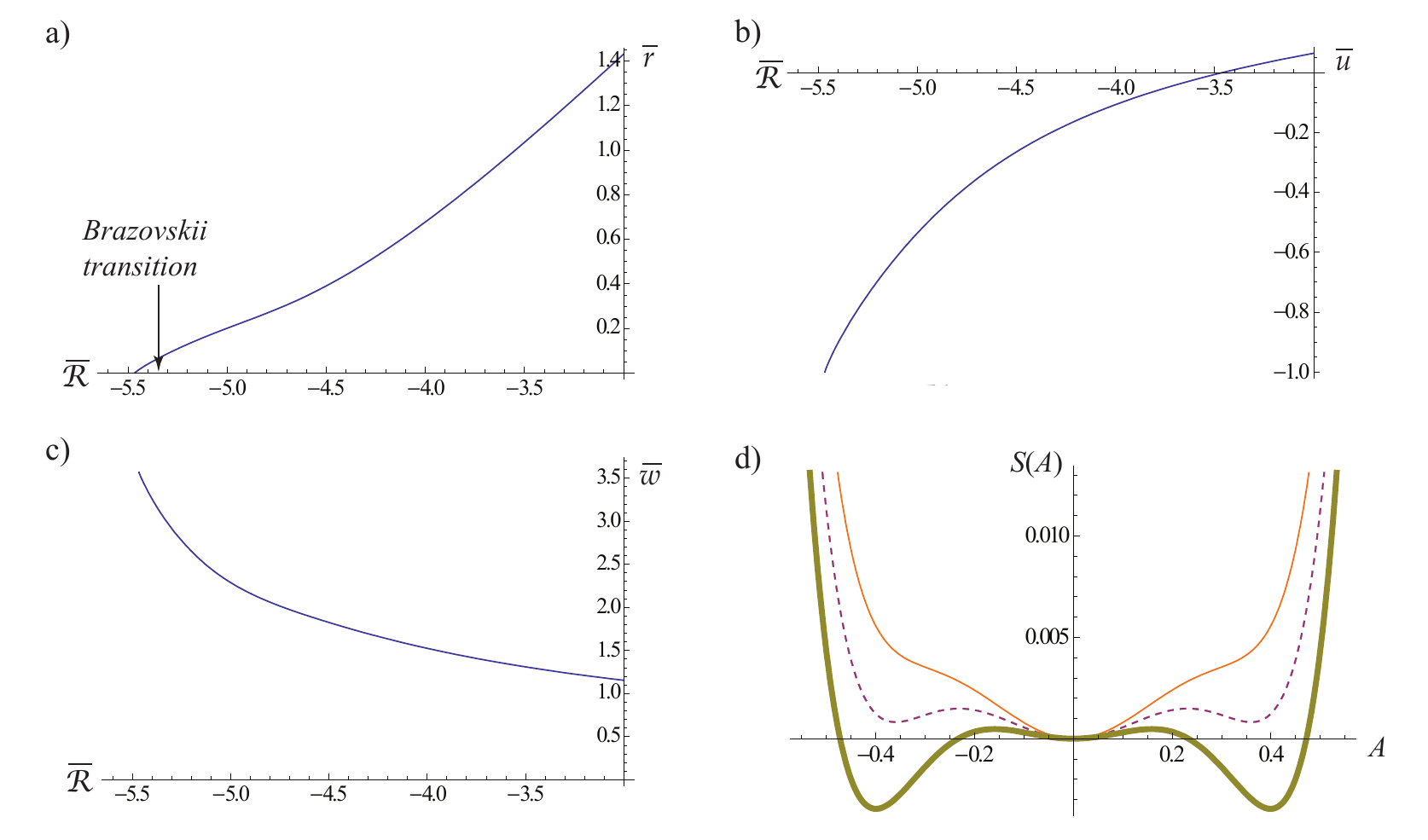}
	\caption{(a)-(c) Dependence of coarse-grained, fluctuation-corrected parameters on the bare control parameter $\mathcal{R}$, which is related to the laser strength, for a fixed value of the bare parameter $\mathcal{U}$. (The bars over the parameters signify that they have been rescaled as described in App.~\ref{app:RGequations}.) These results are obtained by integrating the renormalization-group equations derived in App.~\ref{app:RGequations}.
Panel~(a) shows the flow of the effective ``control parameter''\ $\overline{r}$ (which remains positive).
Panel~(b) shows the flow of the effective interaction parameter $\overline{u}$, which changes sign as discussed in the text. Panel~(c) shows the flow of the emergent six-point coupling $\overline{w}$. Finally, panel~(d) plots the free energy as a function of the order parameter $A$ for three values of $\overline{\mathcal{R}}$, viz. $-5.3$ (thin solid line), $-5.35$ (dashed line), and $-5.4$ (thick line). The first-order phase transition takes place at $\overline{\mathcal{R}} \approx -5.36$. These results are interpreted in terms of microscopic parameters in Sec.~\ref{sec:expt}.}
	\label{fig:mathoutput}
\end{figure*}

As these results are cutoff-dependent, however, one should investigate the quantum Brazovskii action using a more systematic scheme than Brazovskii's, e.g., a renormalization-group scheme such as that developed for the classical Brazovskii problem in Ref.~\cite{swift:bubbles}.
We describe the appropriate quantum adaptation of this renormalization-group scheme in App.~\ref{app:RGequations}.  The corresponding renormalized values of the various (de-dimensionalized) parameters are shown in Fig.~\ref{fig:mathoutput} as functions of the (de-dimensionalized) bare control parameter $\overline{\mathcal{R}}$.
The fact that
$\overline{u}$ first becomes negative for a smaller value of $-\overline{\mathcal{R}}$ than the value at which
$\overline{r}$ goes to zero indicates that the transition remains first order, and this is one of our main results.  This result can, however, be deduced on grounds that are more physically transparent, as we shall now discuss.

\subsection{Analogy with $O(\ncomponent)$ vector model}\label{sec:Onmodel}

At low energies, the only four-point couplings that are relevant, and therefore  survive under coarse graining,
involve either forward- or back-scattering.  Consider order-parameter modes $(m,n)$ that satisfy $\bc$; for these modes the quartic term takes the following form, in which time indices (which follow from locality in time) have been suppressed:
\beq
\mathcal{U}
\left( \sum_{mn}\,\psi_{mn} \psi_{mn} \right)
\left( \sum_{m'n'} \psi_{m'n'}\,\psi_{m'n'} \right)
(1 + \delta_{m,m'}\,\delta_{n,n'}).
\eeq
It is convenient to introduce the notation:
$\theta \equiv n / m$ and $\eta \equiv |\Lambda_0 - (m + n)|$,
which together provide an alternative labeling of the mode $(m,n)$.
In terms of these labels, the action takes the form
\bea
&&S=
\int d\omega \sum_{\eta \theta} (\mathcal{R}
+ \omega^2 + \eta^2) \psi_{\eta \theta}\,\psi_{\eta \theta}
\\
&&
+\mathcal{U} \sum_{\eta_i} \left[ \sum_{\theta} \psi_{\theta \eta_1} \psi_{\theta \eta_2}\! \right]\!\! \left[ \sum_{\theta'} \psi_{\theta' \eta_3} \psi_{\theta' \eta_4} \! \right] \!\! (1 + \delta_{\theta,\theta'}) \delta_{\sum \eta_i,0}. \nonumber
\eea
If one had ignored the contribution from terms having four equal values of $\theta$ (i.e., the $\delta_{\theta,\theta'}$ term), this action would be an instance of the $O(\ncomponent)$ model in $(1 + 1)$ dimensions, with $\ncomponent$ being the number of cavity modes satisfying the degeneracy condition $\bc$.
In the large $\ncomponent$ limit, terms having $\theta = \theta'$ do not contribute to the renormalization of $\mathcal{R}$: the index $\theta'$ (or $\theta$) is summed over in the relevant diagrams, and the single value $\theta = \theta'$ contributes negligibly to this sum. Therefore, the renormalization of $r$ in the present setting should be the same as that in the $O(\ncomponent)$ model at large $\ncomponent$.  It is known that in $(1 + 1)$ dimensions and for $\ncomponent > 2$, the parameter $r$ is always rendered positive by fluctuations; therefore, criticality is never achieved in the $O(\ncomponent)$ model, and any phase transition that might occur in the $O(\ncomponent)$ model---and hence the present model---must be first order. (This remark also applies to the classical Brazovskii case; the absence of criticality does \textit{not} depend on the weak-coupling approximation that Brazovskii's analysis employed, but follows from the low-energy structure of the free energy.)

Unlike the $O(\ncomponent)$ model, which is isotropic in order-parameter space, the present model \textit{does} undergo a transition (which is first order), owing to the additional contribution to the quartic term involving all four $\theta$'s being equal. In order to show that such a transition is feasible free-energetically, one can turn to the renormalization-group scheme outlined in App.~\ref{app:RGequations}.
%%%

\subsection{Fluctuation-corrected threshold: summary of results}\label{sec:summary}

We now list the fluctuation-corrected values of the threshold for self-organization, for the three cases discussed in previous subsections, in terms of the physically relevant microscopic parameters. In each case, the quantity listed is the fractional change in the threshold pump laser strength, i.e., 
$(\Omega_{\mathrm{th}} - \Omega_{\mathrm{th}}^{\mathrm{mf}}) / \Omega_{\mathrm{th}}^{\mathrm{mf}}$; 
for the regimes in which our analysis is valid, this quantity is generally much smaller than unity.

\begin{subequations}

(i)~For the ideal gas at $T > 0$, the corrected threshold for the Brazovskii transition is given by

\beq
\frac{\Omega_{\mathrm{th}} - \Omega_{\mathrm{th}}^{\mathrm{mf}}} {\Omega_{\mathrm{th}}^{\mathrm{mf}}} 
\simeq
\frac{1}{2} \left[ \alpha \frac{\sbK^2 R^2}{N \sqrt{\chi}} \right]^{2/3}.
\eeq
This expression is somewhat simpler than that given in Ref.~\cite{us} for the \textit{absolute} change in threshold, but is equivalent to it.  (In the present expression, we have explicitly included a factor $\alpha$ that is related to the number of cavity modes that couple appreciably to the atoms.)\thinspace\
The fluctuation correction can be thought of as consisting of two components: 
(i)~$\alpha$ is a geometric factor, introduced in Eq.~(\ref{eq:alpha}); 
(ii)~the other part of the equation is a measure of the number of particles \textit{per correlation area}, 
in the sense that 
(as discussed further in Sec.~\ref{sec:nucleation}) 
the quantity $\sqrt{\chi} / \sbK$ is analogous to a (far-from-criticality) correlation length.

(ii)~For the interacting BEC at $T > 0$, the shifted threshold is given by
\beq
\frac{\Omega_{\mathrm{th}} - \Omega_{\mathrm{th}}^{\mathrm{mf}}} {\Omega_{\mathrm{th}}^{\mathrm{mf}}} 
\simeq
\frac{1}{2} \left[ \frac{\alpha}{\sqrt{\chi}} 
\frac{ \{U - \xi N \} \sqrt{k_B T/\hbar} }{(\zeta_{\mathrm{mf}} N)^{3/2}} \frac{\sbK^2 R^2}{N} \right]^{2/3}.
\eeq
Note that this shift explicitly depends on temperature. 
At sufficiently low temperatures this shift picks up further corrections, 
and eventually crosses over to the quantum Brazovskii result 
discussed in Sec.~\ref{sec:qbraz}, which, in terms of microscopic parameters becomes

\beq
\frac{\Omega_{\mathrm{th}} - \Omega_{\mathrm{th}}^{\mathrm{mf}}} {\Omega_{\mathrm{th}}^{\mathrm{mf}}} \simeq
2.5 \left[ \frac{\alpha U \sqrt{\hbar^2 \sbK^2 / 2 \mass}}{(\hbar \zeta N \chi)^{3/2}} \right].
\eeq
In the fourth case (viz.,~that of an ideal gas at $T = 0$) power-counting suggests that the transition should remain second order, even after cavity-mediated interaction effects are included.  We have not discussed this regime in detail in the present paper; we plan to address its properties and achievability in future work.

\end{subequations}

\subsection{Signatures of criticality}\label{sec:signatures}

The Brazovskii transitions, both classical and quantum, being first order, do not exhibit the power-law dependencies (e.g., of the fluctuation correlation length, the order parameter, and its susceptibility) commonly associated with continuous phase transitions. However, as the Brazovskii transitions are only \textit{weakly} first-order (i.e., they involve small discontinuities in the order parameter), the influence of fluctuations on the atomic and  optical correlations should be experimentally accessible.  In particular, the fluctuation corrections to the density-density correlation function---i.e., Eq.~\ref{eq:r0}---should be experimentally detectable in one of two ways.  A straightforward way to detect these correlations is to release the atoms from the trap, and analyze the correlations in the noise of the spatial density profile of the atomic system. These correlations can then be related to the density-density correlation functions of interest by means of the scheme described in Ref.~\cite{demler04}, which involves the post-processing of absorption images.  An alternative approach---unique to the cavity QED setting, and preferable in that it does not automatically destroy the BEC---is through the correlations of the light emitted from the cavity.  At weak coupling, the \textit{intracavity}-photon correlations are directly related to the atomic density correlations, as follows.  The full action (up to Gaussian order) is of the form
$\omega a^\dagger a + (a^\dagger + a) \rho + \rho G \rho$,
where $G$ is the atomic correlation function given by Eq.~\ref{eq:r0}; one can in principle integrate out $\rho$, thus arriving at the relation $\langle a^\dagger a \rangle \sim 1/(\omega - G^{-1})$. The fluctuation corrections to $G$ are therefore manifest in the correlations of the emitted light.

The weak-coupling approach, just given, has a serious limitation when it comes to describing \textit{quantum} fluctuations: as discussed in the following section, $\hbar \kappa$ acts, in some ways, as an effective temperature for the atoms in the cavity.  Therefore, quantum effects are typically cut off by decoherence on a timescale comparable to $1/\kappa$.  However, $1/\kappa$ is also the timescale on which photons leak out of the cavity.  It would seem to follow, therefore, that effects associated with coherent, \textit{quantum} fluctuations take place on timescales too rapid to be detected via the leakage of light through the mirrors. This analysis, however, neglects the existence of cavity modes that couple relatively weakly to the atoms (so that they barely affect the effective temperature) and have considerably lower finesse. It is plausible, then, that the correlations of these modes can be used to probe the dynamics of the quantum fluctuations of the remaining degrees of freedom of the system.

\section{Nonequilibrium Effects at the Brazovskii transition}\label{sec:neqc}

In this section we reinstate the dissipative effects due to the cavity photon leakage rate $\kappa$, and consider their impact on the self-organization transition. As we shall discuss, the departure from equilibrium implied by a nonzero value of $\kappa$ has three kinds of consequences:
(i)~it cuts off critical fluctuations,
(ii)~it affects the timescale on which the system is able to escape from a metastable state, and
(iii)~it modifies the dispersion of long-wavelength excitations in the ordered state.
In this section we consider effects (i) and (ii); our discussion of effect (iii) is postponed to the next section, in which we discuss the properties of the ordered state.

To accomplish this reinstatement, we follow the standard prescription for expressing an effective equilibrium action in the nonequilibrium formalism (see, e.g., Sec.~4.7 of Ref.~\cite{kamenev}) and then augment the theory with the term $S_{\kappa}$ in Eq.~(\ref{eq:Seffective}).  Thus we arrive at a theory containing a copy of the equilibrium Brazovskii action involving fields on the $\pm$ contours, coupled to one another via $S_{\kappa}$, the consequences of which we shall now address using perturbation theory.

\subsection{Critical effects}
\label{sec:criticaleffects}

In this section we focus on the case of the interacting BEC at zero temperature (cf.~Sec.~\ref{sec:qbraz}), as the analysis is most transparent for this case.  We begin with the dissipative term,
\bea
S_\kappa & = &
\int
\frac{d\omega}{(\omega - \Delta_C)^2 + \kappa^2}
\int d^dx\,d^dx'\,
g_\alpha(\mathbf{x})\,
g_\alpha(\mathbf{x}')
\nonumber \\
& & \quad \times i \zeta \kappa \Delta_C \coth(\hbar \kappa/k_B T)\,
n_1(\mathbf{x}\omega)\,
n_1(\mathbf{x}\omega),
\eea
and re-express the atomic field $\Psi$ in terms of the condensate and non-condensed parts, as follows:
\bea
\Psi_c(\mathbf{x})&=&\sqrt{N_0} + \Phi_c(\mathbf{x}),
\\
\Psi_q(\mathbf{x})&=&\Phi_q(\mathbf{x}).
\eea
In terms of this decomposition, the primary quadratic contribution to the complete action arising from
$S_\kappa$ then becomes
\bea\label{eq:phiqq}
&& i \frac{\zeta N \Delta_C \kappa}{\Delta_C^2 + \kappa^2}
\sum_{mn} \int d\omega \Big(
2 \Phi^{\ast}_{\omega mn, q}\,
  \Phi^{\phantom{\ast}}_{\omega mn,q}
\\
&& \qquad \qquad +
\Phi^{\ast}_{\omega mn, q}\,
\Phi^{\ast}_{-\omega mn,q} +
\Phi^{\phantom{\ast}}_{\omega mn, q}\,
\Phi^{\phantom{\ast}}_{-\omega mn,q} \Big).
\nonumber
\eea
The most salient feature of this term is that its prefactor is frequency-independent. For a system to be at a \textit{quantum} critical point, it is necessary that the prefactor vanish as $\omega \rightarrow 0$ (see, e.g., Ref.~\cite{mitra06} and our App.~\ref{app:effectiveT}). When---as in the present case---this condition fails, the collective dynamics on sufficiently long timescales is classical.
Indeed, the term Eq.~(\ref{eq:phiqq}) is formally identical to the term
\beq
i k_B T\int d\omega\,
\Phi^{\ast}_{\omega,q}\,\Phi^{\phantom{\ast}}_{\omega,q}
\eeq
that arises for an otherwise isolated complex field $\Phi$ coupled to an equilibrium thermal environment that is at temperature $T$ [see, e.g., Eq.~(66) of Ref.~\cite{kamenev}]; one can therefore regard the coefficient
$\hbar \tilde{\kappa} / k_B \equiv \hbar \kappa \zeta N / (\Delta_C k_B)$
in Eq.~(\ref{eq:phiqq}) as an effective system temperature.
In particular, quantum correlations on timescales longer than $1/\tilde{\kappa}$ are washed out by the decoherence arising via  the leakage of photons from the cavity: this effect is analogous to the decoherence due to a \textit{finite temperature} that is known to occur near to a quantum critical point.

This point is---in principle---immaterial, as interaction effects preclude criticality regardless of the value of $\kappa$ (as a consequence of Brazovskii's argument); however, at $T = 0$ the fluctuation-corrected equilibrium control parameter is exponentially small, behaving as
$\exp\big(-\vert\mathcal{R}\vert/\alpha\mathcal{U}\big)$,
and therefore it should be possible to tune the system close enough to criticality that the nonequilibrium suppression of criticality due to dissipation is observable.

\begin{figure}
	\centering
		\includegraphics{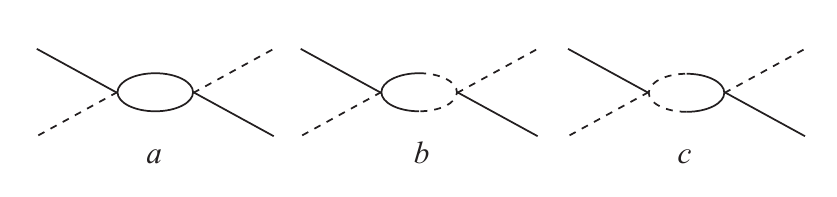}
	\caption{Contributions to the nonequilibrium vertex having external indices $ccqq$. For an introduction to the Keldysh diagrammatic notation see Ref.~\cite{kamenev}.}
	\label{fig:neqdiags}
\end{figure}

We now turn to the issue of influence of the nonequilibrium terms given in Eq.~(\ref{eq:phiqq}) on the effective quartic \textit{interaction} vertex. At tree-level, the interaction vertex only couples terms having Keldysh indices \textit{cqqq} or \textit{cccq}.  However, the nonequilibrium terms generate an effective \textit{ccqq} vertex, via the Feynman diagrams shown in Fig.~\ref{fig:neqdiags}. This vertex brings the factor
\beq
\tilde{u}\equiv \left(\frac{\kappa \zeta N}{\Delta_C} u \right)^{2},
\eeq
in which $u$ is the fluctuation-corrected equilibrium vertex (see Sec.~\ref{sec:cbraz}). The most notable feature of this vertex is that it does \textit{not} change sign when $u$ does. In the regime considered in the present work, we have that $\kappa\ll\Delta_C$, and hence $\tilde{u}$ provides a subdominant correction to $u$, and therefore cannot prevent the net vertex from changing sign, signaling a first-order transition.  It is possible, however, that in the opposite regime, in which $\kappa\gg\Delta_C$, this correction term would dominate; in this regime, this term might be capable of preventing the Brazovskii transition from taking place at all.

\subsection{Nucleation and state selection}\label{sec:nucleation}

In this section we address the dynamics of the emergence of self-organization associated with the Brazovskii transitions, classical and quantal. As these transitions are first order, one expects them to exhibit regions of two-phase coexistence, in which some parts of the cloud have self-organized and others have not. The time interval that the system takes to approach the steady state, in which the entire system is self-organized, depends on the energetics of critical droplets of the minority phase, which determines their nucleation rate. There are three regimes of interest, distinguished by the primary mechanism responsible for fluctuations:
(i)~near zero temperature in an isolated system (i.e., in the regime where both the system temperature and $\tilde{\kappa}$ are smaller than $U$), quantum tunneling is the primary cause of nucleation;
(ii)~at temperatures that are high compared with $U$, in an essentially isolated system ($k_B T \gg \tilde{\kappa}$), it is thermal activation that is the primary cause; and
(iii)~near zero temperature in the far-from-equilibrium regime (i.e., when $\tilde{\kappa}$ exceeds the temperature and $U$), nucleation is primarily triggered by extrinsic force noise that originates with fluctuations in the photon population in the cavity.  Owing to the formal analogy between $\tilde{\kappa}$ and $T$, discussed in Sec.~\ref{sec:criticaleffects} and App.~\ref{app:effectiveT}, cases~(ii) and (iii) can be treated by similar means. 

In all three cases, an essential ingredient is the energy barrier for thermal (or quantal) nucleation. In many settings this would be easy to read off from the tree-level (i.e., mean-field) Landau free energy (or action); however, in the case at hand, the Landau free energy \textit{does not} predict a first-order transition at tree level.  On the other hand, it is not \textit{prima facie} legitimate to use the \textit{fluctuation-corrected} free energy that was calculated above, as this incorporates fluctuations on \textit{all} length scales, including those larger than the droplet itself. In general, therefore, one must follow a procedure like that due to Hohenberg and Swift~\cite{swift:bubbles}, in which only fluctuations on length-scales smaller than the droplet diameter are self-consistently integrated out.

Motivated by the wish to obtain analytical results, we focus, in the present work, on nucleation kinetics in the regime in which the corrected free energies of the self-organized and uniform phases are sufficiently similar that the critical droplet size is comparable to the system size.  In this regime, the error incurred by using the fully renormalized bulk free energy is expected to be relatively small (as, in this case, most of the renormalizations should already have taken place), and one can legitimately use the fully renormalized parameters computed in Secs.~\ref{sec:cbraz} and~\ref{sec:qbraz}.

\subsubsection{Classical nucleation}\label{sec:classicalnucleation}

\begin{figure}
	\centering
		\includegraphics{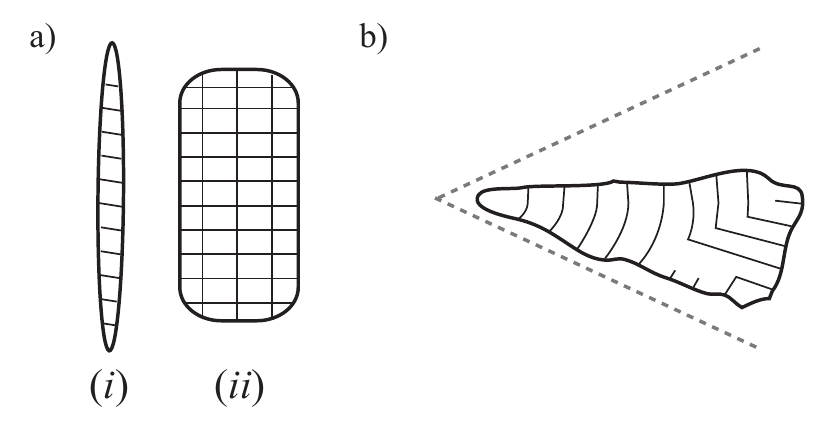}
		\caption{(a) Wulff droplets, corresponding to the TEM$_{00}$ mode ($i$), and to a higher-order mode ($ii$), respectively. The droplets should become less anisotropic (i.e., less ``needle-like'') for higher-order modes; it is, however, possible that the optimal droplets in these cases have more complicated shapes. (b)~Defected droplets, which are favored for $r \approx r_c$, as discussed in the text. For these, the energetic cost of introducing defects inside the droplet is outweighed by the increase in the fraction of the interface that is transverse.}
	\label{fig:nucleate}
\end{figure}

In the classical case, one needs to compute the free-energy barrier to the nucleation of a critical droplet, i.e., the smallest possible droplet intrinsically capable of growing until it encompasses the entire cloud.  The standard procedure for doing this is to identify appropriate saddle-point configurations of the effective free energy. As these saddle-point configurations involve droplets of the ordered phase immersed in the uniform phase, one is interested in spatially varying configurations $\Phi(\mathbf{x})$, and the position coordinates are therefore the appropriate ones to consider. In these coordinates, the Euler-Lagrange equation corresponding to the Brazovskii free energy takes the approximate form:

\beq\label{eq:eom}
\xi_0^4 (\nabla^2 + \sbK^2)^2 \Phi(\mathbf{x}) + [ r - 2 |u| |\Phi(\mathbf{x})|^2 + 3 w |\Phi(\mathbf{x})|^4 ] \Phi(\mathbf{x}) = 0,
\eeq
in which $\xi_0 \sim \sqrt{\chi} / \sbK $ is an effective healing length for the crystalline order parameter.
In order to find the saddle-point configuration, we look for solutions of the Euler-Lagrange equation that obey the boundary conditions that $\Phi = 0$ near the edge of the cloud and $\Phi(\mathbf{x}) = A\,\Xi_{\tilde{m} \tilde{n}} (\mathbf{x})$ near the middle of the cloud, where $A = \pm \sqrt{(2u + \sqrt{4 u^2 - 12 r w}) / 6 w}$ is the value of the order parameter in the ordered state, and $(\tilde{m},\tilde{n})$ is the mode into which the atoms are self-organized.
In the \textit{conventional} Brazovskii problem, the difference in energy between a configuration including a droplet of area $\mathcal{A}$ and the uniform state is given by

\beq
\label{eq:FEdifference}
F_{\mathrm{drop}} = (F_0 - F_A) \mathcal{A} + \sigma_\perp L_\perp + \sigma_\parallel L_\parallel,
\eeq
in which $F_0$ ($ = 0$) and $F_A$ are, respectively, the free energy densities of the uniform and self-organized states; $\sigma_\perp$ ($\sigma_\parallel$) is the energy cost of an interface perpendicular (parallel) to the lamell\ae; and $L_\perp$ ($L_\parallel$) is the length of the interface that lies perpendicular (parallel) to the lamell\ae.
The concentric-cavity geometry differs from that considered in Ref.~\cite{swift:bubbles} in that it is not translationally invariant in its radial direction: both the atomic density and the mode functions depend on $\mathbf{x}$ and so, therefore, do the parameters in the Brazovskii model.  As our purpose in the present section is to focus on order-of-magnitude estimates, we shall neglect this complication.
A further difference between the present case and the conventional Brazovskii problem is that the modes are checkerboard-shaped rather than lamellar (see Ref.~\cite{us}, and also Sec.~\ref{sec:basicproperties}).  Thus, a generic interface has some aspects of both transverse and longitudinal character, and the optimal droplet shape varies from mode to mode. In what follows, we focus on the (physically most relevant) modes for which $m \ll n$, and consider droplets of the form sketched in Fig.~\ref{fig:nucleate}.

Returning to Eq.~(\ref{eq:FEdifference}), we see that the free-energy difference per unit area, $\Delta F$ is of order

\beq\label{eq:deltaf}
\Delta F \approx - \frac{r_c - r}{r_c} \frac{u^3}{27 w^2},
\eeq
in which $r_c$ is the value of the control parameter at which the equilibrium transition occurs. As argued in Ref.~\cite{swift:bubbles}, the \textit{width} of a longitudinal interface is approximately $\xi_0 (\sbK \xi_0 / 2r)$, whereas that of a transverse interface is $\xi_0$. Near the phase transition, $r \approx (\alpha \mathcal{U})^{2/3}$, which is small relative to $\sbK \xi_0$ for weak coupling $\mathcal{U}$; hence, longitudinal interfaces are larger than transverse ones. Furthermore, the interface energy cost per unit area is given by the interface width multiplied by the quantity $f_0 = u^3 / (27 w^2)$, which is related to the curvature of the free-energy landscape about the minimum corresponding to the self-organized state. Thus, the total interface energy cost is given by the expression

\beq
\frac{u^3}{27 w^2} \xi_0 \left(L_\perp + \frac{\sbK \xi_0}{2r} L_\parallel \right).
\eeq
Because transverse interfaces cost less energy at weak coupling,
the optimal droplet shape is needle-like, as shown in Fig.~\ref{fig:nucleate}.

To determine the free-energy barrier for droplet nucleation, we use the well-known Wulff construction (see, e.g., Ref.~\cite{ll5}); in the present case, this amounts to considering bubbles of dimension
$\propto \sigma_\parallel$
in the \textit{perpendicular} direction and
$\propto\sigma_\perp$ in the \textit{parallel} direction. Thus, the free energy of a configuration including a droplet is:

\beq
F_{\mathrm{drop}} = \gamma \xi_0^2 \frac{\sbK \xi_0}{2r} \left( - \gamma \, \frac{r_c - r}{r_c} + 2 \right) \frac{u^3}{27 w^2},
\eeq
where $\gamma$ is a variational parameter that sets the overall scale of the droplet. By minimizing $F_{\mathrm{drop}}$ with respect to $\gamma$, we see that the critical bubble is that for which $\gamma = r_c / (r_c - r)$; thus, the free-energy barrier for the thermal nucleation of droplets of the ordered state is

\beq
F_{\mathrm{drop}} = \Delta F \xi_0^2 \frac{\sbK \xi_0}{2r}.
\eeq
The nucleation rate follows directly, being given by $r \exp(-F_{\mathrm{drop}} / k_B T)$ or, in the case that nucleation is due to external noise [i.e., case~(iii)], by $r \exp(-F_{\mathrm{drop}} / \hbar \tilde{\kappa})$.
For a discussion of the relevant experimental parameters, see Sec.~\ref{sec:expt}.

\subsubsection{Qualitative features of the defect morphology}

Near coexistence (i.e., for $r \approx r_c$), the critical droplet is arbitrarily large and---even for a highly anisotropic droplet---the energetic cost of the longitudinal interface (which scales linearly with bubble size) becomes greater than that of introducing localized defects, arranged so that the surface of the bubble is made as transverse as can be.  A possible arrangement of such defects is shown in Fig.~\ref{fig:nucleate}. It was argued in Ref.~\cite{swift:bubbles} that such defects are energetically favorable only for $|(\mathcal{R} - \mathcal{R}_c)/\mathcal{R}_c)| \leq (U / \zeta N)^{10/27}$, which is a narrow range compared with the thermal-fluctuation-dominated regime, which obtains for $|(\mathcal{R} - \mathcal{R}_c)/\mathcal{R}_c)| \leq 1$.

\subsubsection{Quantum tunneling}

In the case of quantum tunneling rather than thermal barrier crossing, the argument of Sec.~\ref{sec:classicalnucleation} must be modified in two ways. First, the expression for the tunneling rate should be given by the form $\omega_0 \exp(-S_0)$, where $\omega_0$ is a characteristic collective frequency (e.g., being proportional to the value of the renormalized control parameter $r$ in the disordered phase) and $S_0$ is the appropriate instanton action~(see, e.g., Ref.~\cite{altland}).  A crude approximation to $S_0$ is the product of the width (which is of order $A$) and height of the energy barrier; thus the difference between the initial and final values of the order parameter would act as an effective inverse temperature.  Second, one must use the coarse-grained values of $r$, $u$, and $w$ from the \textit{quantum} rather than the classical model, i.e., from Sec.~\ref{sec:qbraz} rather than Sec.~\ref{sec:cbraz}.

\section{Properties of the crystalline state}\label{sec:elasticity}

In this section we examine various properties of the crystalline state exhibited by the coupled atom-light system, including its basic structure, elementary excitations, and topological defects.

	\subsection{Basic properties}\label{sec:basicproperties}
		
		\subsubsection{Concentric cavity}
		
\begin{figure}
	\centering
		\includegraphics{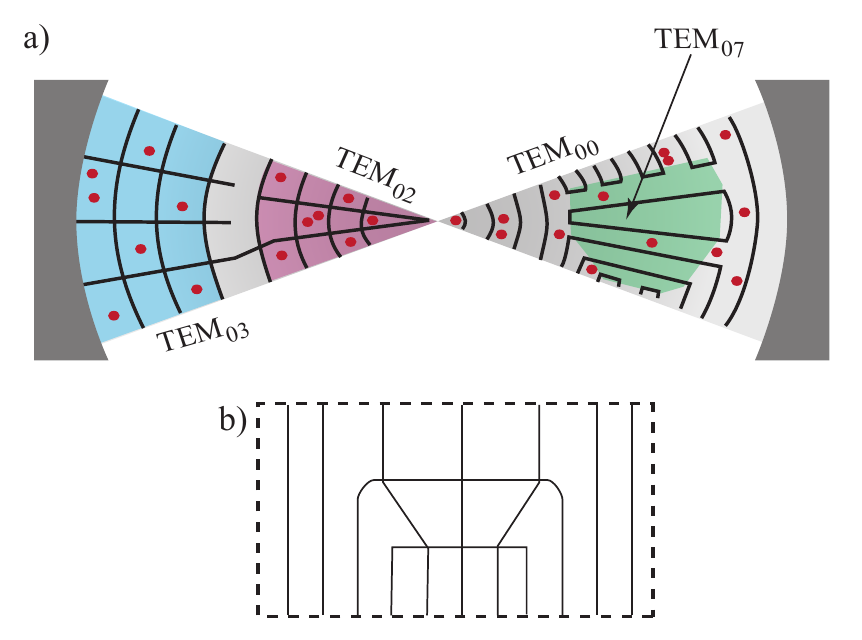}
	\caption{Elementary excitations of the self-organized state in the concentric cavity. (a) Domains that have self-organized into distinct modes can be separated by analogs of grain boundaries (left half of panel) or by continuous textures (right half of panel). (b)~Excitations that are analogous to the splay mode in smectic-A liquid crystals (see Sec.~\ref{sec:phonons}). Lines indicate nodes of the cavity electromagnetic field. The curved wavefronts along the radial direction have been drawn as flat lines to emphasize that the sketched feature is small-scale, relative to the size of the cavity.}
	\label{fig:ordstb}
\end{figure}

Soft condensed matter systems that undergo the Brazovskii transition commonly exhibit one-dimensional, lamellar patterns. The present realization does not, owing to the influence of boundary conditions on the optical mode structure, and therefore on the possibilities for atomic crystallization.  Instead, in the concentric cavity, the ordered states follow the two-dimensional optical mode patterns, which may be visualized as distorted checkerboard patterns, as shown in Fig.~\ref{fig:ordstb}.  Locally, the atomic density selects amongst the cavity modes by crystallizing into the \lq\lq odd or even squares\rq\rq\ of the selected mode (cf.~Fig.~\ref{fig:ordstb} and the discussion in Sec.~\ref{sec:expectations}). In physical realizations, the corresponding states of crystallization would not be exactly degenerate, as optical modes having stronger angular variation (i.e., larger $m$) are of lower finesse; on the other hand, repulsive interactions have a stronger impact on atoms that are crystallizing into modes of lower $m$~\cite{Note6}.  Such effects can readily be accounted for within our model, via the introduction of fictitious fields that would bias the system towards crystallizing into certain modes. In practice, the most straightforward way to include such effects is by adding the terms

\beq
S_{\mathrm{fict}} = \sum_{mn} h_{mn}\, \rho_{mn}^2
\eeq
to the Brazovskii action, which would raise (or lower) the threshold laser power in a mode-dependent way.

		\subsubsection{Other multimode cavities}
		
A well-known example of a multimode cavity is the confocal cavity, in which all the even TEM modes are degenerate~\cite{siegman}.  (It is also possible to make multimode cavities in which every $p^{\rm th}$ TEM mode is degenerate.)\thinspace\  These cavities have the practical advantage over the concentric cavity that their stability criteria are easier to fulfill (e.g., their mode structures are more robust with respect to mirror misalignment).\thinspace\ To the extent that it is legitimate to think of such cavities as each having a continuous family of degenerate modes (i.e., provided they possess a large number of modes that are both degenerate and not heavily suppressed by diffractive losses), the self-organization transition in these cavities should belong to Brazovskii's universality class, and our analysis of the transition itself should extend to these models.  Where confocal cavities are likely to differ from concentric ones is in the geometry of the ordered states and of their defects, which is much more involved in the confocal case because of its less-evident symmetry structure.

A feature common to most multimode geometries is that it is possible to tune the system across the point at which the modes are degenerate by gradually changing the mirror spacing. Thus, one can explore the crossover between the multimode physics discussed in the present work and the single-mode physics realized in Ref.~\cite{baumann}. 

    \subsection{Phonons and nonequilibrium elasticity}\label{sec:phonons}

\begin{figure}
	\centering
		\includegraphics{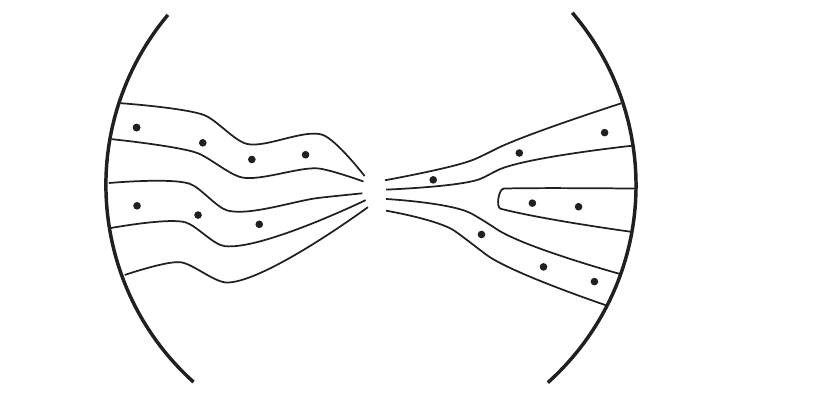}
	\caption{Case of the large-solid-angle concentric cavity, in which the atom-light system possesses a continuous symmetry associated with the relative phase between the $+m$ and $-m$ components of each mode function. This symmetry, when broken by the self-organized atomic cloud, leads to the existence of both phonon excitations (shown in the left panel of the figure) and true edge dislocations (shown in the right panel of the figure).}
	\label{fig:ripples}
\end{figure}

Manifestly, the concentric cavity geometry does not possess translational invariance in the radial direction; hence, there are no translational Goldstone modes in the radial direction.  The geometry that we have primarily considered in this paper does not possess translational invariance in the angular direction either, because of the hard-wall boundary conditions that we imposed on the mode functions at the edge of the cavity, when computing the mode structure (see Sec.~\ref{sec:modestructure}).  For the usual experimental situation, in which the cavity mirrors cover a relatively small solid angle, this is the relevant case.  In the opposite regime, in which each cavity mirror occupies most of a hemisphere, one \textit{would} essentially recover translational invariance in the angular direction; concomitantly, there would be phonons, corresponding to the \lq\lq rippled\rq\rq\ atomic arrangement shown in Fig.~\ref{fig:ripples}. For large cavities, such excitations would have a linear, phononlike spectrum, with a speed of sound related to the order parameter for crystallinity.

The concentric cavity geometry does, however, possess an analog of \textit{rotational} invariance, in that the energy is unchanged if one reorganizes the crystallization of atoms in mode $(m,n)$ into crystallization in a degenerate mode $(m',n')$. For a large cavity, having many modes, this symmetry is effectively continuous.  Consequently, there are low-energy excitations involving the gradual variation of $m$ and $n$ across the cavity; an example of such an excitation is sketched in Fig.~\ref{fig:ordstb}. The physics of these layer-wandering excitations is analogous to that of the splay mode of smectic liquid crystals~\cite{chaikin}; as in the liquid-crystal case, the effective elastic energy for these excitations takes the Landau-Peierls form:

\beq
F_{\mathrm{el}} = K_1 \int d^2 x \, (\nabla_\perp^2 \theta)^2 + \cdots,
\eeq
where $\theta$ ($ \equiv n / m$) parametrizes the macroscopically occupied mode, and the ellipses indicate terms involving higher powers of the gradient operator.  Near the transition, the wandering rigidity $K_1$ is proportional to the square of the equilibrium order parameter and also to the fourth power of the healing length $\xi_0$ in Eq.~(\ref{eq:eom}).

We note in passing that the nonequilibrium character of the phase transition affects the spectrum of phonon and wandering excitations at very long wavelengths.  For example, the effective retarded Green function for phonons, which in general has the form
\beq
G_R(\omega,m,n) \approx
\frac{1}{\omega^2 - K_{\mathrm{eff}} (m + n)^2 + 2 i \omega \tilde{\kappa}}\,,
\eeq
has purely imaginary poles when $\sqrt{K_{\mathrm{eff}}} (m + n) \leq \tilde{\kappa}$;
modes satisfying this criterion are diffusive rather than propagating.
This idea, which was discussed in Refs.~\cite{littlewood06, littlewood07} in the context of excitonic condensates, has nontrivial consequences for, e.g., the spatio-temporal decay of correlations in sufficiently large systems.

    \subsection{Defects}\label{sec:defects}

In addition to the low-energy splay excitations (which are analogous, in some ways, to phonons), the ordered state can also have gapped excitations or defects, which in the present case are analogous to grain boundaries (see left panel of Fig.~\ref{fig:ordstb}). When the ordered states on the two sides of such a boundary are of opposite parity, as in the figure, the boundary wipes out a fraction of a row of crystalline order. Its energetic cost is therefore approximately $L \sigma_\perp$, where $L$ is the length of the boundary and $\sigma_\perp$ is the interface energy discussed in Sec.~\ref{sec:nucleation}. These defects are analogous to conventional topological defects in the sense that, for certain configurations of the order parameter at the boundary of some region in the cavity (e.g., at the edges of the cavity), the system is forced to have at least one defect somewhere inside this region. These defects are not, however, directly related to the existence of continuously broken symmetries in the system.

Other kinds of topological defects might also be realizable. For instance, in the large-solid-angle case discussed above in Sec.~\ref{sec:phonons}, genuine edge dislocations, of the kind illustrated in Fig.~\ref{fig:ripples}, may arise. Another possibility is a \textit{texture} of the kind sketched in the right-hand panel of Fig.~\ref{fig:ordstb}: such a texture would be analogous to a closed lamella in the conventional Brazovskii case. It is not clear, however, that such textures are experimentally \textit{feasible}, as they would require that the system self-organize into a high-$m$ mode in some region of the cavity.

\begin{figure}
	\centering
		\includegraphics{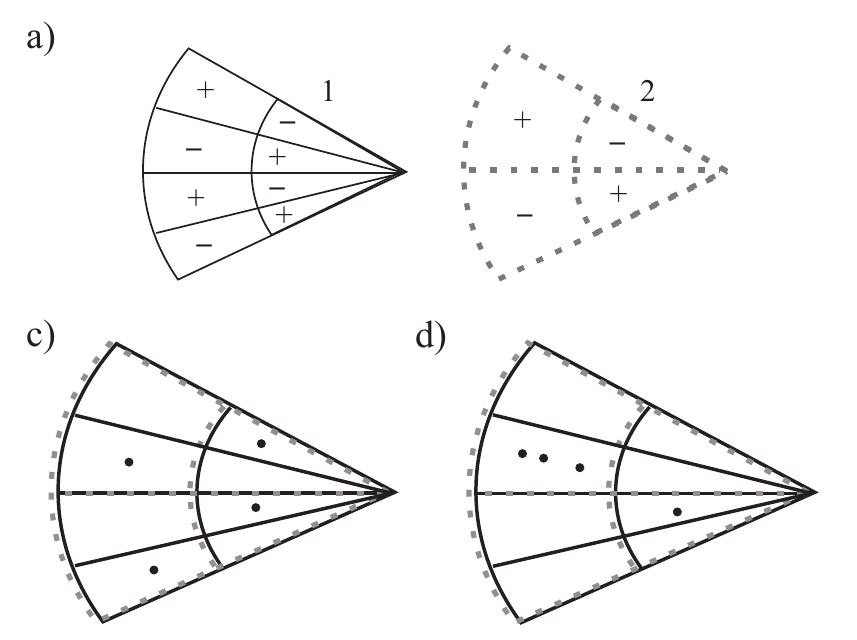}
	\caption{Proposed scheme for detecting supersolid order.
(a)~Profiles of two cavity modes: Mode~1 (into which the atoms self-organize) and Mode~2 (which can be used to detect phase coherence, as discussed in Sec.~\ref{sec:supersolidity}).
The two modes are degenerate; Mode~2 possesses more nodes along the $z$ direction (i.e., perpendicular to the plane of the figure). The $\pm$ signs describe the phases of the electromagnetic fields in the two modes relative to some reference (e.g., the pump laser) in various regions of the cavity.
%(b)~Phases of the two modes (relative to some reference, e.g., the pump laser) in various parts of the cavity.
(b)~Atomic configuration in which the atoms emit constructively into Mode~1 and destructively into Mode~2.
In the insulating phase, this is the typical configuration, as the number of atoms per site is fixed; hence, there is suppressed emission into Mode~2.
(c)~Atomic configuration in which the atoms emit constructively into both Mode~1 and Mode~2. Such configurations, which involve multiple occupancy, occur in the superfluid phase but are suppressed in the insulating phase; hence, the amount of light emitted into Mode~2 is a measure of superfluidity.}
	\label{fig:ssdet2}
\end{figure}

    \section{Supersolid aspects of the self-organized state}\label{sec:supersolidity}

The spatially ordered state of a BEC in a multimode cavity is a \lq\lq supersolid\rq\rq\ in the following sense. It possesses emergent forms of both crystalline and superfluid order: i.e., it spontaneously breaks two \textit{continuous} symmetries, the rotational invariance of space (to the extent that the cavity admits an effectively continuous family of modes) and the $U(1)$ invariance associated with the phase of the condensate wavefunction.  The properties, and even the existence, of supersolids have recently been active issues in condensed-matter research~\cite{shevchenko, toner08, supersolids, davis:superglass, reppy10}.  Amongst traditional condensed-matter systems, the primary candidate for exhibiting supersolidity is solid $^4$He, which was conjectured to have a supersolid phase in the late 1960s~\cite{andreev:lifshitz, chester}.  Shortly thereafter, Leggett~\cite{leggett} predicted that a supersolid would exhibit ``nonclassical rotational inertia'' when rotated sufficiently slowly, owing to the quantization of angular momentum of a rotating superfluid. Evidence for this phenomenon was reported in Ref.~\cite{supersolids}; the interpretation of this and subsequent experiments is, however, still controversial.  It has been proposed, for instance, that rather than indicating bulk supersolidity, the missing moment of inertia arises because of superfluidity in dislocation cores~\cite{shevchenko, toner08}, because of elastic effects arising from the presence of $^3$He impurities~\cite{west09}, as a by-product of glassiness~\cite{davis:superglass}, and so on.

One must distinguish between two types of questions regarding supersolids:
(i)~whether solid $^4$He, or any other neutral substance having \lq\lq realistic\rq\rq\ interparticle interactions, is supersolid in the sense defined above, and
(ii)~what properties a supersolid would possess, should one exist.
In the condensed-matter context, attempts to address question (ii)~have been vitiated by the uncertainty about whether the substance being studied is in fact supersolid, whereas attempts to address question (i)~have been hampered by  imperfect understanding of the characteristic phenomenology of supersolids.
The advantage of ultracold atomic realizations of supersolidity, such as the present one, is that one can explore question~(ii) without first addressing question~(i), as the existence of both superfluid and crystalline order is relatively easy to establish.  The existence of \textit{crystallinity} can be deduced via the superradiant emission of light (see Sec.~\ref{sec:signatures}), whereas that of superfluidity may be explored using a range of standard techniques (see, e.g., Ref.~\cite{baumann}).

Furthermore, it is in principle possible in the present setting to test for supersolidity in a manner that enables one to distinguish between scenarios in which the \textit{bulk} of the sample is a supersolid, and those that involve phase separation of some kind---e.g., scenarios in which BEC is restricted to, e.g., dislocation lines.  This can be accomplished via an analysis of the spatial correlations of the light emitted from the cavity.  Such a technique is an adaptation of that developed by Ref.~\cite{ritsch07} to explore the superfluid-insulator transition in an optical lattice.  It uses the fact that the number of particles per site is not fixed in a superfluid; therefore, even if the emission into a particular cavity mode is zero on average because of destructive interference between the contributions from even and odd sites, local atomic number fluctuations would render this destructive interference imperfect and would lead to a nonzero photon population, which can be detected in the light leaking out of the cavity.  This idea is sketched in Fig.~\ref{fig:ssdet2}; for calculational details, we refer to Ref.~\cite{ritsch07}.

		\subsection{Coupling the superfluid order parameter to the solid order parameter}

A well-known manifestation of supersolidity is the nonclassical behavior of the the moment of inertia~\cite{leggett}, which results from the requirement that the macroscopic wavefunction be single valued.  This effect can be explored directly in the present setting, e.g., by imparting angular momentum to the BEC via an auxiliary laser beam that carries orbital angular momentum.  Furthermore, one can study the implications of the presence of crystalline order for the superfluid transition~\cite{pmg:toner} by increasing the density of the atomic cloud in the cavity until it undergoes Bose-Einstein condensation.  A third possibility is to displace the superfluid from the center of the overall dipole trap and observe its relaxation~\cite{mckay}; this should provide information about the coupling between the Anderson-Bogoliubov mode of the superfluid and the excitations (both phononlike and topological) of the crystal. In particular, it should be possible to tune the cavity geometry \textit{across} a multimode geometry by adjusting the mirror spacing, as discussed in Sec.~\ref{sec:basicproperties}, thus altering the spectrum of crystalline excitations.

It should be emphasized that all these experiments depend crucially on the cavity's being a \textit{multimode} one, and on the broken spatial symmetry being at least approximately continuous. In a single-mode cavity, in which the broken symmetry is of the discrete (i.e., even/odd) type (and, moreover, the interactions are effectively infinite-ranged), the ``solidity'' is of a different kind; in particular, there can be neither Goldstone modes nor topological defects in the solid.

			\subsection{Supersolid-``Mott'' transition}\label{sec:SItrans}

As discussed in, e.g., Ref.~\cite{leggett}, the ``normal solid'' state that competes with a supersolid is analogous to a Mott insulator with regard to its transport properties.  In the setting of self-organized atom-light crystals, for laser intensities well above threshold one expects the emergent lattice potential to be sufficiently deep to cause the supersolid state to have undergone a transition into a nonsuperfluid state.  This state can be either a Mott insulating state (which would be a normal solid) or a Bose glass state (which, too, would have many of the properties of a normal solid, particularly a non-quantized moment of inertia).  The latter possibility arises even in the absence of extrinsic disorder because vacancies and dislocations in the self-organized lattice might dynamically generate disorder.  The Mott insulator state (which is incompressible) and Bose glass state (which is compressible) should be distinguishable via, e.g., their large-scale spatial density profiles.

\begin{figure}
	\centering
		\includegraphics{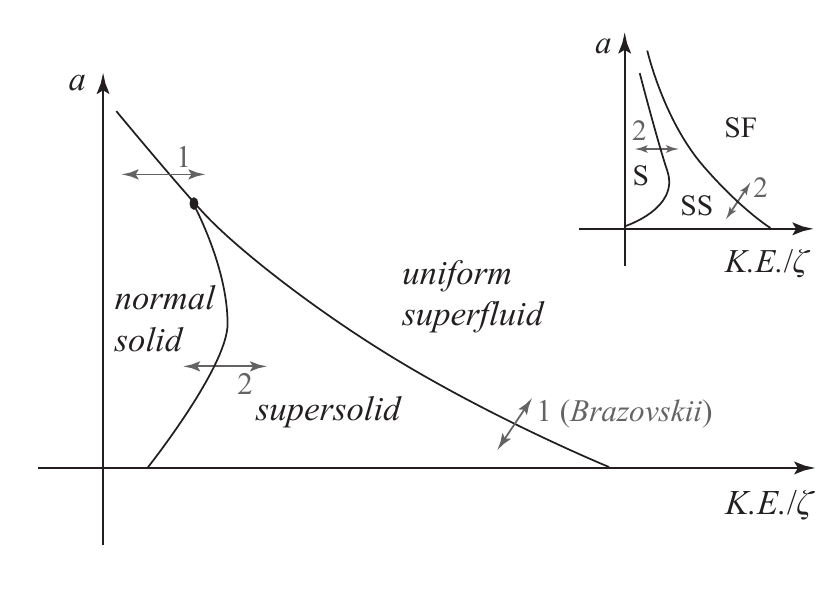}
		\caption{Schematic zero-temperature (i.e., quantum) phase diagram for a BEC in a concentric cavity, with the control parameters being the atomic scattering length $a$ and the inverse effective atom-cavity coupling $\zeta^{-1}$ (or equivalently the inverse laser intensity $\Omega^{-2}$).  For weak, repulsive interactions, the superfluid first undergoes self-organization via the Brazovskii transition, thus forming a supersolid.  If the laser intensity is increased further, the supersolid undergoes a transition into a normal solid (i.e., a Mott insulator). However, for strong, repulsive interactions, the uniform BEC can lose phase coherence concurrently with the first-order self-organization transition.  This situation is to be contrasted with that for the case of a single-mode cavity (inset), in which there should always be a supersolid (SS) region separating the uniform fluid (SF) and normal solid (S) regions. First- and second-order transitions are marked (1) and (2) respectively.}
	\label{fig:phasediag}
\end{figure}

The BEC-to-Mott transition has recently been addressed---for the case of a single-mode cavity---in Ref.~\cite{morigi10}.  The case of the concentric cavity has several features in common with the single-mode cavity case; there is, however, one key difference, which follows from the difference in the character of the self-organization transitions in these two cases.  Consider the phase diagram in terms of the two physically adjustable parameters, viz., the pump laser strength $\Omega$ (or, equivalently, the effective coupling constant $\zeta$)
and the scattering length of the atoms, $a$ (which can be tuned, e.g., by approaching a Feshbach resonance).
In the case of a \textit{single-mode} cavity, the self-organization transition is continuous; therefore, by tuning the laser to sufficiently near threshold, the emergent lattice depth can be made arbitrarily small.  Accordingly, regardless of how large the scattering length might be, there is always a region in which the self-organized lattice is \textit{too} shallow to support a Mott insulator.  Put differently, there is always a region of the supersolid phase between the liquid (i.e., the uniform BEC) and the normal solid. By contrast, self-organization in a \textit{concentric} cavity occurs by means of a first-order, Brazovskii transition. The emergent lattice depth therefore jumps discontinuously to some nonzero value at the self-organization transition; if this minimum lattice depth is greater than that required to support a Mott insulator, it is possible to have a \textit{direct} liquid-to-Mott transition, without an intermediate supersolid phase~\cite{Note7}.
The phase structure of our atom-cavity system that results from the foregoing considerations is summarized in the schematic phase diagram shown in Fig.~\ref{fig:phasediag}.  Curiously, it has the same morphology as that predicted for the corresponding liquid-to-solid transitions of $^4$He~\cite{stoof} (but with temperature and pressure serving as the control parameters in the $^4$He case).

\section{Experimental Feasibility}\label{sec:expt}

In this section we discuss typical parameter values for which the Brazovskii transition, and the resulting self-organized state, should be observable in a \textit{multimode} cavity.  Recently, the self-organization transition was observed in a \textit{single-mode} cavity using Bose-Einstein-condensed $^{87}$Rb atoms.
The cavity was characterized by by the cavity QED parameters $(g,\kappa,\gamma) = 2\pi \times (10,1,3)$~MHz; and
the pump laser was detuned from the atomic resonance by $\Delta_A \approx 2\pi \times 10^{13}$~Hz~\cite{baumann}.
An appreciable photon population was observed in the cavity for $\Delta_C$ up to $2\pi \times 40$~MHz.
For these parameters, the mean-field value of the self-organization threshold $\Omega_{\mathrm{mf}}$
is approximately $ 2\pi \times 2$~GHz.
For a concentric cavity having these parameters, the range in frequency space of the fluctuation-dominated regime for the \textit{classical} self-organization transition is approximately $2\pi \times (2-20)$~MHz.  (The precise value depends on the healing-length parameter $\chi$, but only as $\chi^{-1/3}$; hence these results are not very sensitive to the choice of $\chi$.)
As previously noted~\cite{us}, this range is greater than the frequency width associated either with spontaneous decay or with the intensity-noise of a typical laser (converted to frequency units). For the \textit{quantum} transition, the fluctuation-dominated regime (for $^{87}$Rb at unit filling far from a Feshbach resonance) would be of order $0.1-1$ MHz; the width of this regime can, however, be extended by tuning the interparticle interaction through a Feshbach resonance. 

The nucleation rates, both thermal and quantal, can be expressed---for the case of the interacting system---as $U \exp[-\mathfrak{K} (\Omega_{\mathrm{th}} - \Omega_{\mathrm{th}}^{\mathrm{mf}}) /(\Omega - \Omega_{\mathrm{th}})]$. In this expression, $\Omega$ is the pump laser strength and $\mathfrak{K}$ is a number of order unity; the expression is only valid when the exponent exceeds unity, i.e., for $\Omega$ sufficiently near $\Omega_{\mathrm{th}}$. For $U$ far from a Feshbach resonance, this expression implies that, for the regime discussed in this work, in which nucleation proceeds via the formation of large, well-defined droplets, the average nucleation timescale should typically exceed the lifetime of the experiment (discussed in the next paragraph). It might therefore prove necessary to enhance $U$ by means of a Feshbach resonance in order to explore the physics of nucleation. Note that this does not imply that the self-organized state is \textit{inaccessible} away from a Feshbach resonance: rather, one expects that the laser strength would have to be increased (decreased) well past threshold before the system entered (left) the self-organized state. In other words, the process of self-organization should exhibit significant hysteresis. 

For the experimental parameters just given, Baumann et al.~\cite{baumann} found that the lifetime of the self-organized BEC (i.e., the timescale on which atom loss destroys the BEC) was approximately $10$~ms.
For detunings $\Delta_C > \kappa$, atom loss is largely due to spontaneous scattering, which occurs at a rate $R_\gamma$ proportional to $\gamma \Omega^2 / \Delta_A^2$.
The corresponding timescale is long enough to enable the observation of the family of phenomena discussed in the present work.  In order to perform a similar experiment with cavities in the \textit{weak}-coupling regime (i.e., for smaller $g$), one would have to ensure that the spontaneous decay rate $R_\gamma$ stays below $1$ kHz.
This would involve satisfying two conditions:
$g^2 N / \Delta_C > 10^3$ and
$\kappa \ll \Delta_C$.
One could do so by increasing the number of atoms in the BEC
or by increasing the finesse of the cavity mirrors (which would reduce $\kappa$).

\section{Systems of coupled layers and the origins of frustration}\label{sec:glassiness}

We have discussed how an equilibrium atomic cloud, confined by the pump laser to a plane near the equatorial plane of the cavity, spontaneously crystallizes globally into one of a family of degenerate quasi-checkerboard arrangements. Now let us consider an atomic cloud confined to a single plane \textit{away} from the equator of the cavity. In this case, spontaneous crystallization still occurs but, as we shall now explain, the particular checkerboard arrangement into which the atoms crystallize varies \textit{statically} across the plane---energetics demands, e.g., that the center and edge of the cloud crystallize in distinct arrangements. This is a consequence of \textit{frustration}: satisfying local energetic preferences introduces \lq\lq fault zones\rq\rq\ between locally ordered regions.

\begin{figure}
	\centering
		\includegraphics{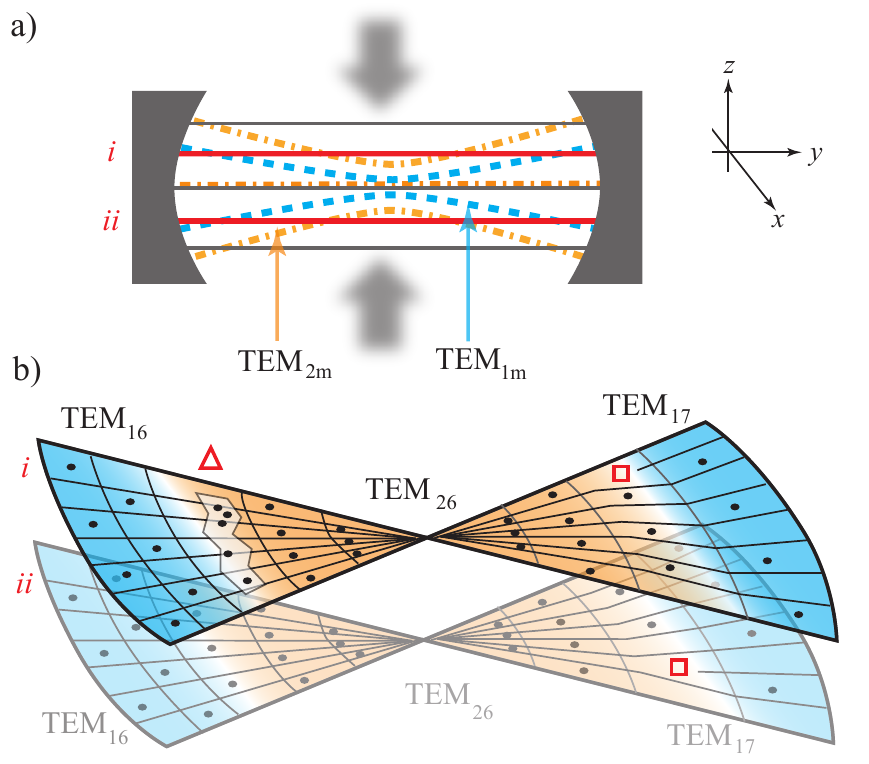}
	\caption{% 
Schematic illustration of the implications of frustration. 
Atoms are loaded into sheets~(i) and (ii), shown as thick lines in panel~(a), 
which are an integer number of pump-laser wavelengths apart.
The dashed and dashed-dotted curves are, respectively, antinodal regions of the modes $\mathrm{TEM}_{1m}$, 
which have low intensity near the centers of sheets~(i) and (ii), and $\mathrm{TEM}_{2m}$, 
which have low intensity away from the centers of sheets~(i) and (ii).
Near the center of each sheet, atoms crystallize into the $\mathrm{TEM}_{2m}$ modes; 
away from the center, they crystallize into the $\mathrm{TEM}_{1m'}$ modes.  
Within a sheet, regions may be separated by faults in the ordering, as illustrated in panel~(b).
For example, on the left side the fault has the form of a discommensuration (see Sec.~\ref{sec:glassiness}).  
By contrast, the fault on the right side is a grain boundary.
Between layers, the opposing parity of adjacent modes leads to frustration, which precludes ordering, as in the regions indicated by a $\triangle$ and a $\Box$.  
Grain boundaries (denoted by $\Box$) are more localized faults, and are therefore less costly, energetically, than discommensurations ($\triangle$).% 
}
\label{fig:frustra}
\end{figure}

In our analysis of equatorial-plane atomic distributions (see Fig.~\ref{fig:ordstb}), we were able to focus on the family of degenerate modes $\mathrm{TEM}_{lm}$ having $l = 0$.  To generalize our analysis beyond the equatorial plane, we must consider all modes that meet the degeneracy condition $l + m + n = \sbK R$.
Consider the situation illustrated in Fig.~\ref{fig:frustra}a, first focusing on the non-equatorial sheet marked~(i). Near the center of this sheet, crystallization into modes with $l = 1$ is suppressed because such modes have low intensity, whereas crystallization into $l = 2$ modes is favored because they have maximal intensity; away from the center, the opposite is true. The change in $l$ forces a change in $m$ or $n$, owing to the degeneracy condition, so the mode functions in the sheet must change across an interfacial zone between the $l = 1$ and $l = 2$ favored regions. Therefore, either a dislocation, associated with a change in $m$, or an abrupt change in lattice periodicity (i.e., a discommensuration), associated with a change in $n$, is expected. (This picture assumes that, as is always the case near threshold, the self-organized lattice is not strong enough to trap the entire atomic distribution at the center or the edge of the sheet. The kinetic energy cost of localization, as well as the cost in repulsive energy, will act to spread the atomic cloud out.)

Now consider a situation in which two symmetrically disposed sheets on opposite sides of the equator are populated with atoms, as in sheets~(i) and (ii) of Fig.~\ref{fig:frustra}.
Atoms in sheet~(i) and those in sheet~(ii) are coupled via the cavity modes.
Because $l = 2$ ($l = 1$) mode functions are symmetric (antisymmetric) about the equatorial plane, the atoms in the $l = 2$ ($l = 1$) arrangement in sheet~(ii) occupy the same (opposite) checkerboard as those in sheet~(i). If there are no dislocations, atoms in the interfacial zone remain disordered, because it is impossible for the atoms to satisfy both desiderata [or, equivalently, because the corresponding cavity modes interfere destructively in sheet~(i) and constructively in sheet~(ii)]. The introduction of dislocations enables the system to order in part of the interfacial zone, as shown in the right hand side of Fig.~\ref{fig:frustra}b, and is therefore preferred.

The full many-layer, many-mode system is expected to experience the same kinds of disordering effects as the idealization sketched above: i.e., one expects systems slightly above threshold to develop locally crystalline phases separated by zones riddled with faults.

\section{Concluding remarks}\label{sec:conclusions}

In this work, we have shown that the self-organization of BECs in multimode cavities is accompanied by a range of effects, such as fluctuation-driven nonequilibrium first-order transitions (both classical and quantal), topological defects, rigidity, frustration, and supersolidity.  We have developed both a nonequilibrium formalism for exploring such systems in general, and an effective equilibrium description valid in the regime of greatest interest, viz., the quantum phase transition undergone by a Bose-Einstein condensate.  We have outlined, moreover, how these formalisms may be used to compute the correlation functions of the photons emitted from the cavity, as well as those of the atoms, and how such correlations may be detected experimentally. Finally, we have suggested realistic values of experimental parameters that might be used to realize self-organization and its attendant phenomenology in the laboratory.

A major outstanding question that we hope to address in future work is whether the self-organized state in the layered three-dimensional geometry described in Sec.~\ref{sec:glassiness} exhibits glassy dynamics. Furthermore, we see three promising avenues for future work extending the ideas developed in this paper.
The first is to consider pumping with light sources that exhibit either thermal or quantum noise, instead of considering classical laser light.
Such light sources are likely to lead to nontrivial effects because of the interaction between the input noise and the environmental noise due to photon loss.
The second avenue is to apply the current formalism to related instabilities of atoms in cavities:
e.g., the phenomena of collective atomic recoil lasing (CARL)~\cite{slama} and ``excess noise''~\cite{domokos:noise}.
Such phenomena are well understood at the mean-field level, but the present formalism makes it straightforward to address the effects of fluctuations.
Finally, it might be of interest to explore the physics of many fermions trapped in multimode cavities, especially the prospects for their Cooper pairing.
There are two possible ways to arrange this: the first is to exploit the previously mentioned analogy (see Sec.~\ref{sec:expectations}) between photons and phonons in order to achieve photon-mediated Cooper pairing; the second is to use the phonons of the emergent crystalline phase as the ``glue'' that would bind the Cooper pairs.

\begin{acknowledgments}

We thank Helmut Ritsch and Brian DeMarco for discussions. This material is based upon work supported by the U.S.~Department of Energy, Division of Materials Sciences under Award No.~DE-FG02-07ER46453, through the Frederick Seitz Materials Research Laboratory at the University of Illinois at Urbana-Champaign (SG), by the U.S.~National Science Foundation under Awards DMR 09-06780 (PMG) and PHY 08-47469 (BLL), and by the U.S.~Air Force Office of Scientific Research under award FA9550-09-1-0079 (BLL).

\end{acknowledgments}

\appendix

\section{Toy equilibrium model}\label{app:toymodel}

\def\spd{{\phantom{\dagger}}}
In this Appendix we elaborate on the connection between quantum phase transitions in quasi-equilibrium cavity QED settings and the conventional theory of quantum phase transitions as level-crossings (as discussed, e.g., in  Ref.~\cite{sachdev}), by means of a toy model. This model has the following components:
(i)~a laser mode $a_L$ of frequency $\omega_L$,
(ii)~a cavity mode $a_C$ of wavevector $k$ and frequency $\omega_C = \omega_L + \Delta_C$, and
(iii)~an anharmonic atomic phonon $b$ of wavevector $k$ and frequency $\omega_p$.
The Hamiltonian for the model is given by:

\bea
H & = & \hbar \omega^{\spd}_L a^\dagger_L a^{\spd}_L + \hbar \omega^{\spd}_C a^\dagger_C a^\spd_C + \hbar \omega_p b^\dagger b + \lambda (b^\dagger b)^2 \nonumber \\
&& \qquad\qquad
+ \hbar \Gamma (a^\dagger_L a^\spd_C + \mathrm{h.c.})  (b^\dagger + b),
\eea
where $\Gamma$ is the photon-phonon coupling constant, and $\lambda$ is a parameter describing the strength of the anharmonicity of the phonon. In what follows we shall rescale all energies by $\hbar \omega_p$ in order to de-dimensionalize them; the dimensionless parameters will be denoted as $\tilde{\omega}_L$ etc. As $H$ commutes with the total number of photons $\mathcal{N} (\equiv a^\dagger_L a_L + a^\dagger_c a_c)$, one can diagonalize it in a space of fixed total photon number $\mathcal{N}$. (For sufficiently large $\mathcal{N}$, the difference between number and coherent states is irrelevant.)\thinspace\
Let us attempt to find the ground state for a particular value of $\mathcal{N}$. 
It is helpful to simplify further and treat the phonons as being ``classical'' by taking the commutator $[b^\dagger,b] = 0$; this is equivalent to rewriting $b = \sqrt{m \omega/2\hbar} [x + i (p/m\omega)]$ and taking the $m \rightarrow \infty$ limit. In this limit, all terms depending on $p$ in the Hamiltonian are suppressed; thus the Hamiltonian can be rewritten, in terms of $x \propto (b + b^\dagger)$, as 

\bea
H & = &
\tilde{\omega}^\spd_L ( a^\dagger_L a^\spd_L + a^\dagger_C a^\spd_L) +
\tilde{\Delta}^\spd_L \, a^\dagger_C a^\spd_L \nonumber \\
&& \quad +  x^2 + \tilde{\lambda} x^4 + 2 \tilde{\gamma} x (a^\dagger_C a^\spd_L + \mathrm{h.c.}).
\eea
In this expression the prefactors accompanying $x$ have been absorbed as appropriate into the various coupling constants. The first term is simply a constant times $\mathcal{N}$, and can be ignored for our purposes.
As $x$ is also now a good quantum number, it is now possible to diagonalize $H$ in a manifold of fixed $\mathcal{N}$ and $x$. This can be achieved via canonical transformation from from $a_{L/C}$ to $A_{1/2}$:
\begin{subequations}
\begin{eqnarray}
a_L & = & \alpha A_1 + \sqrt{1 - \alpha^2} A_2, \\
a_C & = & - \sqrt{1 - \alpha^2} A_1 + \alpha A_2,
\end{eqnarray}
\end{subequations}
where

\beq
\alpha^2 = \frac{1}{2} \left(1 \pm \frac{\tilde{\Delta}_C}{\sqrt{4 \tilde{\gamma}^2 x^2 + \tilde{\Delta}_C^2}} \right).
\eeq
In terms of $A_1$ and $A_2$, the Hamiltonian $H$ has the form

\beq
H = \Delta_C \mathcal{N} + x^2 + \tilde{\lambda} x^4 + \frac{4 \tilde{\gamma}^2 x^2 - \tilde{\Delta_C}^2}{\sqrt{\tilde{\Delta}_C^2 + 4 \tilde{\gamma}^2 x^2}} (A_1^\dagger A_1 - A_2^\dagger A_2).
\eeq
Expanding in powers of $x$, one finds that the coefficient of $x^2$ in the Hamiltonian is now given by

\beq
1 + (A_1^\dagger A_1 - A_2^\dagger A_2) \frac{6 \tilde{\gamma}^2}{\tilde{\Delta}_C}.
\eeq
For a fixed $\mathcal{N}$, the smallest value that the last term can attain is

\beq
- \frac{6 \tilde{\gamma}^2}{\tilde{\Delta}_C} \mathcal{N}.
\eeq
Thus, when $6 \tilde{\gamma}^2/\tilde{\Delta}_C \leq 1/ \mathcal{N}$, the coefficient of $x^2$ is always positive; hence, the ground state always lies in the $x = 0$ manifold. However, for $6 \tilde{\gamma}^2/\tilde{\Delta}_C > 1/ \mathcal{N}$, the ground state is that in which $A_2^\dagger A_2 = \mathcal{N}$, and

\beq
x^2 = \frac{\frac{6 \tilde{\gamma}^2}{\tilde{\Delta}_C} \mathcal{N}-1}{2 \tilde{\lambda}}.
\eeq
Hence, as $\tilde{\gamma}$ is increased, the ground-state value of $x$ (for fixed $\mathcal{N}$) becomes nonzero, singularly.  In other words, the lowest level of the $x = 0$ manifold and that of an $x \neq 0$ manifold cross at the critical value of $\tilde{\gamma}$: such a level crossing is known as a ``quantum phase transition''~\cite{sachdev}. Although the phenomenon of self-organization in a multimode cavity involves substantially more than a single photon and a single phonon, its essential character in the equilibrium limit, for a classical laser, is also that of a ground-state level-crossing occurring at some fixed photon number $\mathcal{N}$.

\section{Determining the healing length parameter $\chi$}\label{app:chi}

The parameter $1/\chi$, which sets the healing length for crystallinity, is a measure of how weak the atomic coupling $\zeta_{lmn}$ to $l \neq 0$ modes is, relative to the coupling to $l = 0$ modes. The value of $\chi$ is
determined by the following effects:
(1)~the atoms, being confined near the equatorial plane of the cavity, couple most strongly to modes that have the highest amplitude there, and thus to the lowest-order modes along the $z$ direction (i.e., $g$ is in effect a function of $l$ once $\Xi(\mathbf{x})$ is projected onto the equatorial plane);
(2)~higher-order modes along the $z$ direction have lower finesse and hence couple more weakly to the atoms (i.e., $\kappa$ is a function of $l$);
(3)~the effective laser-cavity detuning, $\Delta_C - g^2 N / \Delta_A$, is a function of $l$ because $g$ is;
and (4)~for a \textit{nearly} concentric (or confocal, planar, etc.) cavity---the experimentally relevant case---$\Delta_C$ is a function of $l$ because higher-order modes have lower resonant frequencies. Thus, $\zeta$ is in general a complicated function of $l$.

For the specific case of a concentric cavity, one can generalize the mode structure given in Sec.~\ref{sec:modestructure} by imposing, e.g., finite-well rather than hard-wall boundary conditions on $\theta$. In this case, it is clear that both $g_l / g_0$ and $\kappa_l / \kappa_0$ (in which the subscripts denote the appropriate value of $l$) must go approximately as $(1 - \mathrm{const} \times l^2)$, where the constant is approximately $1 / l_{\mathrm{max}}^2$, i.e., the number of higher-order modes that are of sufficiently high finesse to couple significantly to the atoms. Thus, $\chi$ can be taken to be approximately $1/l_{\mathrm{max}}^2$, up to a factor of order unity; the physically relevant quantities $\xi_0$ and $\Omega_{\mathrm{th}}$, which depend on $\chi^{1/3}$ and $\chi^{1/4}$ respectively, should not be sensitive to this neglected factor.

Note that this analysis ignores effect~(4). According to Eq.~(19.24) of Ref.~\cite{siegman}, this term would lead to a frequency shift $\Delta_C(l) / \Delta_C(0) \propto -l$, with a proportionality constant depending on the distance from concentricity. Thus, $\zeta$ should have a \textit{positive} linear term in $l$, in addition to the negative quadratic term; the potential effect of such a term is to favor some family of $l = l_0 \neq 0$ modes, even at the equator, over $l = 0$ ones.
Similar considerations apply to other geometries, such as the confocal or planar cavities.

\section{Renormalization-group flow equations for the quantum Brazovskii model}\label{app:RGequations}

In this Appendix we outline our derivation of the renormalization-group (RG) flow equations for the quantum Brazovskii model discussed in Sec.~\ref{sec:qbraz}. Our procedure parallels that discussed in App.~A of Ref.~\cite{swift:bubbles} for the \textit{classical} Brazovskii model; that work, in turn, was based on techniques developed by Shankar~\cite{shankar} for the Fermi liquid. The objective of the renormalization-group procedure is to arrive at a spatially coarse-grained effective theory in terms of modes for which $m + n \approx \Lambda_0$. This is done by progressively integrating out modes for which $|m + n - \Lambda_0| \geq \rgscale$, where $\rgscale$ is referred to as the renormalization-group ``scale\rlap.''
The microscopic, or ``bare\rlap,'' theory has an RG scale that is associated with the physical high-energy cutoff (e.g., $\Delta_A$ in the cavity QED case); this scale is progressively decreased by the  integrating out of ``shells'' of modes, i.e., modes for which $\rgscale_{\mathrm{new}} \leq |m + n - \Lambda_0| \leq \rgscale_{\mathrm{old}}$, thus yielding effective theories involving progressively fewer modes.
As one integrates out these shells of modes, the theory maintains its basic structure, but the various coupling constants flow; therefore, the coupling constants are functions of $\rgscale$.  For example, the microscopic values of the parameters, $\mathcal{R}$ and $\mathcal{U}$, are their values at a value of $\rgscale$ determined by $\Delta_A$, whereas the fully coarse-grained parameters $r$ and $u$ are those corresponding to $\rgscale = 0$. At each step in the RG procedure, we integrate out all the \textit{frequency} components associated with that spatial shell; in making this choice [which has the advantage of preserving the causality structure of the microscopic theory (as mentioned in Ref.~\cite{mitra06})] we are following Refs.~\cite{mitra06, shankar}.

\begin{figure}
	\centering
		\includegraphics{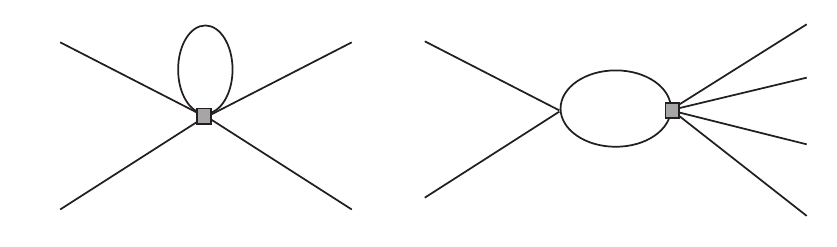}
	\caption{Feynman diagrams involving the six-point vertices associated with the couplings $w_i$ ($i = 1,2,3$), to one-loop order. The six-point vertices are denoted as grey squares. The diagram shown on the left renormalizes the four-point vertices; that on the right renormalizes the six-point vertices.}
	\label{fig:appdiag}
\end{figure}

We begin with the observation that the action can be written as follows:

\beq
S = S_2 + S_4 + \cdots,
\eeq
where

\bea
S_2 & = & \frac{(2\pi)^d}{2} \int d1\,r(1) \, \Phi(1) \,\Phi(1), \\
S_4 & = & \frac{(2\pi)^d}{4!} \int d1\,d2\,d3\,d4\,u(1234) \, \Phi(1)\, \Phi(2)\, \Phi(3)\, \Phi(4)
\nonumber\\
& & \qquad\qquad\times\delta(1 + 2 + 3 + 4),
\nonumber
\eea
and notation of the form $d1$ is to be interpreted in the following way:

\beq
\int d1 \equiv \frac{1}{(2\pi)^d} \int_{|m + n - \Lambda_0| < \rgscale} \Lambda_0^{d - 1} d\eta \int_{-\infty}^{\infty} d\omega \int d\theta,
\eeq
where $\eta$ and $\theta$ are, respectively, the quasi-radial and quasi-angular variables discussed in Sec.~\ref{sec:Onmodel}, and are treated here as continuous.
We now perform the RG transformation, which involves two steps:
(i)~integrating out all modes satisfying 
$\rgscale / b \leq |m + n - \Lambda_0| \leq \rgscale$ for $b = 1 + l$ with $l \ll 1$; and
(ii)~rescaling the spatial coordinates and the fields in the action so that the new action is similar in form to its predecessors under the RG transformation.  Under this rescaling, $\eta \rightarrow b \eta$ and $\Phi \rightarrow \Phi / b$.  Under the combined effects of integrating out the shell of modes (at one-loop order, via the diagram in Fig.~\ref{fig:brazdiag}) and the rescaling, the coefficient of the quadratic term transforms as follows:

\beq
r({\rgscale}) \longrightarrow r({\rgscale/b}) = b^2 \big(r({\rgscale}) + \Delta_2\big),
\eeq
where

\beq
\Delta_2 \equiv \alpha\,u_{1}({\rgscale}) 
\int_{\rgscale/b}^{\rgscale} d\eta\,
\frac{1}{r({\rgscale}) + \eta^2}.
\eeq
By setting $b = 1 + l$ and differentiating with respect to $l$, 
we arrive at a differential version of the RG equation for $r$, viz.,

\beq
\frac{dr}{dl} = 2r(\rgscale) + 
\frac{\rgscale\,\alpha u_{1}(\rgscale)}{\sqrt{r(\rgscale) + \rgscale^2}}.
\eeq
We have introduced the notation $u_{1}$ instead of $u$ because, as we shall now see 
(and as we anticipated in Sec.~\ref{sec:cbraz}), 
the renormalized value of $u$ depends on whether the four $\theta$s entering the vertex are identical or not.
The \textit{generic} vertex, in which the four values of $\theta$ are not all identical, is associated with the coupling $u_{1}$; the special vertex, in which the four values of $\theta$ are all identical, is associated with the coupling $u_{2}$. Similarly, we shall denote the six-point vertices respectively as $w_{1}$, $w_{2}$, and $w_{3}$, depending on whether two, four, or all six of the incoming propagators have identical values of $\theta$. The RG equations for $(u_{1},u_{2})$ can be derived via the same procedure as those for $r$; they read as follows:

\bea
\frac{du_1}{dl} & = & 2 u_1 - 
\frac{\alpha u_1^2 \rgscale}
{r(\rgscale) + \rgscale^2} + 
\frac{\alpha w_1 \rgscale}
{\sqrt{r(\rgscale) + \rgscale^2}}, 
\\
\frac{du_2}{dl} & = & 2 u_2 - 
2 \frac{\alpha u_1^2 \rgscale}{r(\rgscale) + \rgscale^2} + 
  \frac{\alpha w_2 \rgscale}{\sqrt{r(\rgscale) + \rgscale^2}}.
\eea
Similar equations can be derived for the three $w$ couplings; these involve the diagrams in Fig.~\ref{fig:brazdiag}d and Fig.~\ref{fig:appdiag}.  Before writing these down, however, we introduce the following convenient changes of variables, which serve to de-dimensionalize the RG equations (following Ref.~\cite{swift:bubbles}):

\bea
\newrgscale                       & \equiv & \rgscale(\alpha\,\mathcal{U})^{-1/2}\,e^{-l}, \\
\overline{r}(\newrgscale)         & \equiv & r(l) (\alpha\,\mathcal{U})^{-1}\,e^{-2l}, \\
\overline{u}_{1,2}(\newrgscale)   & \equiv & u_{1,2}(l)\,\mathcal{U}\,e^{-2l}, \\
\overline{w}_{1,2,3}(\newrgscale) & \equiv & w_{1,2,3}(l)\,\mathcal{U}^{-1}\,\alpha^2 e^{-2l}.
\eea
In terms of the new variables, the full set of RG equations is as follows:
\bea
\frac{d\overline{r}}{d\newrgscale}   & = & - \frac{\overline{u}_1}{(\rgdenom)^{1/2}}, \\
\frac{d\overline{u}_1}{d\newrgscale} & = & \frac{\overline{u}_1^2}{\rgdenom} - \frac{\overline{w}_1}{(\rgdenom)^{1/2}}, \\
\frac{d\overline{u}_2}{d\newrgscale} & = & \frac{2 \overline{u}_1^2}{\rgdenom} - \frac{\overline{w}_2}{(\rgdenom)^{1/2}}, \\
\frac{d\overline{w}_1}{d\newrgscale} & = & - \frac{2 \overline{u}_1^3}{(\rgdenom)^{3/2}} + \frac{3 \overline{u}_1 \overline{w}_1}{\rgdenom}, \\
\frac{d\overline{w}_2}{d\newrgscale} & = & - \frac{4 \overline{u}_2^3}{(\rgdenom)^{3/2}} + \frac{\overline{u}_2 \overline{w}_2}{\rgdenom} + \frac{4 \overline{u}_1 \overline{w}_1}{\rgdenom}, \\
\frac{d\overline{w}_3}{d\newrgscale} & = & - \frac{12 \overline{u}_1^3}{(\rgdenom)^{3/2}} + \frac{8 \overline{u}_1 \overline{w}_2}{\rgdenom}.
\eea
The appropriate microscopic values (i.e., initial conditions for these equations) are as follows: 
$(\overline{r}(\infty), \overline{u}_{1,2}(\infty), \overline{w}_{1,2,3}(\infty)) = (\overline{\mathcal{R}}, 1, 0)$. 
Numerically integrating the equations yields the coarse-grained parameters, 
which are plotted in Fig.~\ref{fig:mathoutput} and discussed in the main text.

\section{Effective temperatures}\label{app:effectiveT}

In this Appendix we briefly explain why, and for what purposes, $\tilde{\kappa}$ in Sec.~\ref{sec:neqc} behaves as an effective temperature.  The basic result we shall review is as follows: consider a quadratic Keldysh action that has a $q-q$ component of the (low-frequency) form

\beq\label{eq:efft1}
i \sum_\nu \Gamma_\nu \int d\omega\,\phi_{\nu,q}(\omega)\,\phi_{\nu,q}(-\omega),
\eeq
for some real set of parameters $\Gamma$. (The generic set of indices parameterized by $\nu$ can describe positions, momenta, mode indices, etc.)\thinspace\ The long-time dynamics of such a theory can then be described by a Langevin equation, having a white noise term of strength $\Gamma$.

In what follows, we shall suppress the $\nu$ index; the argument, which is adapted from Ref.~\cite{kamenev}, can be made independently for each $\nu$.  First, note that Eq.~(\ref{eq:efft1}) is expressed in the time domain as

\beq
i \Gamma\int d\omega\,\phi_q(t)\,\phi_q(t).
\eeq
This term appears in the system's partition function 
$Z\equiv\int D\phi_q(t)\,D\phi_c(t)\,\exp iS$, 
in the form
\beq\label{eq:app4eq3}
Z = \int D\phi_q(t)\,\exp\left( - \Gamma \int dt\,\phi_q(t)\,\phi_q(t) \right) \times \cdots,
\eeq
in which the ellipses denote factors resulting from other terms in the action.  Eq.~(\ref{eq:app4eq3}) above can be rewritten, by means of a Hubbard-Stratonovich transformation, as

\beq
\int D\phi_q(t)\,D\xi(t)\,
\exp\left\{ - \int dt\,\left( \frac{\xi(t)^2}{\Gamma} - 2i \xi(t)\,\phi_q(t)\right) \right\}\times \cdots .
\eeq
Once this is done, the full partition function, which also includes terms \textit{linear} in $\phi_q$ (from the retarded and advanced components of the Keldysh action) can be written as follows:

\bea
&& \!\!\!\!\!\!\! Z = \int D\xi(t)\,D\phi_c(t) \exp\left( - \frac{1}{\Gamma} \int dt \xi(t)^2\right) \\
&&
% \,\,\,
\times\int D\phi_q(t)\,
\exp\left\{ i\!\!\int\!\! dt\,dt' [\phi_c(t')\,G(t',t) - \xi(t)]\phi_q(t) \right\} \nonumber,
\eea
where $G$ represents some (unspecified) integral kernel that couples $\phi_c$ and $\phi_q$. 
Note that the $c-c$ term in the action is absent, via causality, as discussed in Ref.~\cite{kamenev}.
If one now integrates out $\phi_q$, one finds that the partition function is given by

\bea
&& Z = \int D\xi(t)\,D\phi_c(t)\,\exp\left( - \frac{1}{\Gamma} \int dt\,\xi(t)^2\right) \\
&& \qquad\qquad\qquad\times\delta\left[ \phi_c(t)\,G(t,t') - \xi(t') \right]. \nonumber
\eea
Accordingly, the dynamics of the system is described, at long times, by a sum over classical histories in the presence of a Langevin white-noise term $\xi$, the fluctuations of which are given by

\beq
\langle \xi(t)\,\xi(t') \rangle = \Gamma\,\delta(t - t').
\eeq
It follows that the long-time dynamics of the system is \textit{classical} rather than quantal; thus any phase transition that the system undergoes is a thermal rather than a quantum phase transition.  By contrast, for any system that undergoes a true \textit{quantum} phase transition, the coefficient of the $q-q$ component of the Keldysh action \textit{vanishes} at low frequencies, typically as a power law, $|\omega|^\alpha$ (see, e.g., Ref.~\cite{mitra06}). This corresponds to power-law decay of noise correlations in the time domain.

A closely analogous argument (see, e.g., Ref.~\cite{kamenev}) shows that, assuming the $q-q$ component of the action is frequency-independent as $\omega \rightarrow 0$, the rate of escape from a metastable state is given by an effective Arrhenius formula with a temperature proportional to $\Gamma$.

%\bibliographystyle{apsrev4-1}
%\bibliography{prabib}
%Merlin.mbs v4.21 2009-07-09.
%

\end{document}